\documentclass[12pt]{article}

\usepackage{graphics,graphicx,fullpage,natbib,multirow}
\usepackage{amsmath,amssymb,verbatim,epsfig}
\usepackage[dvipsnames,usenames]{color}
\usepackage[normalem]{ulem}
\usepackage{soul}
\usepackage{xcolor}

\newtheorem{defin}{\bf Definition}

\def\ga{\mbox{Ga}}

\def\be{\mbox{Be}}

\def\no{\mbox{N}}

\def\un{\mbox{Un}}

\def\E{\mbox{E}}

\def\P{\mbox{P}}

\def\pd{\mbox{PD}}

\def\d{\mbox{d}}

\def\data{\mbox{data}}

\def\bu{{\bf u}}
\def\bv{{\bf v}}

\def\bU{{\bf U}}

\def\bX{{\bf X}}

\def\simind{\stackrel{\mbox{\scriptsize{ind}}}{\sim}}
\def\simiid{\stackrel{\mbox{\scriptsize{iid}}}{\sim}}

\newcommand{\bpi}{\boldsymbol{\pi}}
\newcommand{\btheta}{\boldsymbol{\theta}}
\newcommand{\bvartheta}{\boldsymbol{\vartheta}}

\newcommand{\RB}{\mathbb{R}}

\begin{document}

\baselineskip=24pt

\title{\bf Bayesian nonparametric mixtures of Archimedean copulas}
\author{
{\sc Ruyi Pan, Luis E. Nieto-Barajas, Radu V. Craiu} \\[2mm]
{\sl University of Toronto \& ITAM} \\[2mm]
{\small {\tt ruyi.pan@mail.utoronto.ca, luis.nieto@itam.mx \& radu.craiu@utoronto.ca}}
}
\date{}
\maketitle

\begin{abstract}
Copula-based dependence modeling often relies on parametric formulations.  This is mathematically convenient, but can be statistically inefficient when the parametric families are not suitable for the data and model in focus.  A Bayesian nonparametric mixture of Archimedean copulas is introduced to increase the flexibility of copula-based dependence modeling.  Specifically, the Poisson-Dirichlet process is used as a mixing distribution over the Archimedean copulas' parameter. Properties of the mixture model are studied for the main Archimedean families, and posterior distributions are sampled via their full conditional distributions. The performance of the model is illustrated via numerical experiments involving simulated and real data. 
\end{abstract}

\vspace{0.2in} \noindent {\sl Keywords}: Archimedean copula, Bayesian nonparametrics, mixture model, multivariate dependent model.

\section{Introduction}
\label{sec:intro}

Let $\bX=(X_1,X_2,\ldots,X_p)$ be a continuous $p$-dimensional random vector with joint cumulative distribution function (CDF) $F(x_1,\ldots,x_p)$ and marginal CDFs $F_j(x)$, for $j=1,\ldots,p$. Following \cite{sklar:59}, there exists a unique multivariate copula function $C(u_1,\ldots,u_p)$, with $C:[0,1]^p\to[0,1]$, that satisfies the conditions to be a proper CDF with uniform marginals, such that $F(x_1,\ldots,x_p)=C(F_1(x_1),\ldots,F_p(x_p))$.  

The dependence or association measures between any two random variables $(X_j,X_k)$, independently of their marginal distributions, can be written entirely in terms of the copula. For instance,  Kendall $\tau$ between the $j$-th and $k$-th components of $\bX$ is defined as $\tau=4\E\{C_{jk}(U_j,U_k)\}-1$, or
\begin{equation}
\label{eq:tau}
\tau=4\int_0^1\int_0^1 C_{jk}(u_j,u_k)f_{jk}(u_j,u_k)\d u_j \d u_k-1,
\end{equation}
where $C_{jk}$ is the marginal bivariate copula of $(U_j,U_k)$ and $f_{jk}$ is the corresponding bivariate density \citep[e.g.][]{nelsen:06}. Clearly, flexible models for the copula are beneficial because they capture complex dependence patterns and can return accurate estimates for dependence measures of interest. Moreover, Bayesian methods are perfectly suited to incorporate all the uncertainty incurred from estimating marginals and the dependence structure \citep{levi2018bayesian}.

Copulas are often modeled in statistical applications using parametric families such as those included in the large class of Archimedean copulas \citep[e.g.][]{nelsen:06}.  A Bayesian semiparametric version of an Archimedean copula was introduced by \cite{hoyos&nieto:20}. Nonparametric copulas are, for example, the empirical copula \citep{deheuvels:79}, the sample copula \citep{gonzalez&hoyos:18}, and a Bayesian counterpart of the sample copula \citep{nieto&hoyos:24}. 

Bayesian nonparametric models for multivariate data usually rely on mixtures of a multivariate normal density mixing over the mean vector and/or the covariance matrix via a Bayesian nonparametric model. In particular, \cite{carmona&al:19} used a multivariate normal location mixture for the clustering of mixed-scale data, and \cite{kottas&al:05} used a location-scale mixture of multivariate normals to model multivariate ordinal data. 

In this paper, we consider Bayesian nonparametric mixtures of Archimedean copulas in which the mixing distribution is the two-parameter generalization of the Dirichlet process, called the Poisson-Dirichlet process, introduced by \cite{pitman&yor:97}. This extension is directed towards two important goals. First, it considerably extends the range of dependence patterns that can be modeled using Archimedean copulas, making it useful for capturing complex dependencies. Second, in the case of heterogeneous populations, it clusters the sample based on information contained in the marginals and the dependence structure. Those familiar with Bayesian nonparametric models with Gaussian distributions \citep[e.g.][]{mueller&al:96} will recognize that our model is a generalization, since the Archimedean copulas can accommodate different marginals and will have different tail behavior from that of a bivariate Gaussian distribution, e.g. they are asymmetric.

The content of the rest of the paper is as follows: In Section \ref{sec:acopulas} we introduce notation and briefly review the Archimedean copula family and characterize it by the copula density required in the mixture model. Section \ref{sec:mixmodel} presents the mixture model, a study of its properties, and a guide for sampling the posterior. In Section \ref{sec:numerical}, we investigate the performance of our model using numerical experiments. The paper ends with conclusions and future directions for research.

\section{Archimedean copula densities}
\label{sec:acopulas}

To proceed, we must first introduce some notation. Let $\un(\alpha,\beta)$ denote a uniform density in the continuous interval $(\alpha,\beta)$; $\ga(\alpha,\beta)$ denote a gamma density with mean $\alpha/\beta$; $\be(\alpha,\beta)$ denote a beta density with mean $\alpha/(\alpha+\beta)$; $\no(\mu,\lambda)$ denotes a normal density with mean $\mu$ and precision $\lambda$. The density evaluated at a specific point $x$ will be denoted, for example, in the gamma case, as $\ga(x\mid\alpha,\beta)$.

As mentioned above, a $p$-dimensional copula $C$ is a multivariate cumulative distribution function (CDF) with uniform marginals. One of the richest classes of copulas is the so-called Archimedean family. This family is defined by a continuous, decreasing and convex generator function $\phi$ such that $\phi:[0,1]\to\RB^+$,  $\phi(0)=\infty$, $\phi(1)=0$. Specifically, the copula with generator $\phi$ is defined as 
\begin{equation}
\label{eq:ac}
C(u_1,\ldots,u_p)=\phi^{-1}\{\phi(u_1)+\cdots+\phi(u_p)\}.
\end{equation}

According to \cite{mcneil&neslehova:09}, an Archimedean copula $C$ admits a density $f_C$ if and only if the $(p-1)$th derivative of $\phi^{-1}$, denoted as $\phi^{-1(p-1)}$, exists and is absolutely continuous on $(0,\infty)$. In this case, the density is given by
\begin{equation}
\label{eq:cdens}
f_C(u_1,\ldots,u_p)=\phi^{-1(p)}\left\{\sum_{j=1}^p\phi(u_j)\right\}\prod_{j=1}^p\phi^{(1)}(u_j),
\end{equation}
where $\phi^{(1)}$ denotes the first derivative of $\phi$. If all derivatives $\phi^{-1(j)}$ of $\phi^{-1}$ exist, they must satisfy $$(-1)^{j}\phi^{-1(j)}(u)\geq 0,$$
for $j=1,\ldots,p$. In such a case, it is said that $\phi^{-1}$ is completely monotonic \citep{wu&al:07}.

Using the relationship between derivatives of inverse functions, the copula density, for $p=2$, can be written in terms of derivatives of the generator as 
\begin{equation}
\label{eq:dens2}
f_C(u_1,u_2)=-\frac{\phi^{(1)}(u_1)\phi^{(1)}(u_2)\phi^{(2)}\{C(u_1,u_2)\}}{\left[\phi^{(1)}\{C(u_1,u_2)\}\right]^3},
\end{equation}
where $C(u_1,u_2)$ is given in \eqref{eq:ac} and $\phi^{(j)}$ is the $j$-th derivative of $\phi$.

Generators $\phi$ usually belong to families that are parameterized in terms of a single parameter $\theta\in\Theta$. Therefore, we will use the notation $\phi_\theta(t)$ for the parametric generator and $C(\bu\mid\theta)$ for the copula. We consider here five widely used members of the Archimedean family: Ali–Mikhail–Haq (AMH), Clayton (CLA), Frank (FRA), Gumbel (GUM) and Joe (JOE). The generators associated with each of these families are given in Table \ref{tab:afam}, together with their parameter space and Kendall tau. The first and second derivatives of the generators $\phi_{\theta}(t)$, required to compute the densities of the bivariate copula, as in \eqref{eq:dens2}, are given in Table \ref{tab:derivatives}. 

\begin{figure}
\centerline{\includegraphics[scale=0.4]{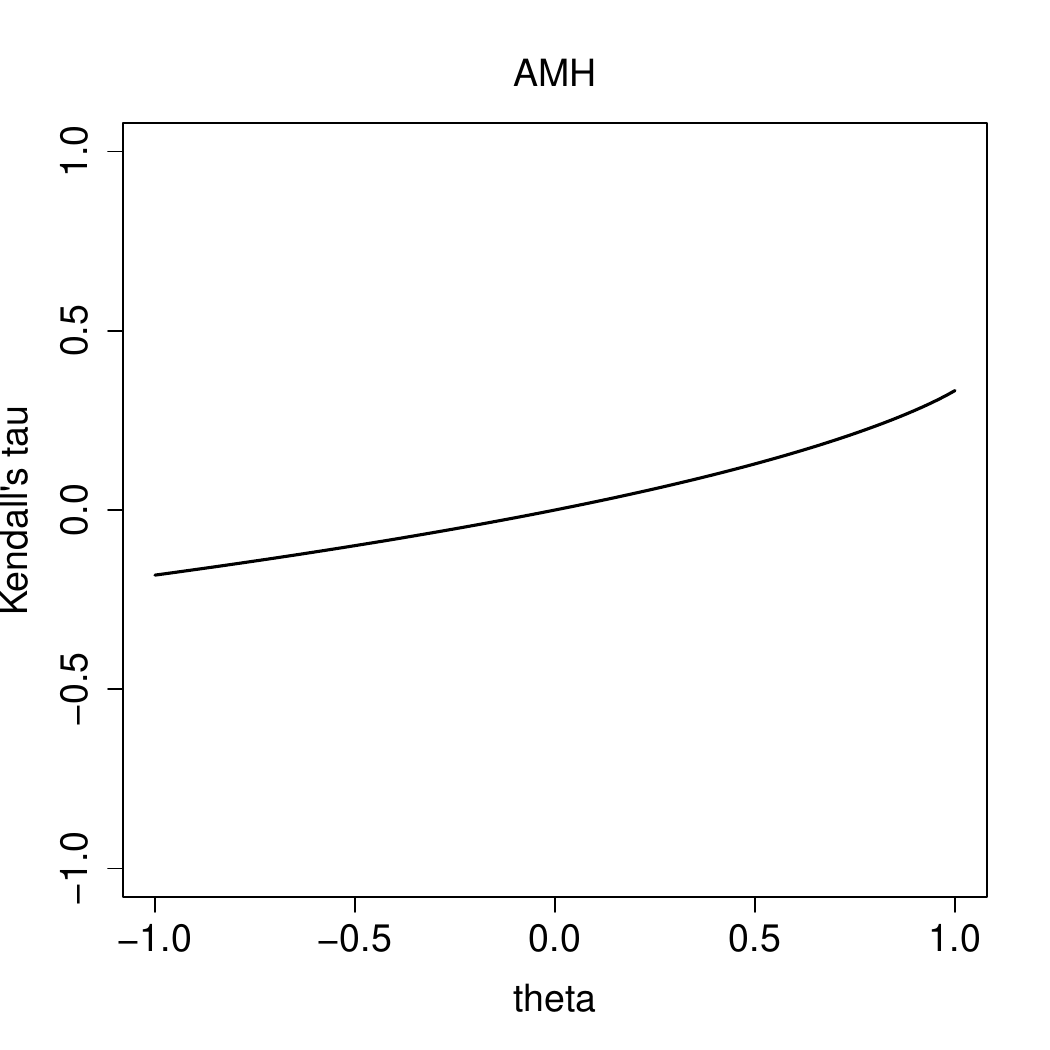}
\includegraphics[scale=0.4]{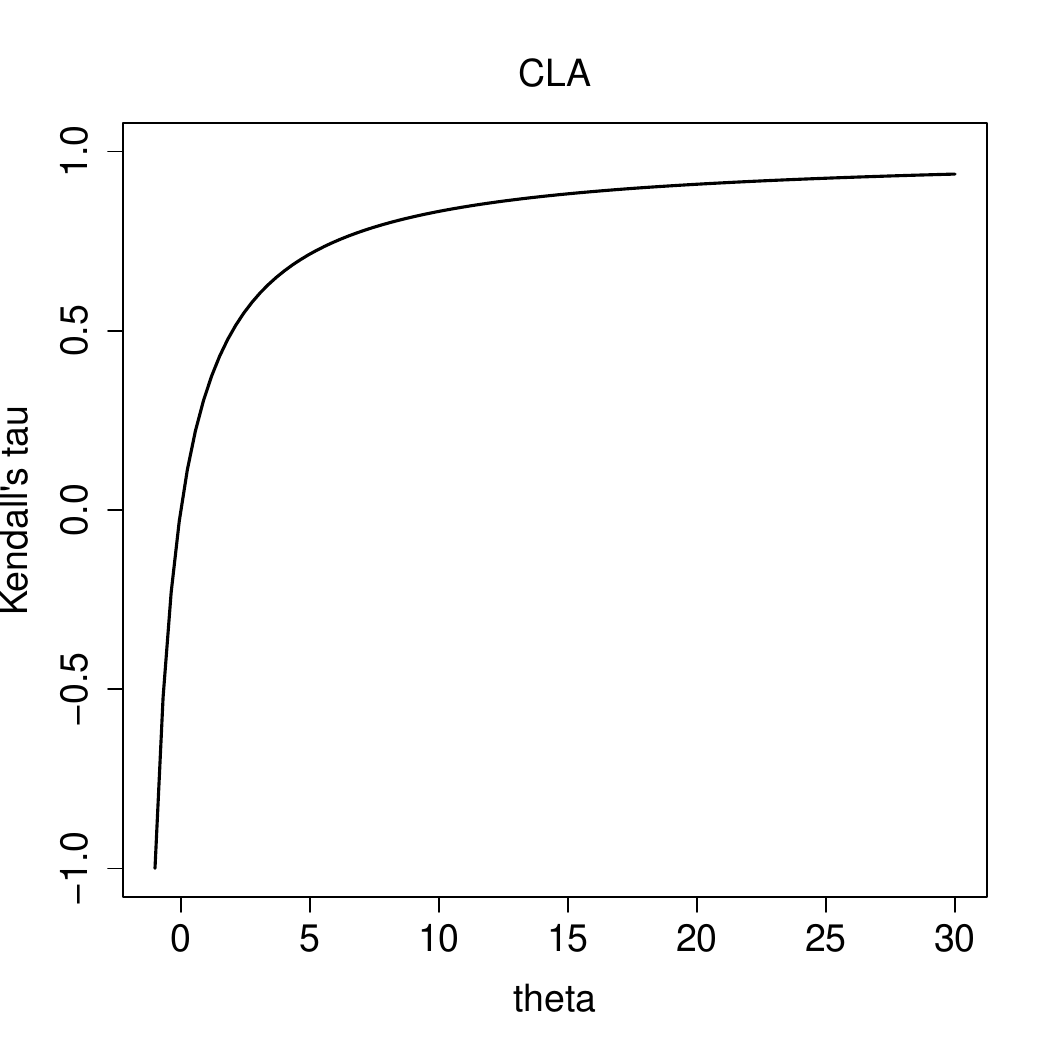}}
\vspace{-2mm}
\centerline{\includegraphics[scale=0.4]{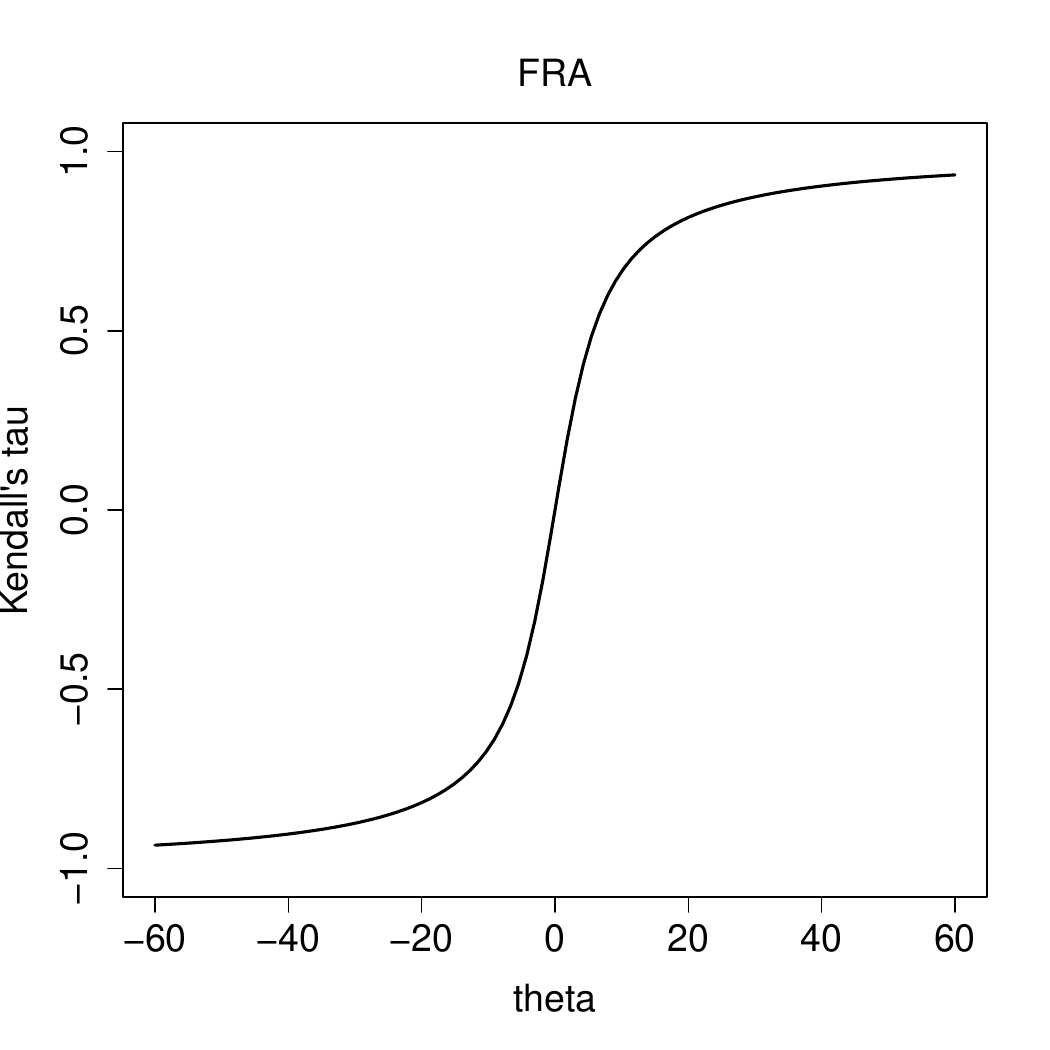}
\includegraphics[scale=0.4]{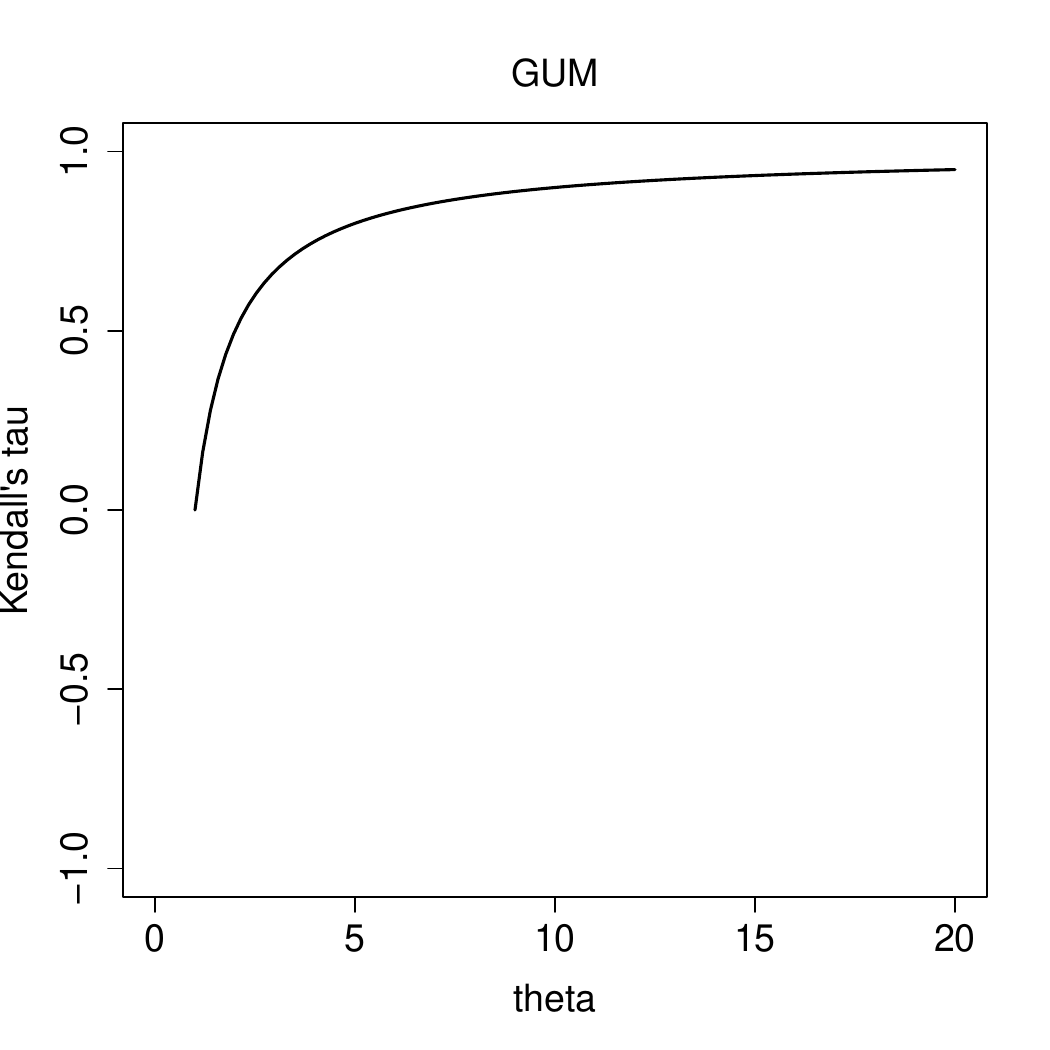}}
\vspace{-2mm}
\centerline{\includegraphics[scale=0.4]{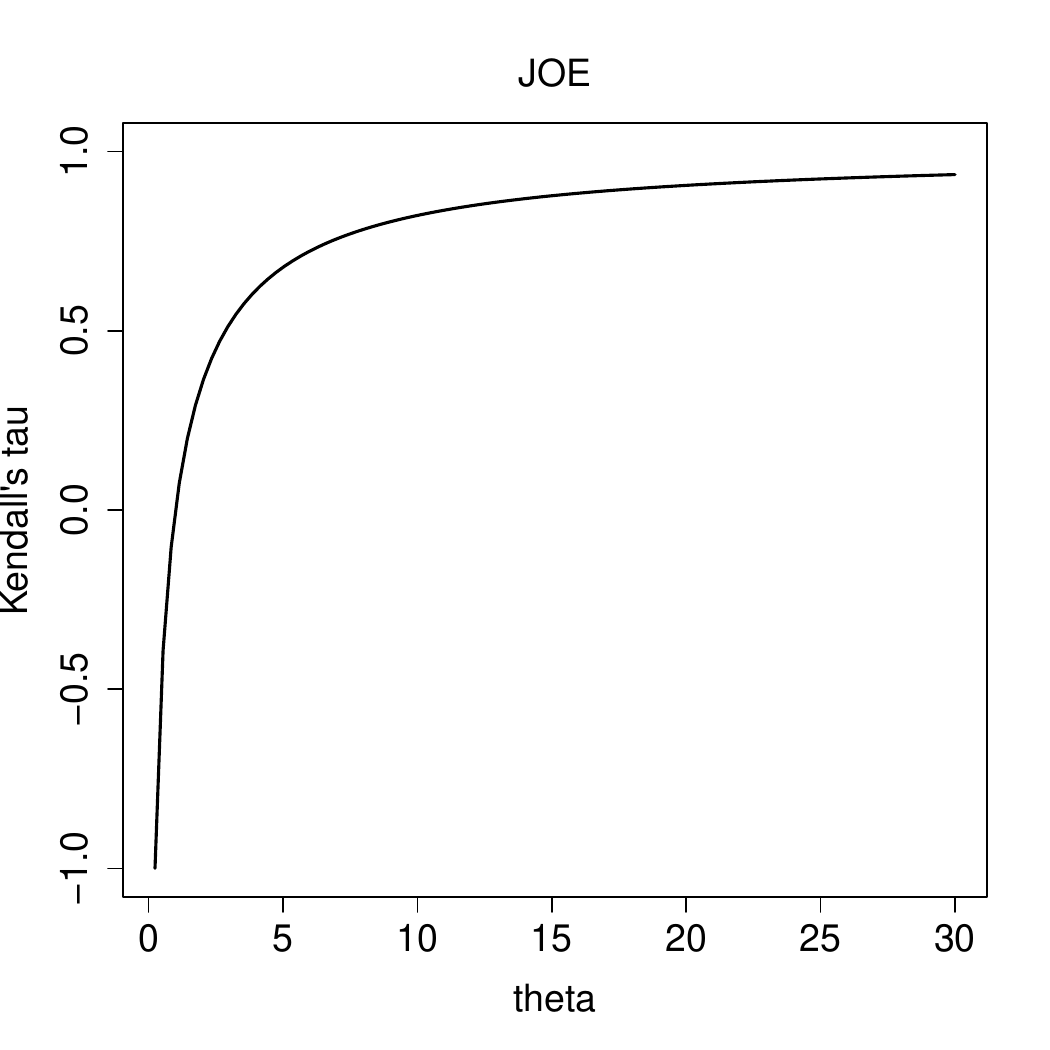}}
\caption{Kendall tau values for the five Archimedean families of Table \ref{tab:afam} as a function of $\theta$.}
\label{fig:ktau}
\end{figure}

One way to assess the difference in the dependence induced by these five Archimedean copula families is by studying their corresponding Kendall tau. In Figure \ref{fig:ktau}, we plot $\tau_\theta$ as a function of $\theta\in \Theta$. Of the five copula families considered, in the case of the Clayton, Frank and Joe classes, the Kendall tau association coefficient spans the entire range $(-1,1)$ as the copula parameter varies in its domain. The other two members induce only constrained associations, with $\tau \in [-0.1817,1/3]$ for the AMH and $\tau \in [0,1)$ for the Gumbel. 

To define Archimedean copulas of dimensions larger than two, we have to carefully identify the parameter space. To be specific, let us consider a setting with $p=3$ variables. Assume that $U_1$ and $U_2$ have positive dependence and $U_1$ and $U_3$ have negative dependence, therefore $U_2$ and $U_3$ must have negative dependence. Since dependence in Archimedean copulas is determined by a single parameter $\theta$, the setting of the previous three variables may not occur in a three-dimensional Archimedean copula. Archimedean copulas assume that the variables are exchangeable, so the dependence between any pair of variables has to be, too. This feature is preserved by the mixture setting that we propose. However, the advantage provided by our construction is that it can model heterogeneity in the dependence structure across the population. This flexibility is accompanied by the ability to group bivariate data according to the information contained in the marginals and the copula.

To avoid the previous problems, the authors that study Archimedean copulas for $p>2$ usually constrain the parameter space $\Theta$ to their positive values. See, for example, \cite{hofert&al:12}, who also present analytical derivatives of order $p$ of the inverse generators $\phi_\theta^{-1}(t)$ for families in Table \ref{tab:afam}, for the positive values of the parameter space $\Theta$.

\section{BNP mixtures}
\label{sec:mixmodel}

\subsection{Model}

Although Archimedean copulas are easy to generalize for multivariate data, the dependence might be too restrictive, since it depends only on a single parameter $\theta$. To equip the model with extra flexibility, we propose to mix the Archimedean copulas via a Bayesian nonparametric prior. 

In particular, we choose the two-parameter extension of the Dirichlet process introduced by \cite{pitman&yor:97}. This Poisson-Dirichlet process is almost surely (henceforth, a.s.) discrete, admits a stick-breaking construction, and can be marginalized to simplify the implementation \citep{ishwaran&james:01}. A probability measure $G$ has a Poisson-Dirichlet prior with scalar parameters $a\in[0,1)$, $b>-a$ and mean parameter $G_0$, denoted as $G\sim\pd(a,b,G_0)$, when
\begin{equation}
\label{eq:pdp}
G(\cdot)=\sum_{k=1}^\infty \omega_k\delta_{\theta_k}(\cdot),
\end{equation}
where $\omega_1=\nu_1$ and $\omega_k=\nu_k\prod_{j<k}(1-\nu_j)$ for $k=2,3\ldots$, with $\nu_k\simind\be(1-a,b+ka)$ independent of the weights, locations, $\theta_k\simiid G_0$ for $k=1,2,\ldots$, and $\delta_\theta$ is the Dirac measure with unit mass at $\theta$. The functional parameter $G_0$ is known as the centering measure since $\E(G)=G_0$. There are two particular cases that can be obtained with the Poisson-Dirichlet prior, the Dirichlet process when $a=0$ \citep{ferguson:73} and the normalized stable process when $b=0$ \citep{kingman:75}.

A Bayesian nonparametric mixture model can be defined by mixing parametric Ar\-chi\-me\-dean copulas $C(\bu\mid\theta)$ and using the Poisson-Dirichlet process as the mixing distribution for the parameter $\theta$, that is, 
\begin{equation}
C(\bu)=\int C(\bu\mid\theta)G(\d\theta)=\sum_{k=1}^\infty\omega_k C(\bu\mid\theta_k),
\end{equation}
where the last equality is obtained from  \eqref{eq:pdp}.

The Bayesian nonparametric mixture copula model can also be defined hierarchically as follows. For $i=1,\ldots,n$  
\begin{align}
\nonumber
(U_{1i},\ldots,U_{pi})\mid\theta_i&\simind f_C(\bu_i\mid\theta_i)\\
\label{eq:bnpmix}
\theta_i\mid G &\simiid G\\ 
\nonumber
G &\sim\pd(a,b,G_0),
\end{align}
where $f_C$ is given in \eqref{eq:cdens}. For each observed multivariate vector $\bU_i=(U_{1i},\ldots,U_{pi})$, we assign a potentially different parameter $\theta_i$. However, since the Bayesian nonparametric prior is a.s. discrete, there could be ties implying $\P(\theta_i=\theta_j)>0$. Therefore, the number of different $\btheta_i$'s could be lower than $n$. Smaller values of $a$ and $b$ produce more ties. 

For the centering measure $G_0$ we consider appropriate densities with support in the parameter space $\Theta$. In general, we denote by $g_0(\theta)$ the density function associated to measure $G_0$. In particular, we take $g_0(\theta)=\un(\theta\mid-1,1)$ for the AMH; $g_0(\theta)=\ga(\theta-k\mid c_\theta,d_\theta)$ for Clayton and Joe, with $k=-1$ and $k=0.238734$, respectively; $g_0(\theta)=\no(\theta\mid\mu_\theta,\lambda_\theta)$ for Frank; and $g_0(\theta)=\ga(\theta-1\mid c_\theta,d_\theta)$ for the Gumbel. These choices of centering measures have no strong impact on posterior inference. 

Since the $\theta_i$'s are conditionally independent given $G$, and $\E(G)=G_0$, then the parameters $\theta_i$ are exchangeable a priori with marginal distribution $\theta_i\sim G_0$ for $i=1,2,\ldots,n$. In particular, \cite{pitman:95} showed that if we integrate out the nonparametric measure $G$, the joint distribution of the $\theta_i$'s is characterized by a generalized Polya urn mechanism with conditional distribution that depends on the density $g_0$ of $G_0$ as
\begin{equation}
\label{eq:prior}
f(\theta_i\mid\btheta_{-i})=\frac{b+am_i}{b+n-1}g_0(\theta_i)+\sum_{j=1}^{m_i}\frac{n_{i,j}^*-a}{b+n-1}\delta_{\theta_{i,j}^*}(\theta_i),
\end{equation}
 with $\theta_{-i}$ being the set of all $\theta_i$'s excluding the $i$th element and $(\theta_{i,1}^*,\ldots,\theta_{i,m_i}^*)$ denoting the distinct values in $\btheta_{-i}$, each with frequencies $n_{i,j}^*$, for $i=1,\ldots,n$, $j=1,\ldots,m_i$. One can immediately see the importance of having $G_0$'s support coincide with $\Theta$. 


It is not difficult to prove that association coefficients like the Kendall tau for a mixture copula requires a double summation. In particular, the Kendall tau for the Bayesian nonparametric mixture model \eqref{eq:bnpmix} is
\begin{equation}
\label{eq:taumix}
\tau=4\sum_{j=1}^\infty\sum_{k=1}^\infty \omega_j\omega_k\iint_{[0,1]^2}C(u,v\mid\theta_j)f_C(u,v\mid\theta_k)\d u\d v - 1.
\end{equation}
For the five Archimedean families discussed earlier, the values for individual components $\tau_{\theta}$ in terms of $\theta$ are given in Table \ref{tab:afam}.

\subsection{Posterior distributions}

The posterior conditional distributions for each $\theta_i$ are given by
$$f(\theta_i\mid\bu,\btheta_{-i})\propto (b+am_i)f(\bu_i\mid\theta_i)g_0(\theta_i)+\sum_{j=1}^{m_i}(n_{i,j}^*-a)f_C(\bu_i\mid\theta_i)\delta_{\theta_{i,j}^*}(\theta_i),$$
for $i=1,\ldots,n$. 

Since the likelihood $f(\bu_i\mid\theta_i)$ does not admit a conjugate prior $g_0(\theta_i)$, we need to use an MCMC sampler to draw from these posterior conditional distributions. We use a Gibbs sampler \citep{smith&roberts:93}, and we rely on Algorithm 8 in \cite{neal:00}. Specifically, we initialize the sampler using random draws $\theta_i$ from the prior $g_0$, for $i=1,\ldots,n$. Then the algorithm proceeds as follows: 
\begin{enumerate}
\item[(i)] For each $i=1,\ldots,n$, sample $r$ auxiliary values $\btheta^{\star}=\{\theta_{m_i+1}^{\star},\ldots,\theta_{m_i+r}^{\star}\}$ from $g_0$. 
\item[(ii)] Draw $\theta_i$, $i=1,\ldots,n$, from 
$$\hspace{-12cm}f(\theta_i\mid\bu,\bv,\btheta_{-i},\btheta^{\star})=$$ 
$$\hspace{5mm}\frac{1}{k_i}\left[\sum_{j=1}^{m_i}\{n_{i,j}^*-a\} f_C(\bu_i\mid\theta_{i,j}^*)\delta_{\theta_{i,j}^*}(\theta_i)+\sum_{j=m_i+1}^{m_i+r}\{(b+a m_i)/r\}f_C(\bu_i\mid\theta_{j}^{\star})\delta_{\theta_{j}^{\star}}(\theta_i)\right],$$
where $k_i=\sum_{j=1}^{m_i}\{n_{i,j}^*-a\}f_C(\bu_i\mid\theta_{i,j}^*)+\sum_{j=m_i+1}^{m_i+r}\{(b+a m_i)/r\}f_C(\bu_i\mid\theta_{j}^{\star})$. 
\item[(iii)] Compute the unique values $(\theta_1^*,\ldots,\theta_m^*)$ in $\btheta$ and re-sample each $\theta_j^*$, $j=1,\ldots,m$ from 
$$f(\theta_j\mid c.c.)\propto g_0(\theta_j)\prod_{\{i:\theta_i=\theta_j^*\}}f_C(\bu_i\mid\theta_j),$$
where $c.c.$ stands for clustering configuration. We suggest performing this sampling using a Metropolis-Hastings (MH) random walk step \citep[e.g.][]{robert&casella:10} as follows. Sample $\theta_j$ from $$h(\theta_j\mid\theta_j^*)=\un(\theta_{j}\mid\theta_{j}^*-\kappa_\theta,\theta_{j}^*+\kappa_\theta)$$ constrained to the parameter space $\theta_j\in\Theta$, and accept it with probability $\min\{1,f(\theta_j\mid c.c.)/f(\theta_j^*\mid c.c.)\}$. 
\end{enumerate}

The hyperparameters $(a,b)$ are essential in determining the number of components in the mixture \eqref{eq:bnpmix}. Instead of giving them a fixed value, we assign a hyperprior distribution. Taking into account the parameter space $a\in[0,1)$ and $b\in(-a,\infty)$, the joint hyperprior is factorized as $f(a,b)=f(b\mid a)f(a)$. Specifically, $b$ given $a$ has a shifted gamma and $a$ has a marginal Beta distribution in the unit interval. In other words, 
$$f(a,b)=\ga(b+a\mid c_b,d_b)\be(a\mid c_a,d_a).$$

This prior distribution for $(a,b)$ is updated with the exchangeable partition probability function (EPPF) induced by the Poisson-Dirichlet process. This was obtained by \cite{pitman:95} and is given by 
$$f(n_1^*,\ldots,n_m^*\mid a,b)=\frac{\Gamma(b+1)}{\Gamma(b+n)}\left\{\prod_{j=1}^{m-1}(b+ja)\right\}\left\{\prod_{j=1}^m\frac{\Gamma(n_j^*-a)}{\Gamma(1-a)}\right\}.$$

The Gibbs sampler is extended to include simulations of the following two conditional distributions. 
\begin{enumerate}
\item[(iv)] Sample $a$ from 
$$f(a\mid b,\data)\propto \left\{\prod_{j=1}^{m-1}(b+ja)\right\}\left\{\prod_{j=1}^m\frac{\Gamma(n_j^*-a)}{\Gamma(1-a)}\right\}\ga(b+a\mid c_b,d_b)\be(a\mid c_a,d_a)$$
by implementing a MH step with a random walk proposal. In iteration $(t+1)$, sample $a\sim\un(\max\{0,a^{(t)}-\kappa_a\},\min\{a^{(t)}+\kappa_a,1\})$ and accept it with probability $\min\{1,f(a\mid b,\data)/f(a^{(t)}\mid b,\data)\}$. 
\item[(v)] Sample $b$ from 
$$f(b\mid a,\data)\propto \frac{\Gamma(b+1)}{\Gamma(b+n)}\left\{\prod_{j=1}^{m-1}(b+ja)\right\}\ga(b+a\mid c_b,d_b)$$
by implementing a MH step with a random walk proposal. In iteration $(t+1)$, sample $b\sim\ga(\kappa_b,\kappa_b/b^{(t)})$ and accept it with probability $\min\{1,f(b\mid a,\data)\ga(b^{(t)}\mid \kappa_b,\kappa_b/b)/f(b^{(t)}\mid a,\data)/\ga(b\mid\kappa_b,\kappa_b/b^{(t)})\}$. 
\end{enumerate}
The algorithm continues to iterate steps (i)--(v) for a period of time after convergence and discards the draws obtained before entering the stationary regime. 

Parameters $\kappa_\theta$, $\kappa_a$, and $\kappa_b$ are tuning parameters that control the acceptance probability in the MH steps. Instead of fixing them, we follow \cite{roberts&rosenthal:09} and adapt them every 50th iteration to achieve a target acceptance rate. We aim for an acceptance probability in the interval $[0.3, 0.4]$ that, according to \cite{robert&casella:10}, contains the optimal value. Specifically, we use batches of 50 iterations, and for each batch $j$, we compute the acceptance rate for each of the parameters $\theta_j^*$, $a$, and $b$, say $AR^{(j)}$. Then for $\kappa_\theta$, increase $\kappa^{(j+1)}=\kappa^{(j)}1.01^{\sqrt{j}}$ if $AR^{(j)}>0.4$ and decrease $\kappa^{(j+1)}=\kappa^{(j)}1.01^{-\sqrt{j}}$ if $AR^{(j)}<0.3$. For the other two $\kappa_a$ and $\kappa_b$, increase $\kappa^{(j+1)}=\kappa^{(j)}e^{\sqrt{j}}$ if $AR^{(j)}<0.3$ and decrease $\kappa^{(j+1)}=\kappa^{(j)}e^{-\sqrt{j}}$ if $AR^{(j)}>0.4$.
We use $\kappa_\theta^{(1)}=0.1$, $\kappa_a^{(1)}=1$, and $\kappa_b^{(1)}=1$ as starting values.

\subsection{Fitting measure and clustering selection}
\label{ssec:predgof}

We evaluate the fit of the model by computing the logarithm of the pseudo-marginal
likelihood (LPML), which is a measure of a model's predictive performance.  LPML is defined as $LPML=\sum_{i=1}^n\log(CPO_i)$, where the conditional predictive ordinate (CPO) statistic \citep{geisser&eddy:79} is the predictive density for $i$-th observation given the remaining data,  $CPO_i=f(\bu_i\mid\bu_{-i})$. It is well known \citep[e.g.][]{nieto&contreras:14} that $CPO_i$ can be estimated with a Monte Carlo sample $\theta_i^{(l)}$ for $l=1,\ldots,L$ as 
$$\widehat{CPO}_i=\left[\frac{1}{L}\sum_{l=1}^L\frac{1}{f_C(\bu_i\mid\theta_i^{(l)})}\right]^{-1}.$$

In addition, Kendall tau in \eqref{eq:taumix} can be used to summarize the dependence in the data and is approximated using
$$\widehat{\tau}=\frac{1}{L}\sum_{l=1}^L\tau_{\theta_0^{(l)}},$$
where $\theta_0^{(l)}$ is sampled from the conditional distribution 
$$f(\theta_0\mid\btheta)=\frac{b+am}{b+n}g_0(\theta_0)+\sum_{j=1}^{m}\frac{n_{j}^*-a}{b+n}\delta_{\theta_{j}^{*}}(\theta_0).$$

One feature of the Poisson-Dirichlet process mixture models is that the uncertainty about the number of mixture components is quantified by the posterior distribution. Specifically, the number of components is given by the number of distinct parameters $m$, which can be obtained using the MCMC sample. When an estimate of the number of components is of interest, one can use a zero-one loss function and report the mode; however, there are multiple cluster configurations with the same number of components defined by the mode. In order to select a single clustering configuration and produce further inferences, we propose to use the search algorithm presented in \cite{dahl&al:22} and implemented in the R package \texttt{salso}. The default loss function of the package is a variation of information; however, for the examples worked here, the binder loss seems to perform better for clustering. Finally, conditional on the chosen clustering configuration, we perform a post-MCMC sampling of the unique parameter values $(\theta_1^*,\ldots,\theta_{\tilde{m}}^*)$ and compute the weights assigned by the model to each of the mixture components. In particular, we compute the posterior mean of $(n_j^*-a)/(b+n)$, with respect to the posterior distribution of $(a,b)$, where $n_j^*$ is the number of data points assigned to the cluster $j$, for $j=1,\ldots,\tilde{m}$.

\section{Numerical Experiments}
\label{sec:numerical}

\subsection{Simulation study 1}
\label{ssec:sim2}

We first test our posterior inferential procedure by generating random bivariate data from two-component mixtures, $f(\bu)=\pi f_C(\bu\mid\vartheta_1)+(1-\pi)f_C(\bu\mid\vartheta_2)$, of each of the five common Archimedean copulas of Table \ref{tab:afam}. We chose $\pi=1/2$ in all cases and the copula densities $f_C$ were specified by parameter values $(\vartheta_1,\vartheta_2)$ for each family as: $(-0.8,0.8)$ for AMH, $(-0.5,10)$ for Clayton, $(-5,5)$ for Frank, $(5,10)$ for Gumbel, and $(2,10)$ for Joe. We compare analyses for two sample sizes, $n=200$ and $n=500$. For each generated data, we fitted our Bayesian nonparametric Archimedean copula mixture model using each of the five Archimedean family members as mixture densities. This leads to a total of 25 model fits. 

\begin{figure}
\centering
\includegraphics[width=0.8\linewidth]{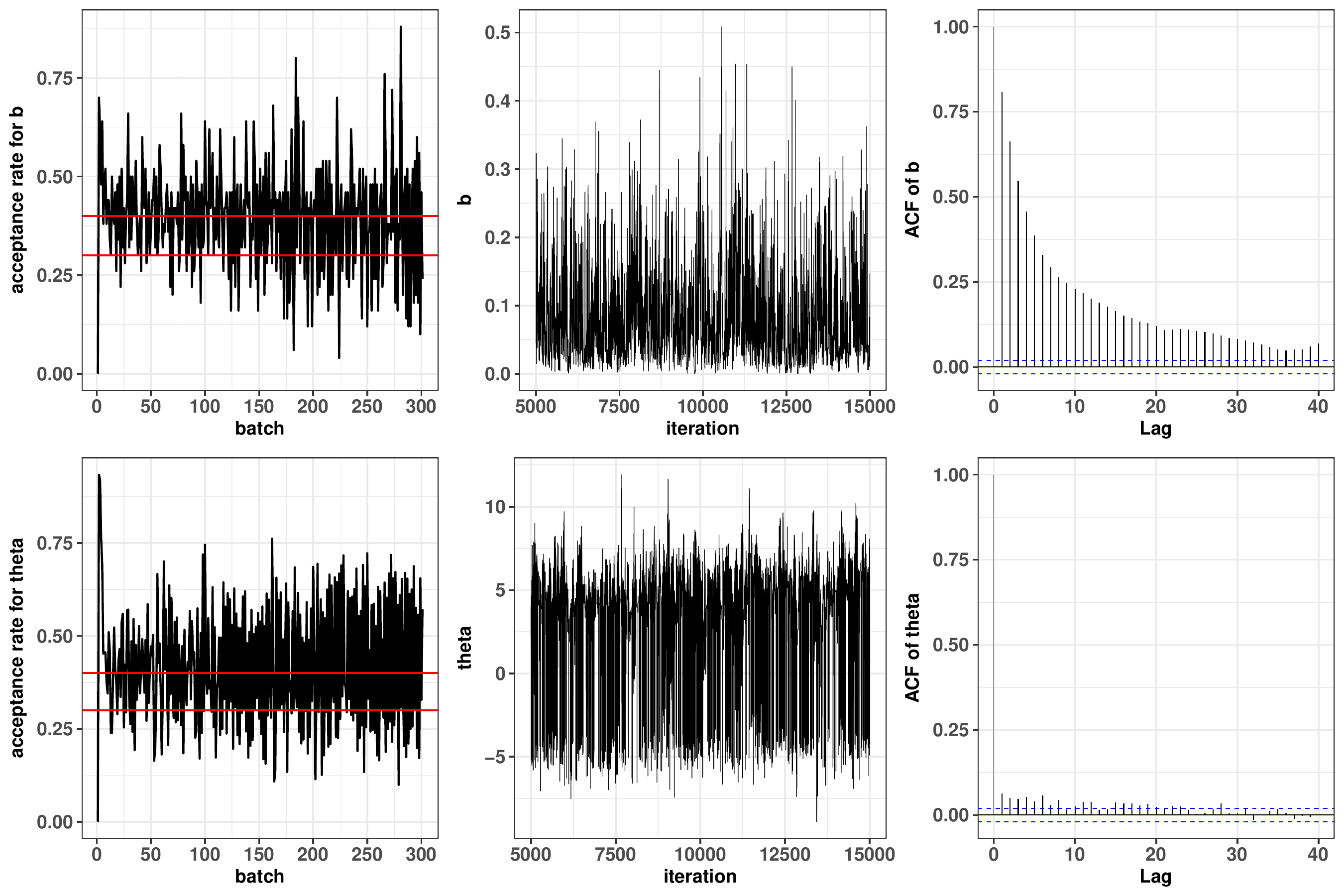}
\caption{Simulation study 1: MCMC diagnostics for data coming from a mixture of Frank copulas and fitted with the same kernel. Acceptance rate (left), trace plot (middle) and ACF (right). Parameter $b$ (top) and $\theta$ (bottom).}
\label{fig:diagnosis}
\end{figure}

Our model is fully specified by determining the centering measure $g_0(\theta)$ and a hyper-prior for the parameters $a$ and $b$. In particular, for the centering measure, we chose: $c_\theta=4, d_\theta=1$ for Clayton, Joe, and Gumbel, $\mu_\theta=0$ and $\lambda_\theta=4$ for Frank. For the precision parameters, we considered informative priors to induce a small number of components: $p_a=1$, $q_a=20$, $c_b=1$ and $d_b=20$. MCMC was run for 15,000 iterations with a burn-in of 5,000. We monitored the acceptance rate for each batch, in the adaptive algorithm, and diagnose the convergence of MCMC by traceplots and autocorrelation functions. This is shown in Figure \ref{fig:diagnosis}. Given the high autocorrelation in the samples for parameter $b$, we decided to thin the chain by keeping one of every 5$^{th}$ iterations. 

To assess the fit of the model, we computed the LPML statistic defined in Section \ref{ssec:predgof}. The 25 values are reported in Tables \ref{tab:lpml200} and \ref{tab:lpml500} for both sample sizes, $n=200,500$, respectively. The way to read these tables is row-wise, where the largest value for each row determines the best fit. The two sample sizes yield fairly similar results. For four of the five families, the best model selected is the one produced by the copula used to generate the data, i.e. the true model.  The exception is the AMH family, where the best fits are obtained by the Clayton and Joe kernels. However, the second/third best model is obtained when using the AMH family. This is because the bivariate copula from the AMH family does not produce a strong dependence; in fact, the parameters chosen induce Kendall tau values $-0.099$ and $0.128$ for each of the two components of the mixture, respectively. The last column in both tables corresponds to the Bayesian semiparametric Archimedean copula (BSA) model that will be discussed later.

\begin{figure}
\centerline{
\includegraphics[scale=0.15]{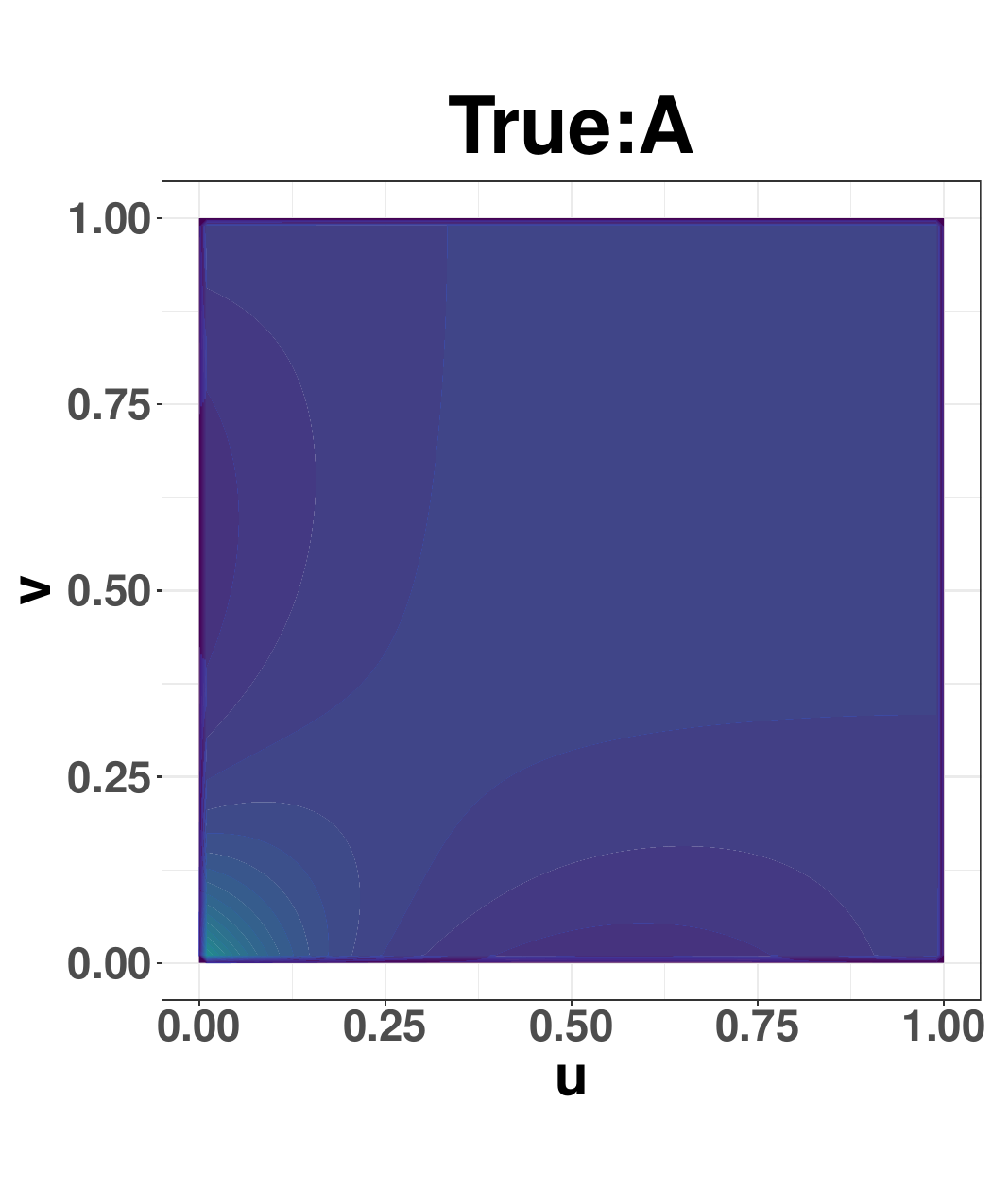}
\includegraphics[scale=0.15]{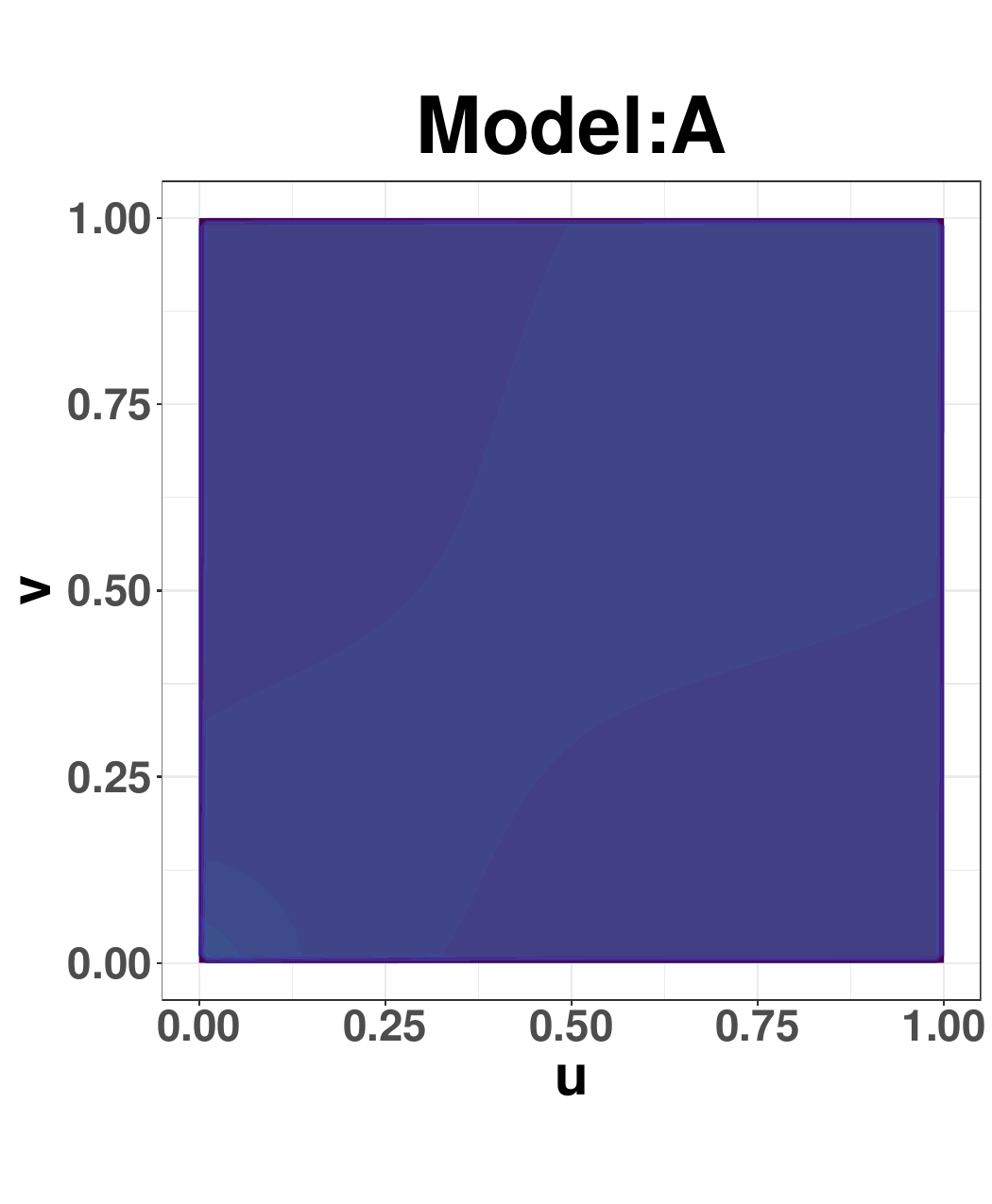}
\includegraphics[scale=0.15]{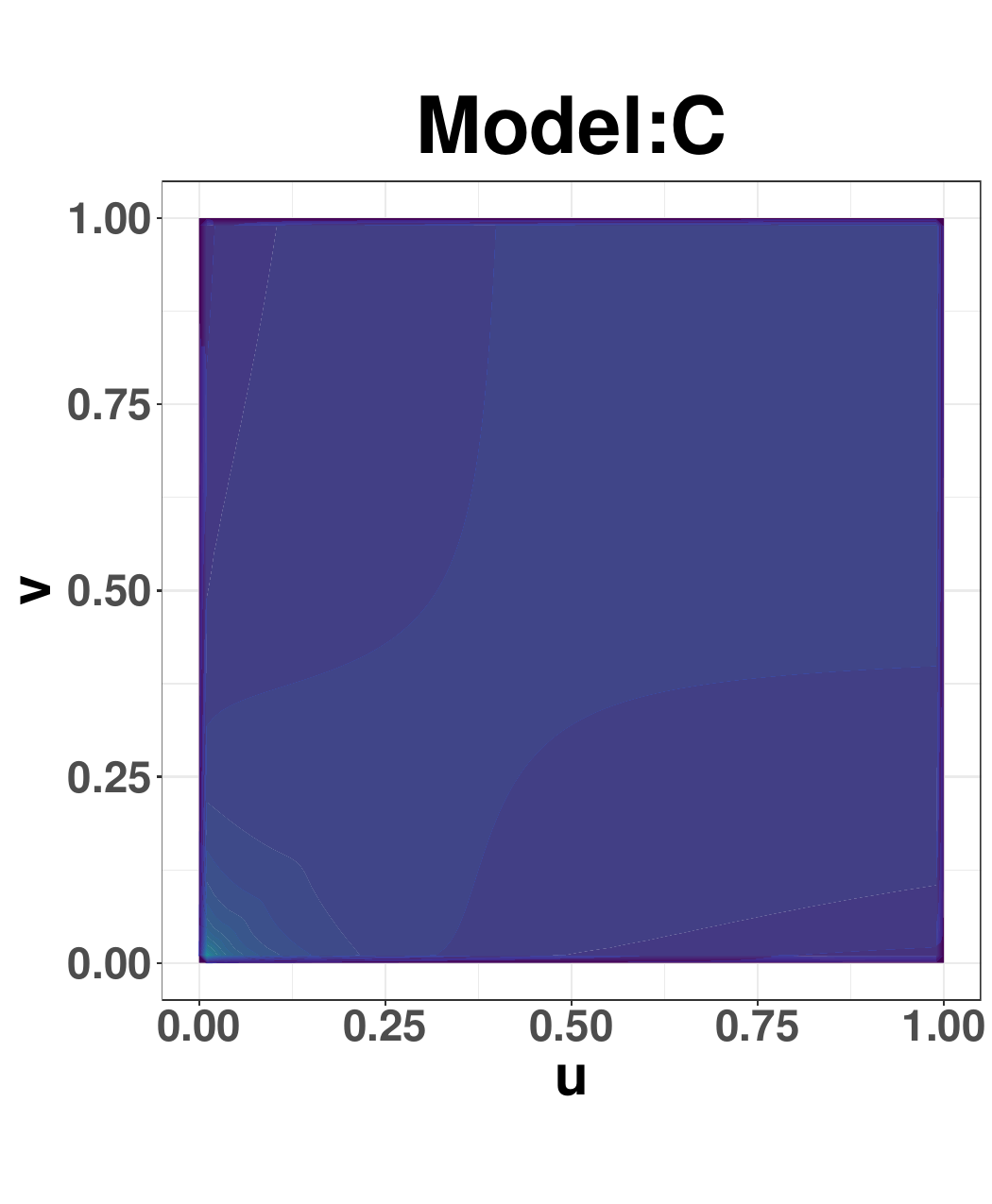}
\includegraphics[scale=0.15]{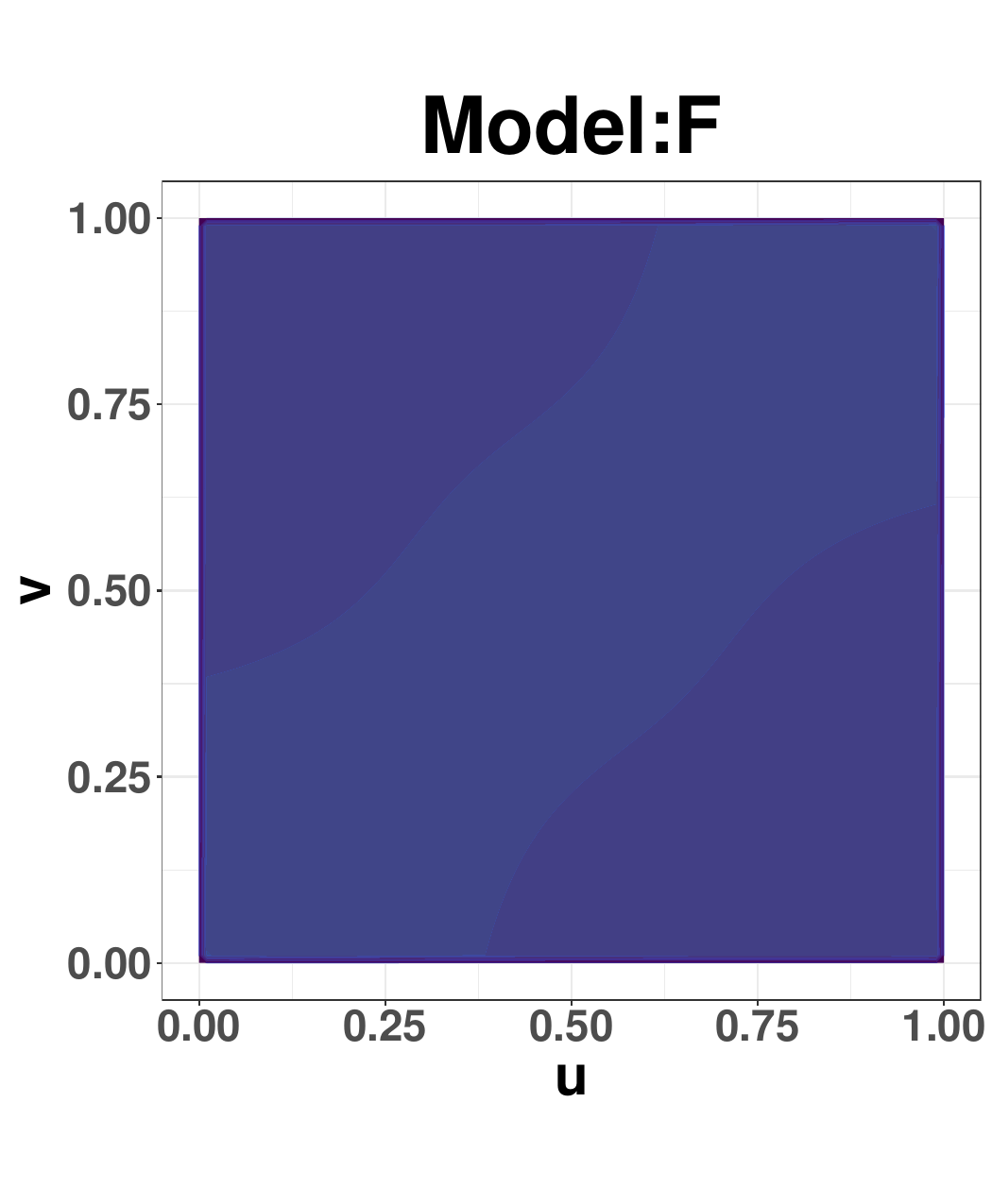}
\includegraphics[scale=0.15]{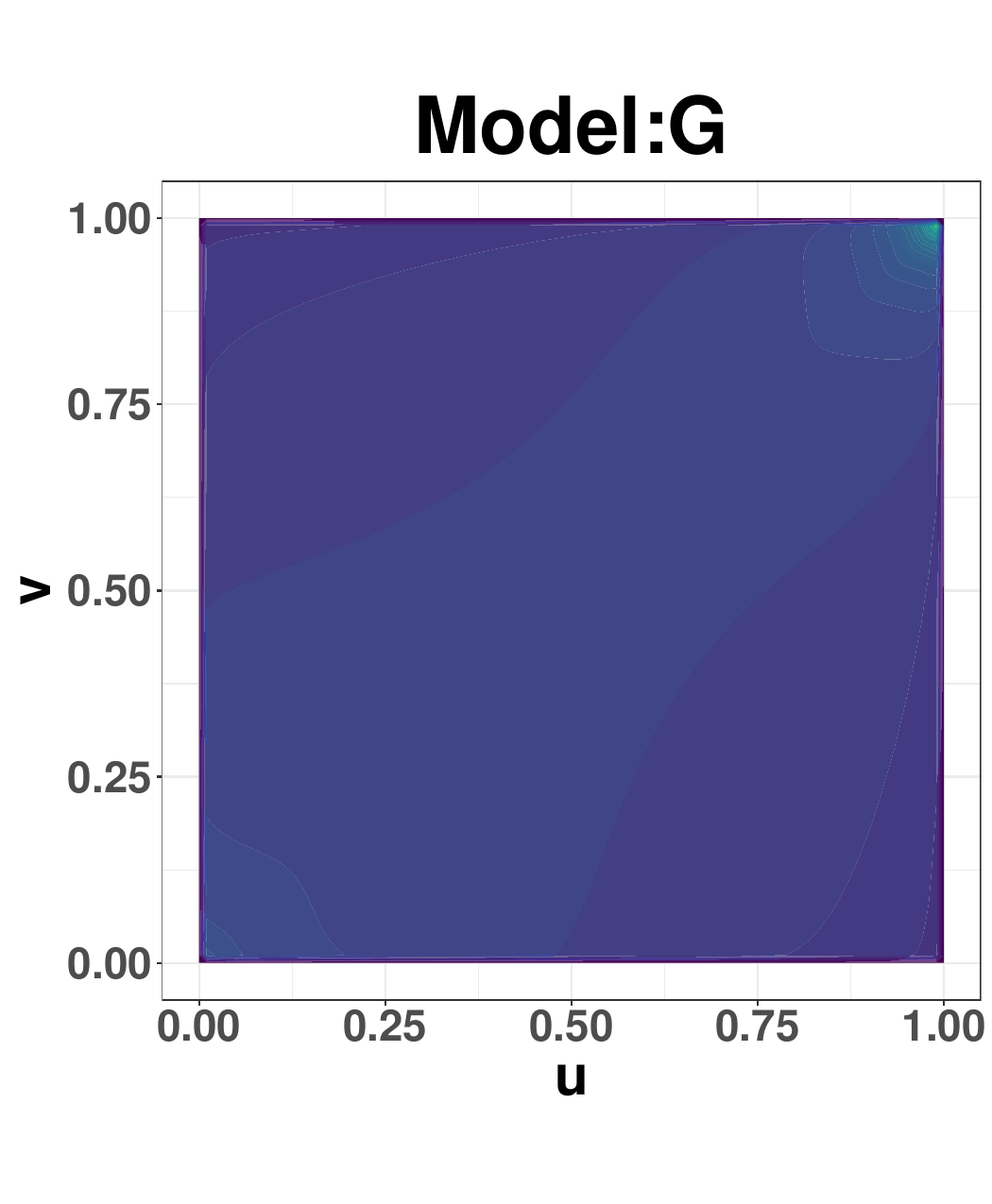}
\includegraphics[scale=0.15]{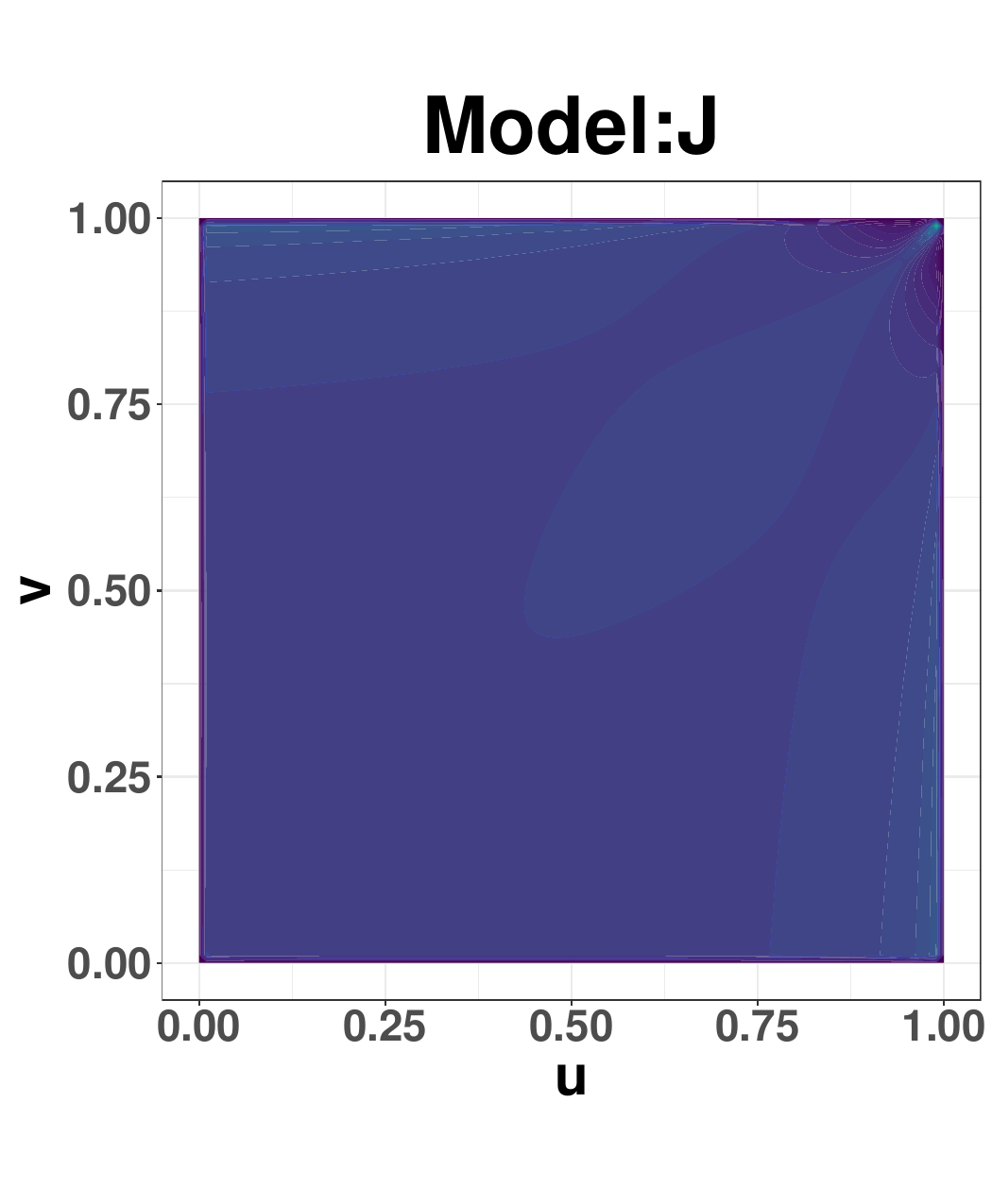}
}
\centerline{
\includegraphics[scale=0.15]{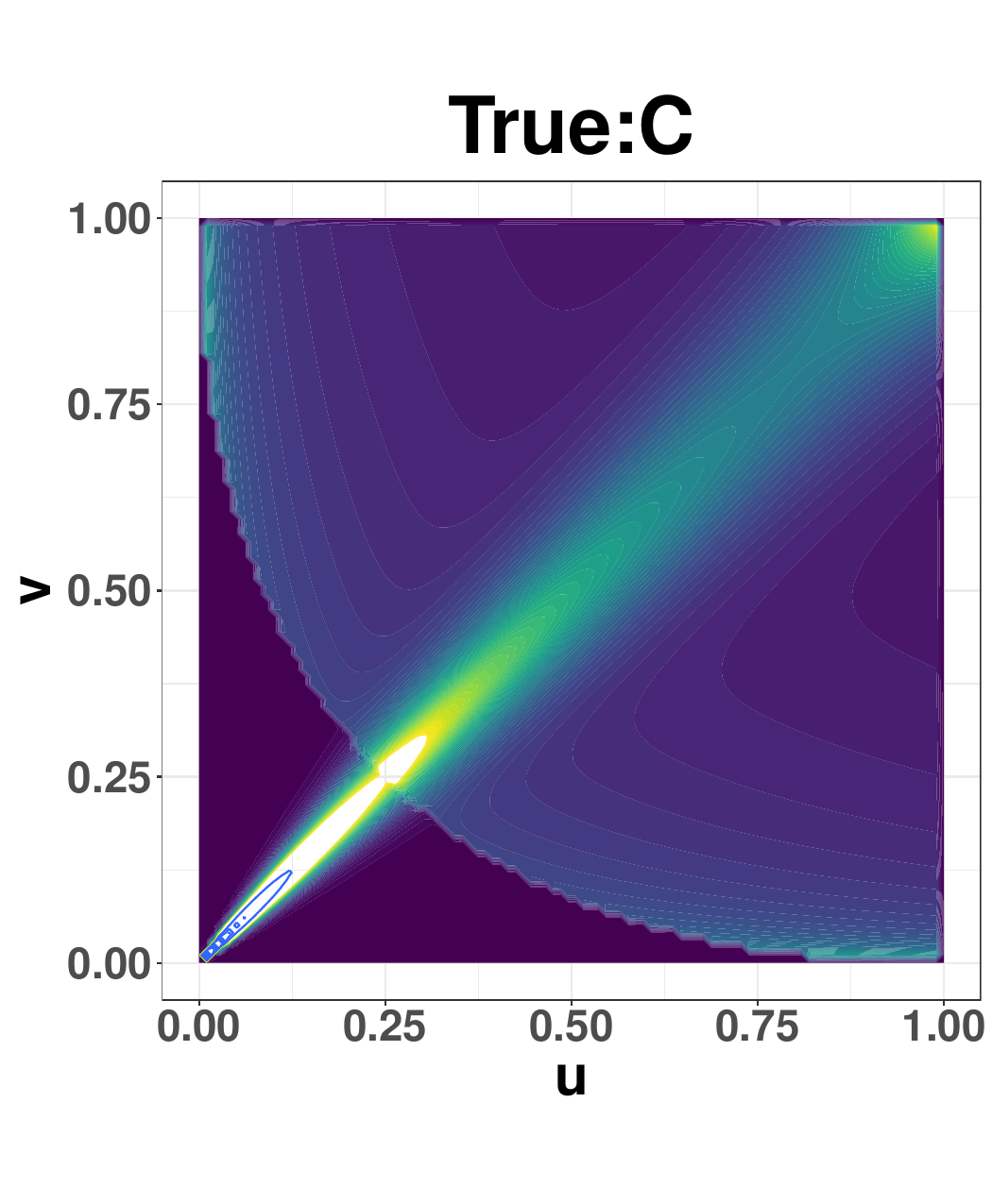}
\includegraphics[scale=0.15]{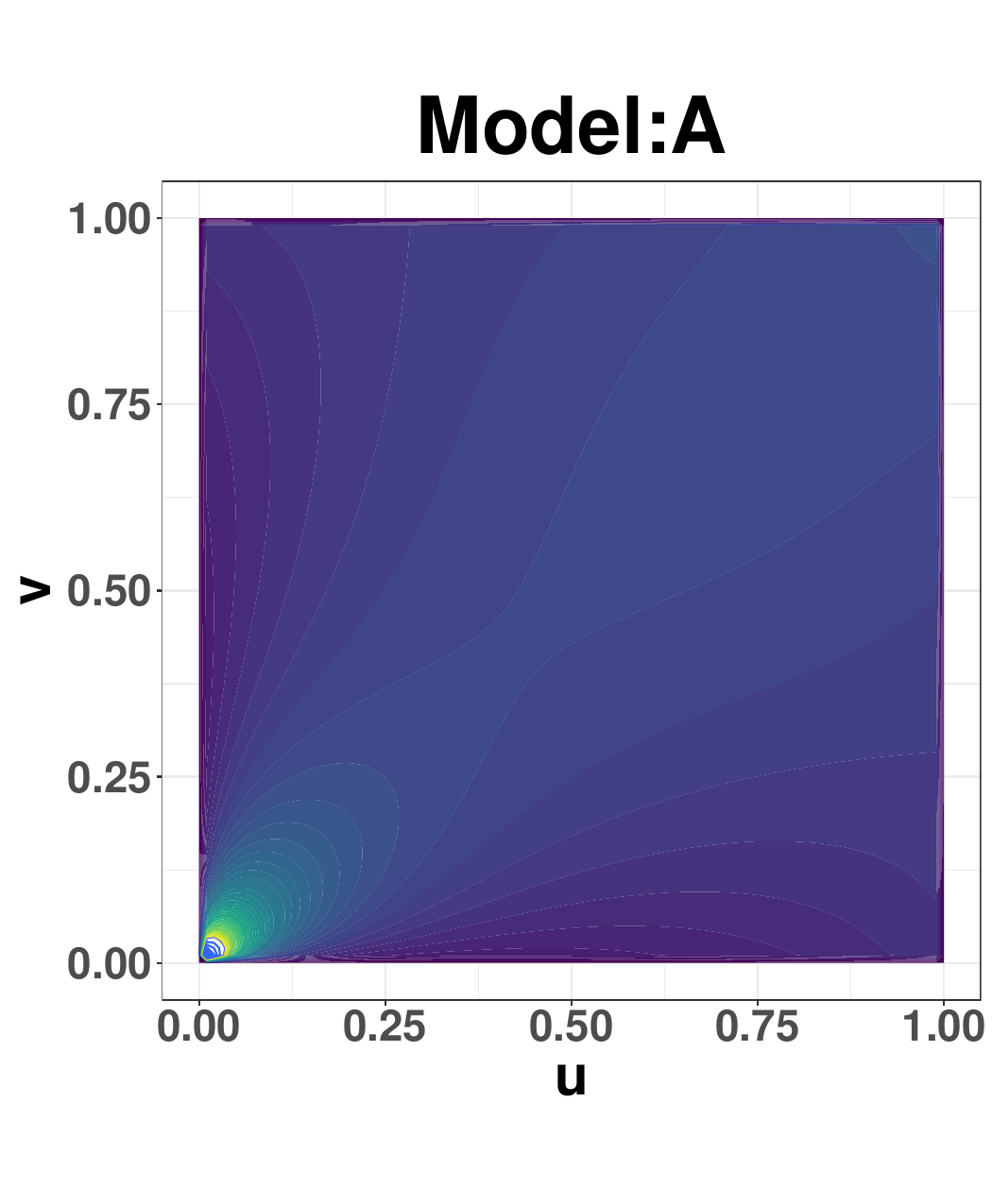}
\includegraphics[scale=0.15]{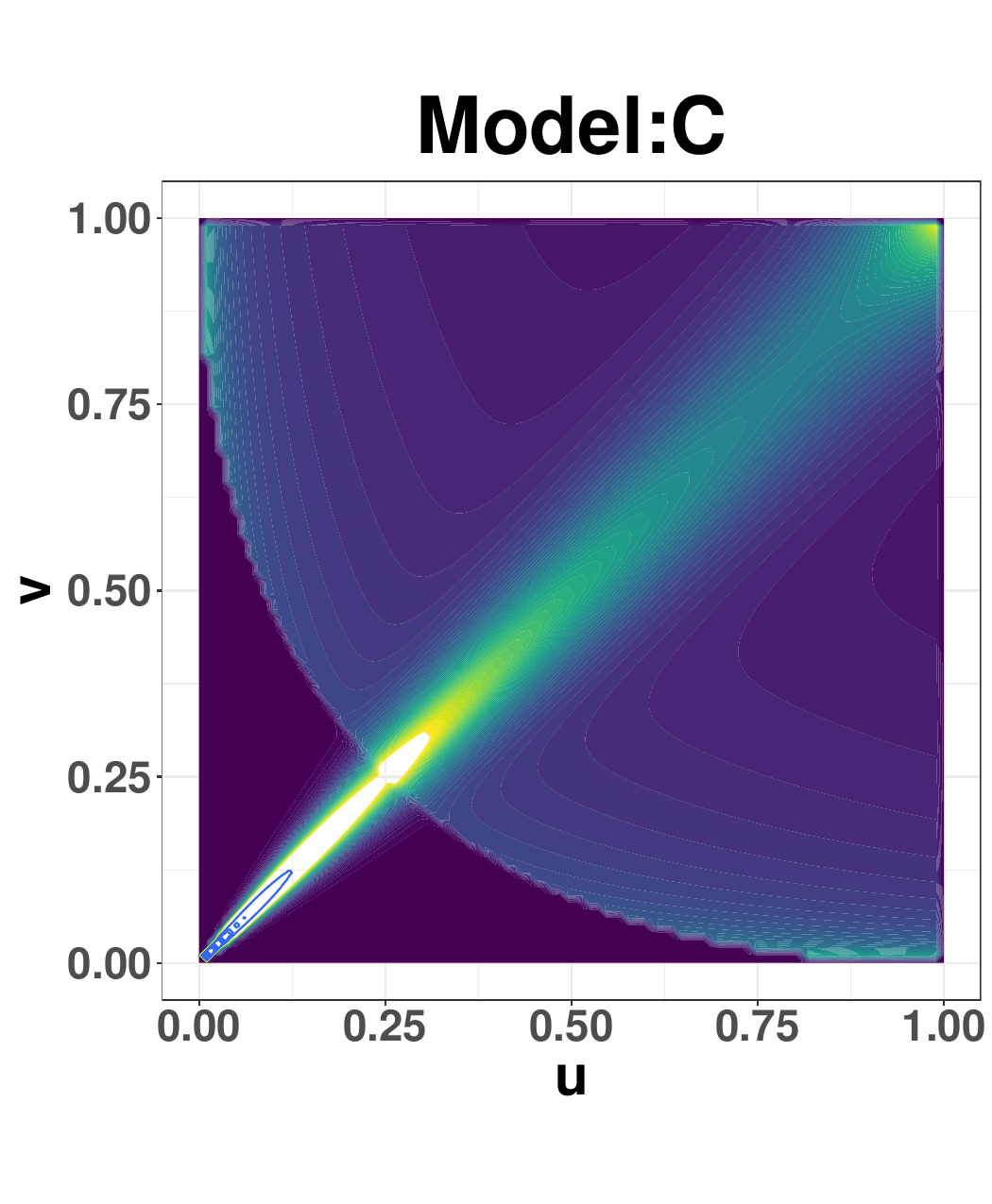}
\includegraphics[scale=0.15]{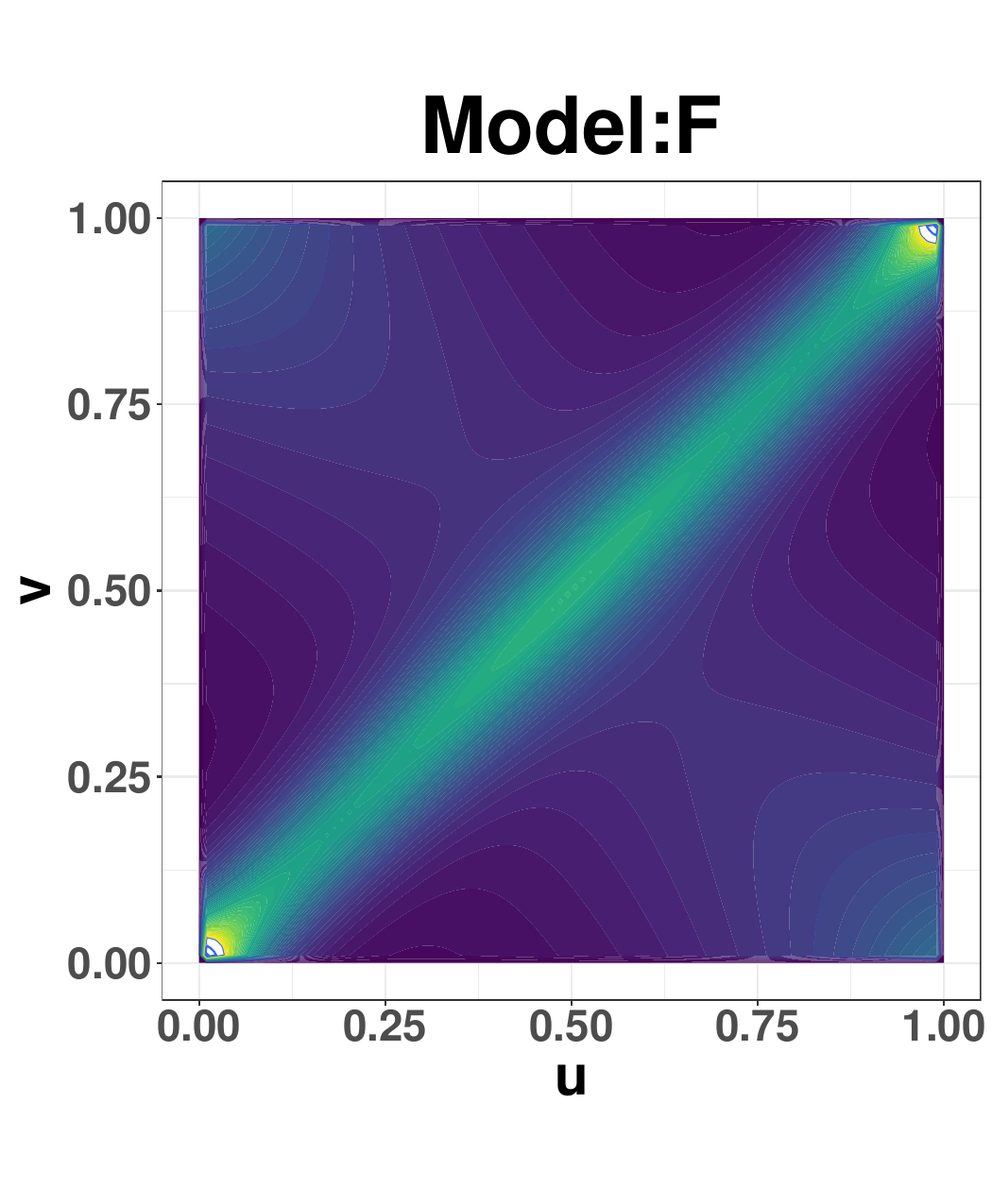}
\includegraphics[scale=0.15]{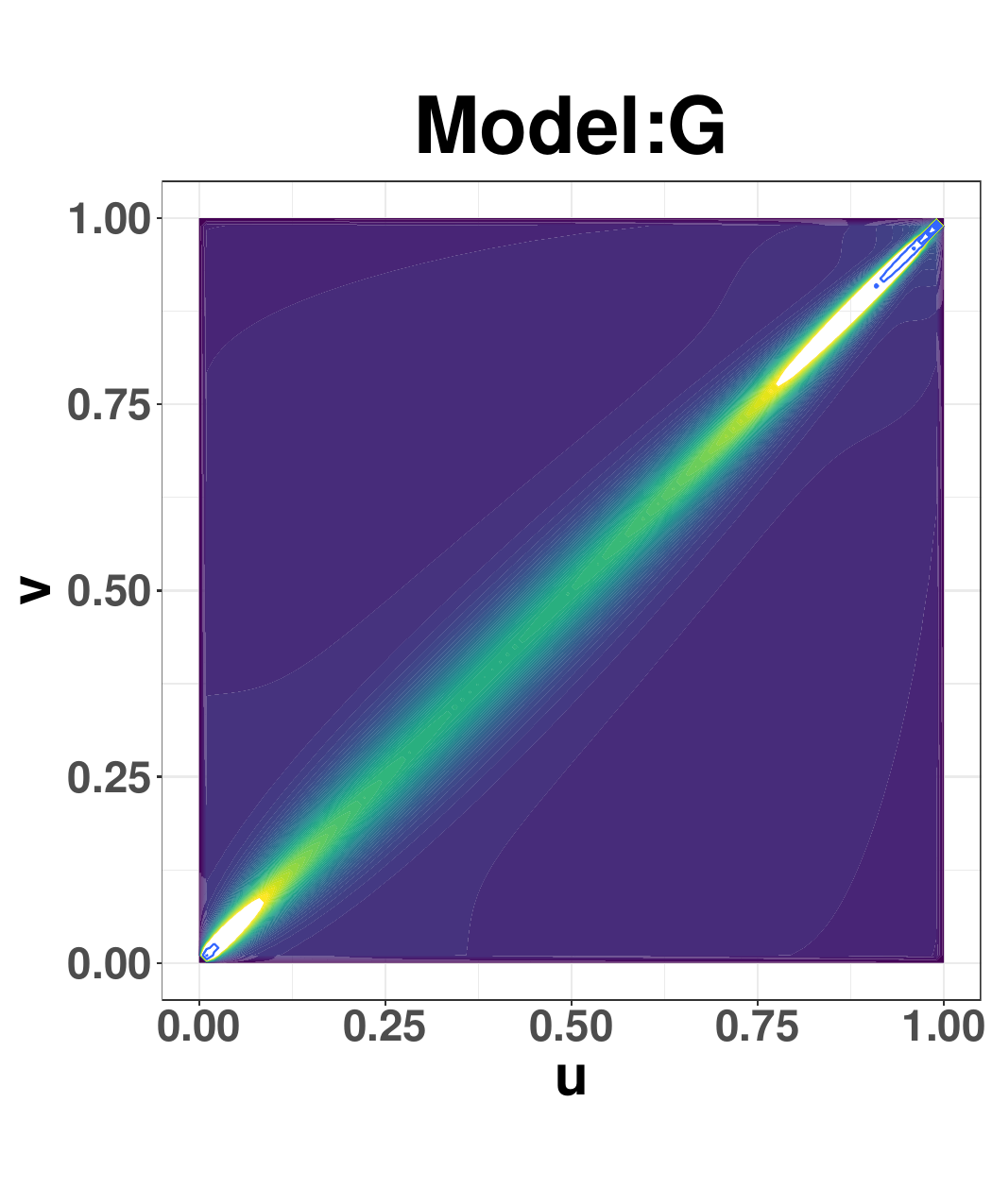}
\includegraphics[scale=0.15]{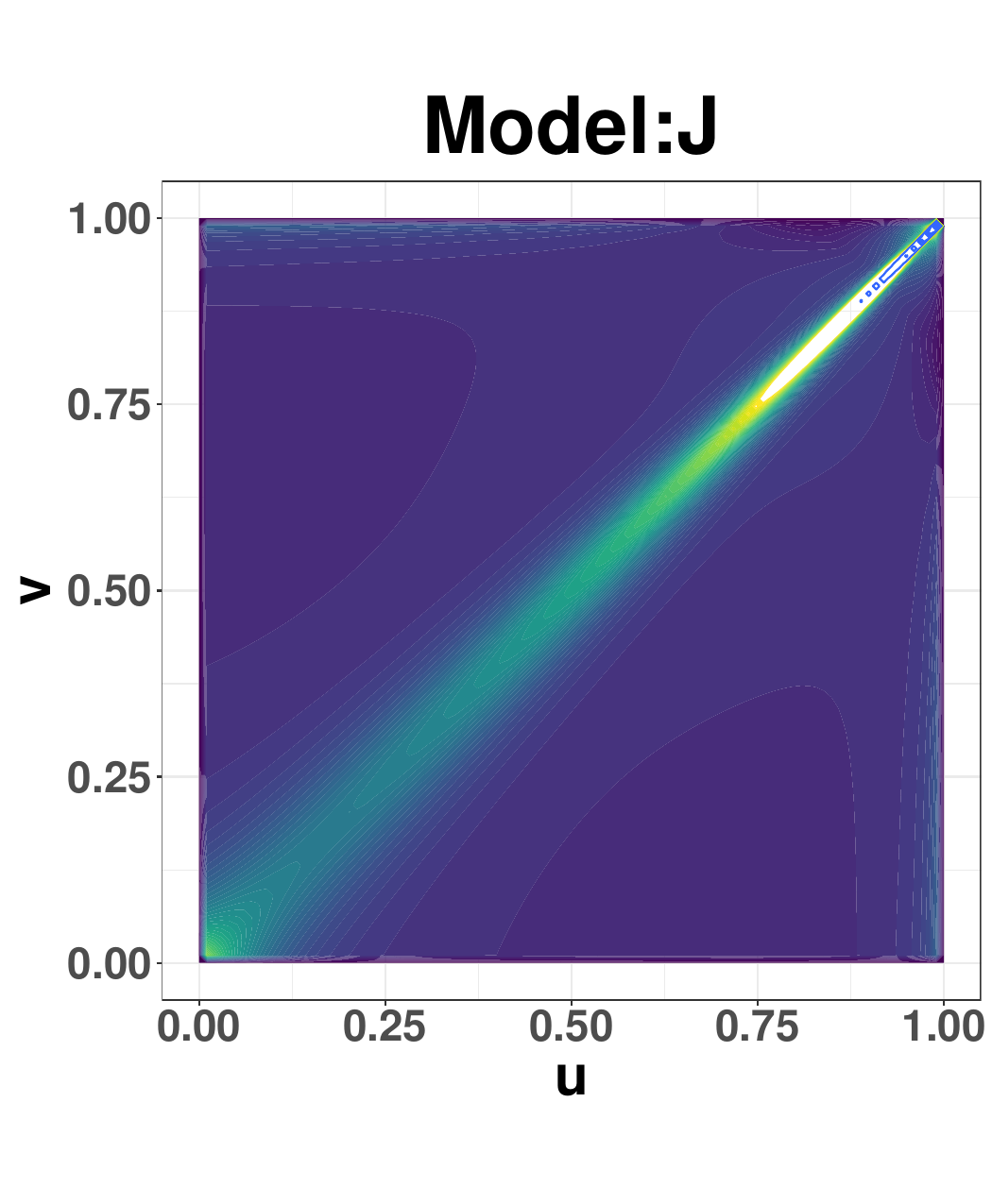}
}
\centerline{
\includegraphics[scale=0.15]{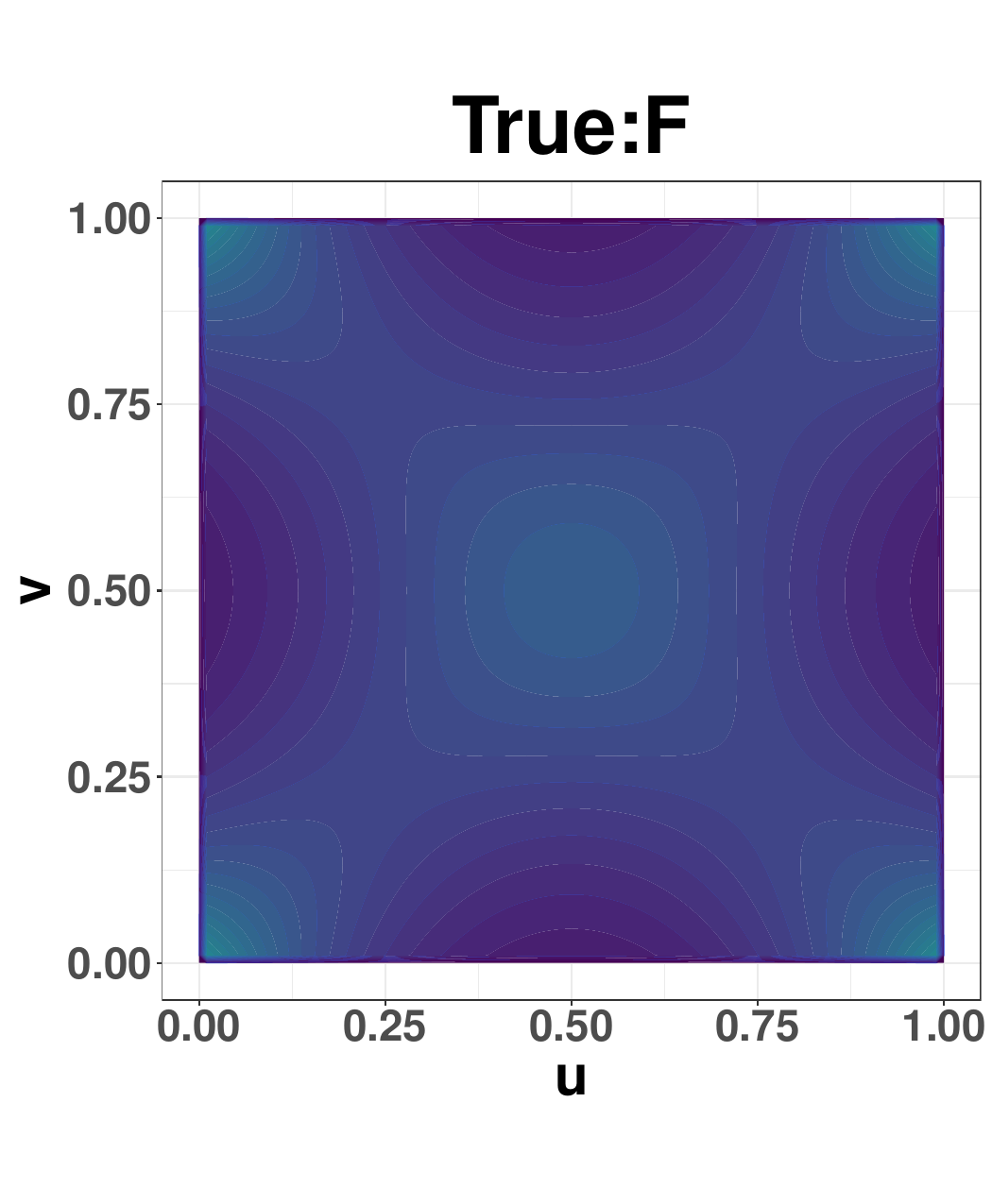}
\includegraphics[scale=0.15]{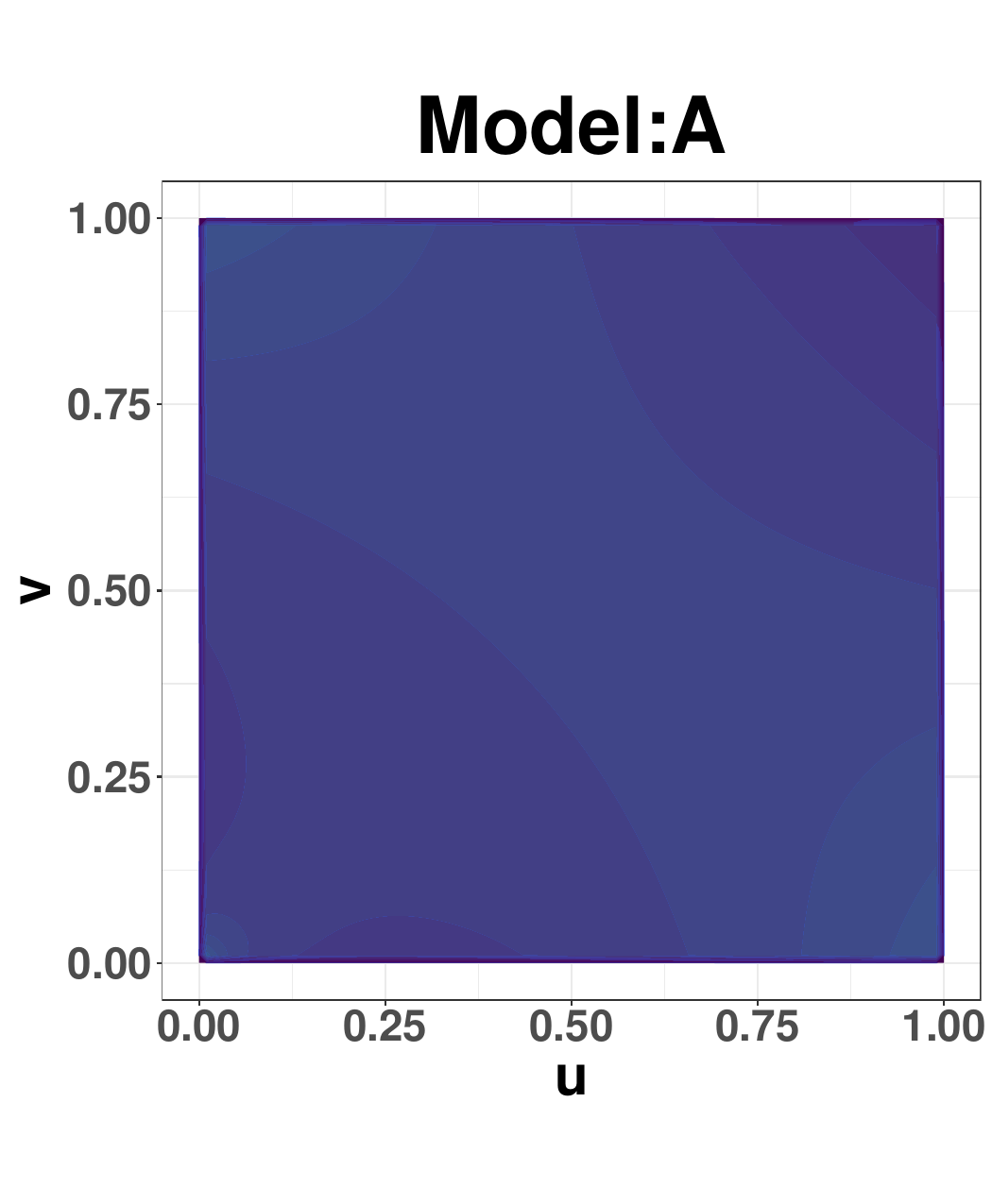}
\includegraphics[scale=0.15]{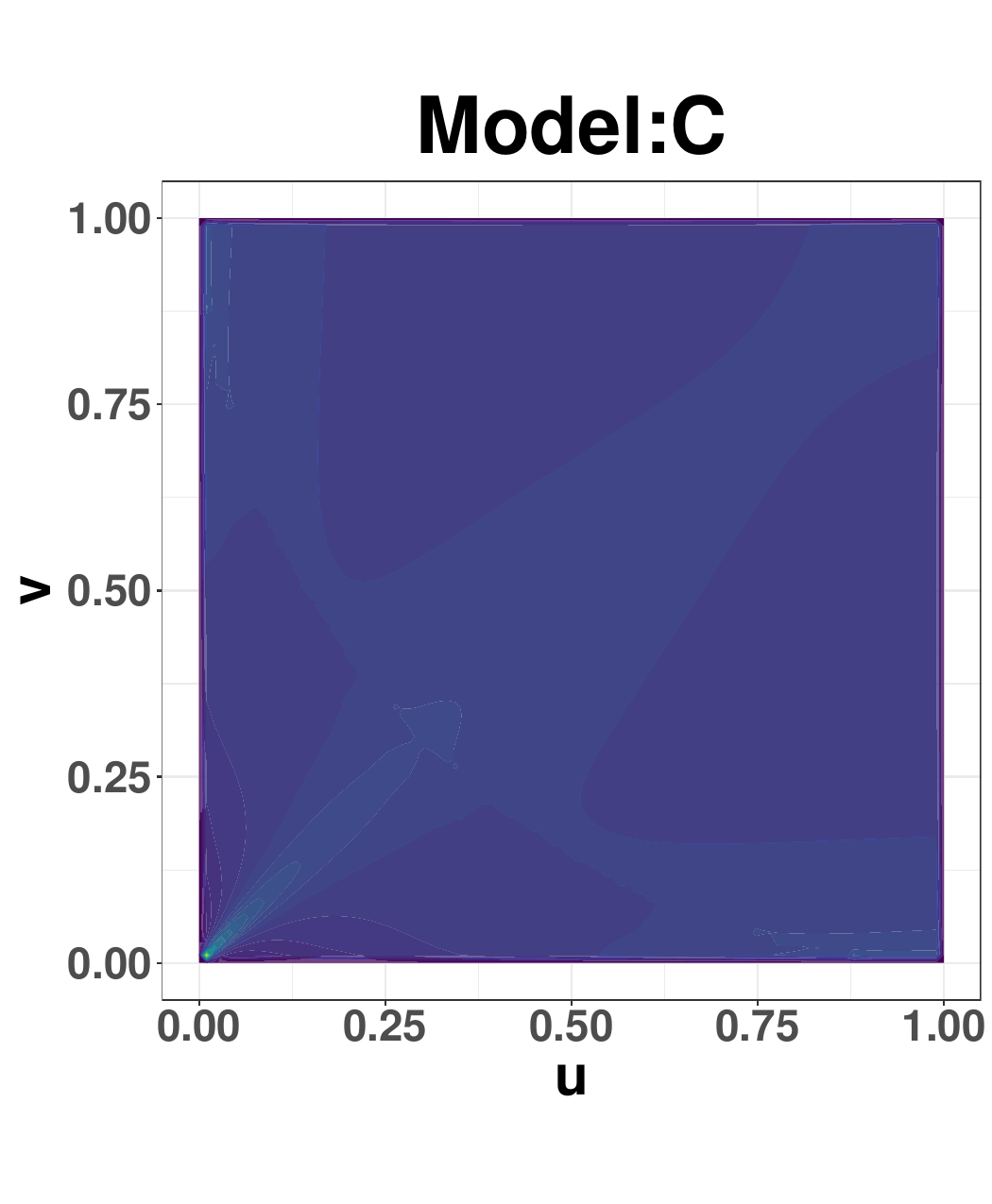}
\includegraphics[scale=0.15]{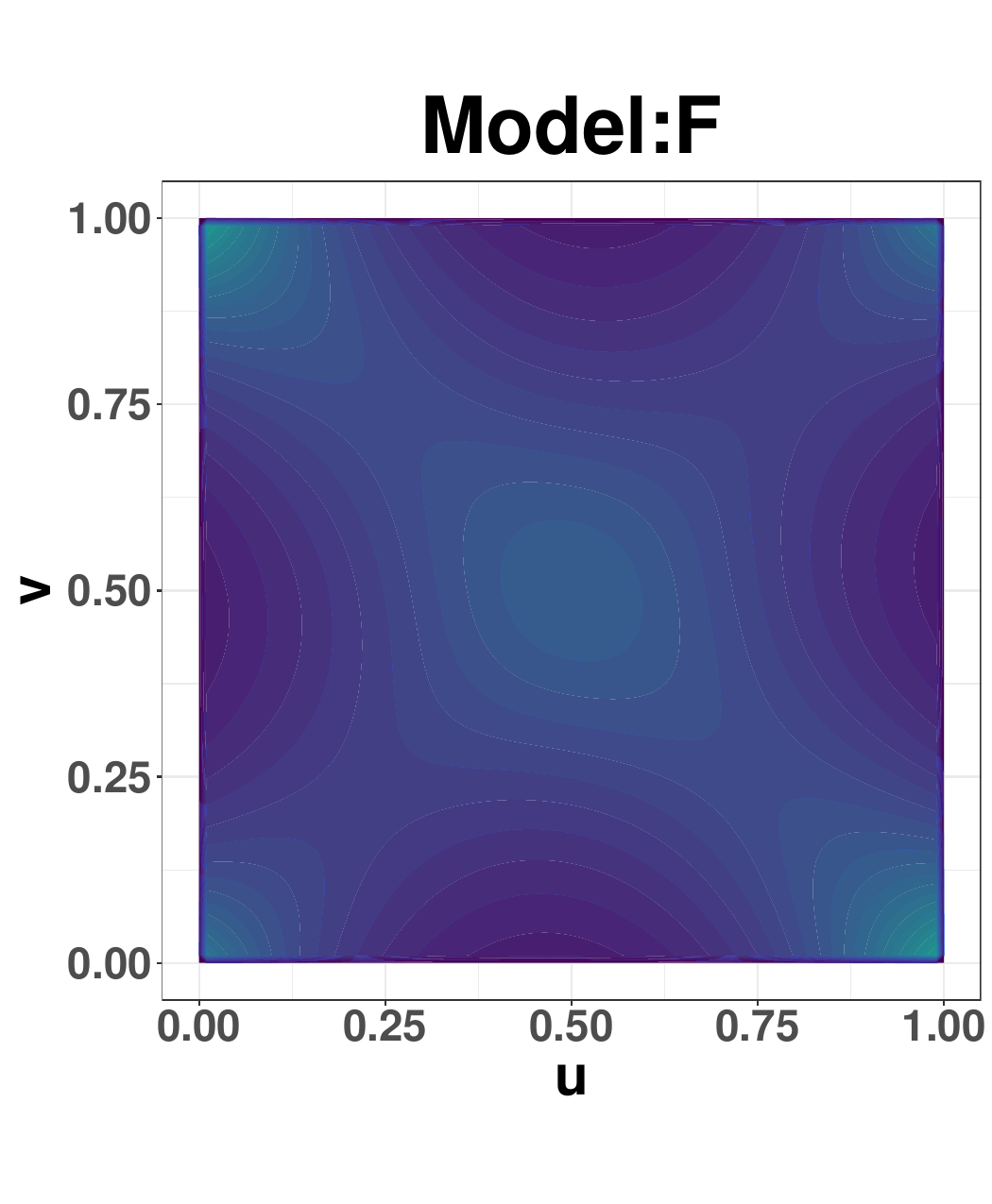}
\includegraphics[scale=0.15]{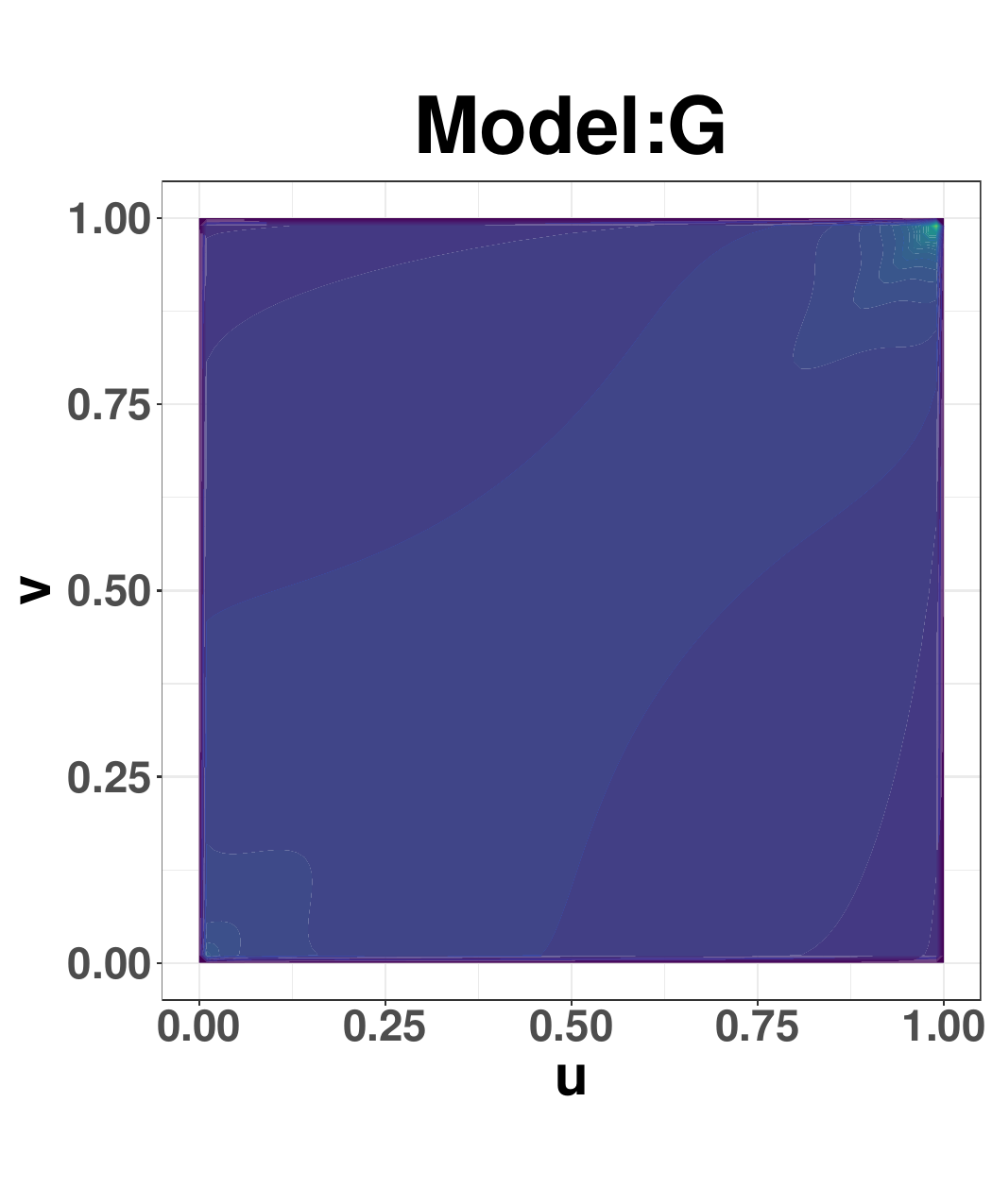}
\includegraphics[scale=0.15]{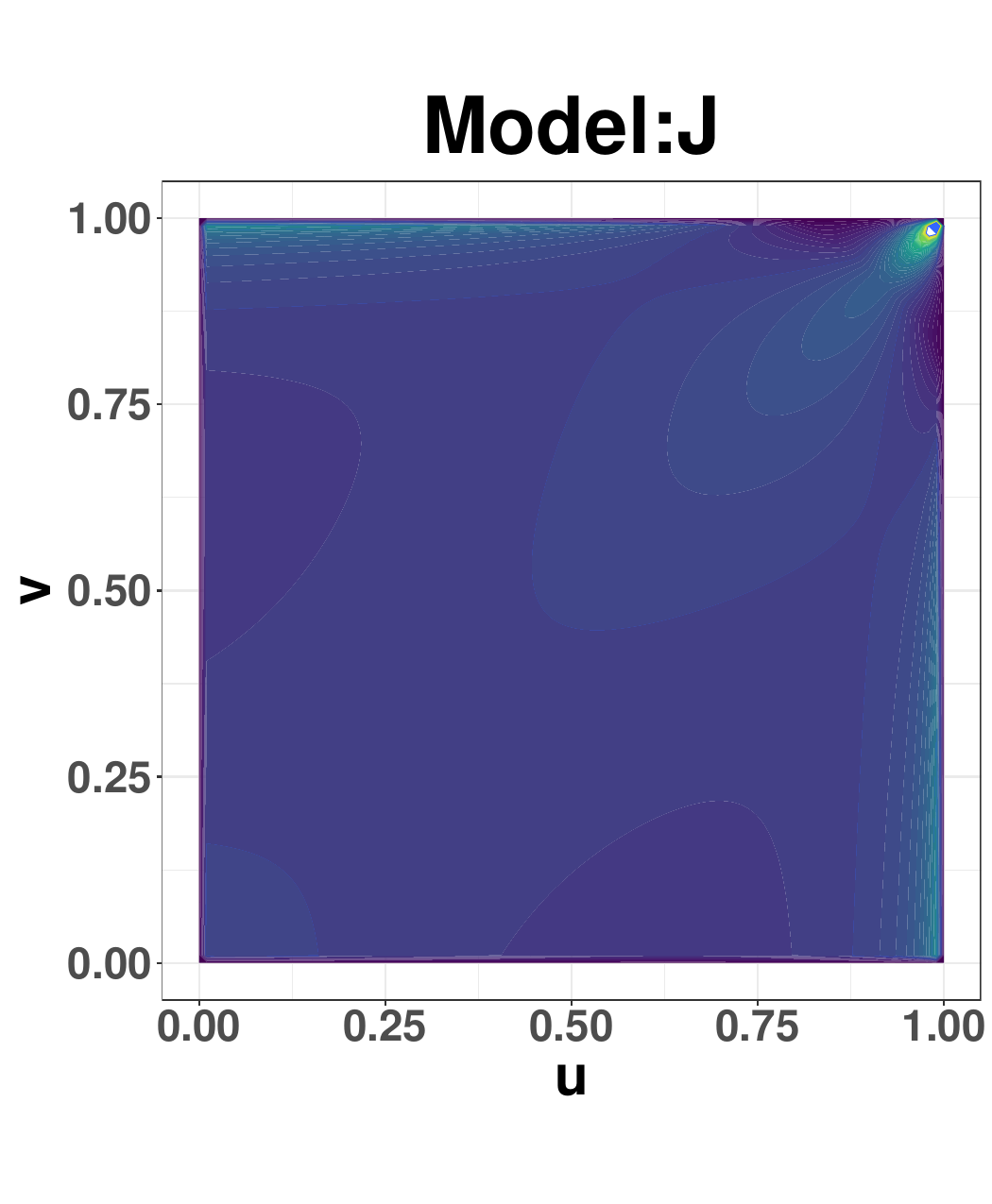}
}
\centerline{
\includegraphics[scale=0.15]{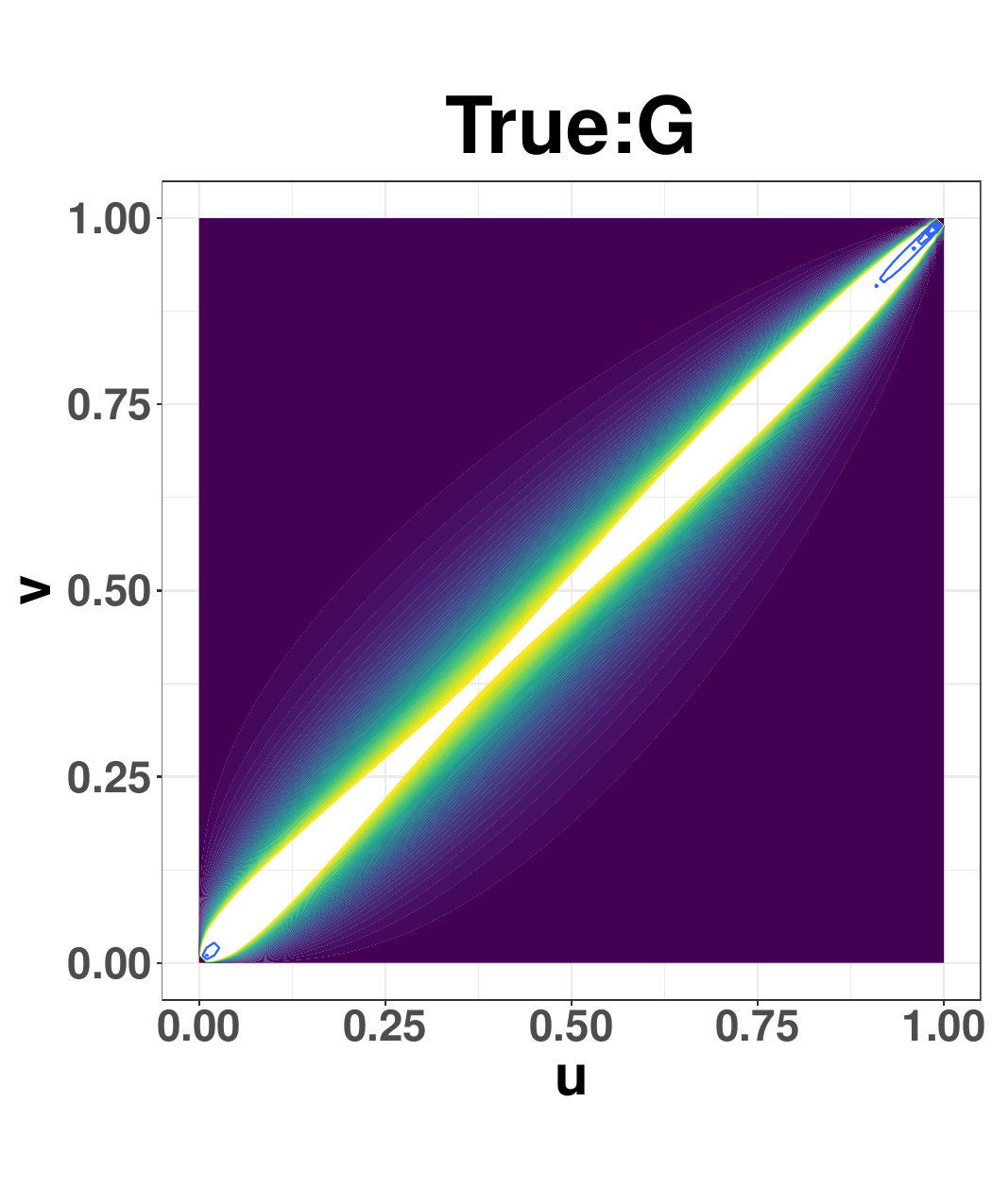}
\includegraphics[scale=0.15]{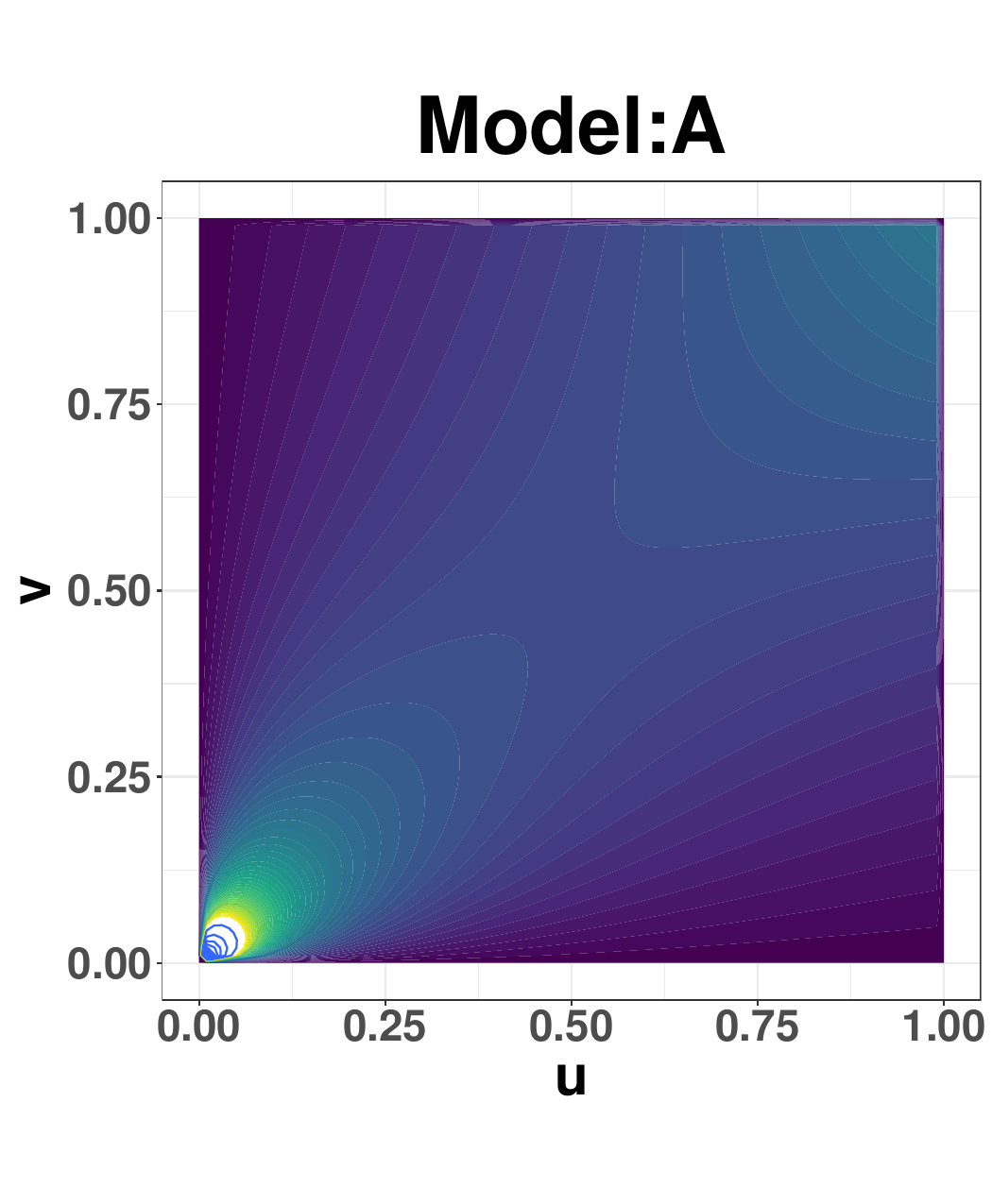}
\includegraphics[scale=0.15]{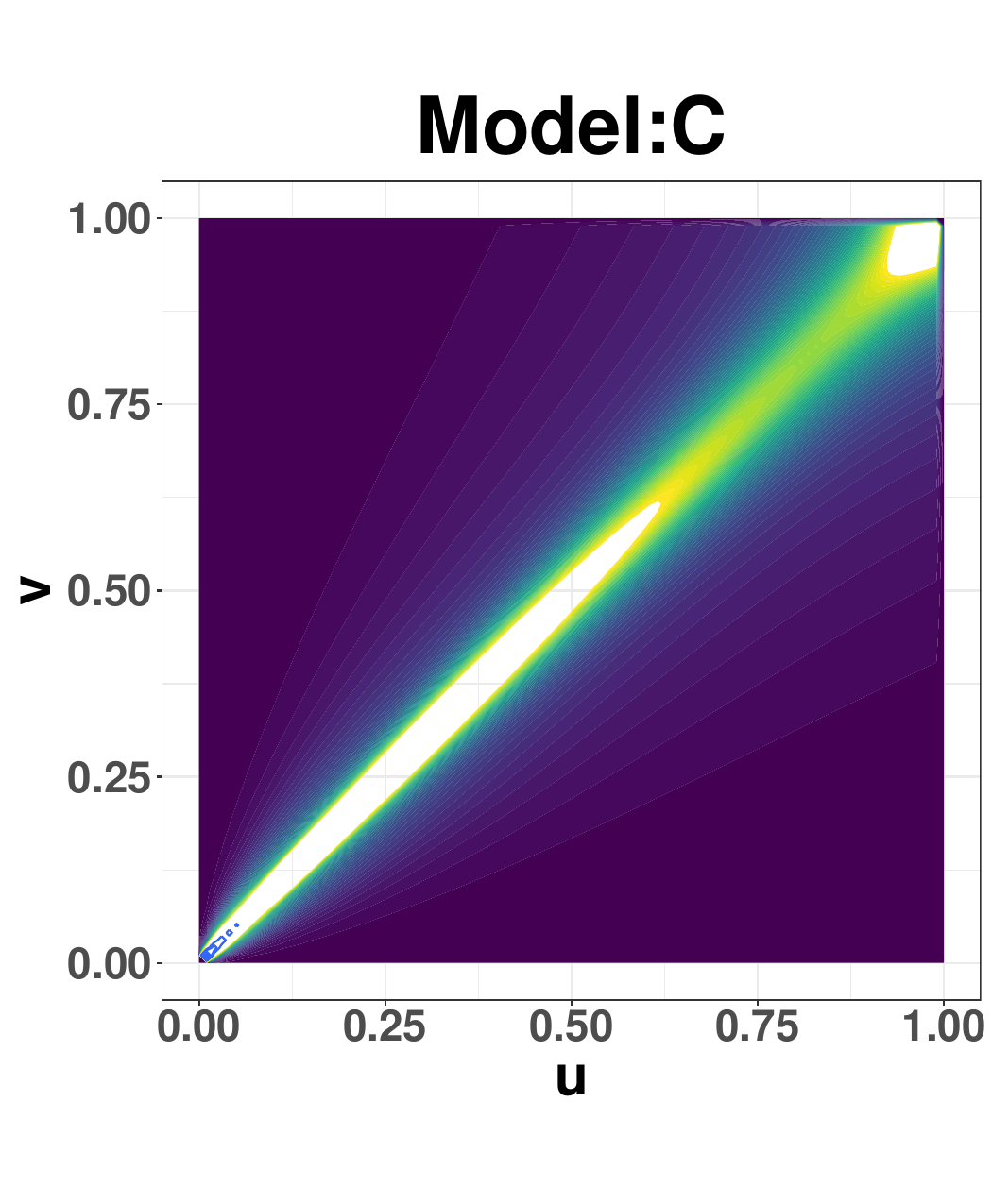}
\includegraphics[scale=0.15]{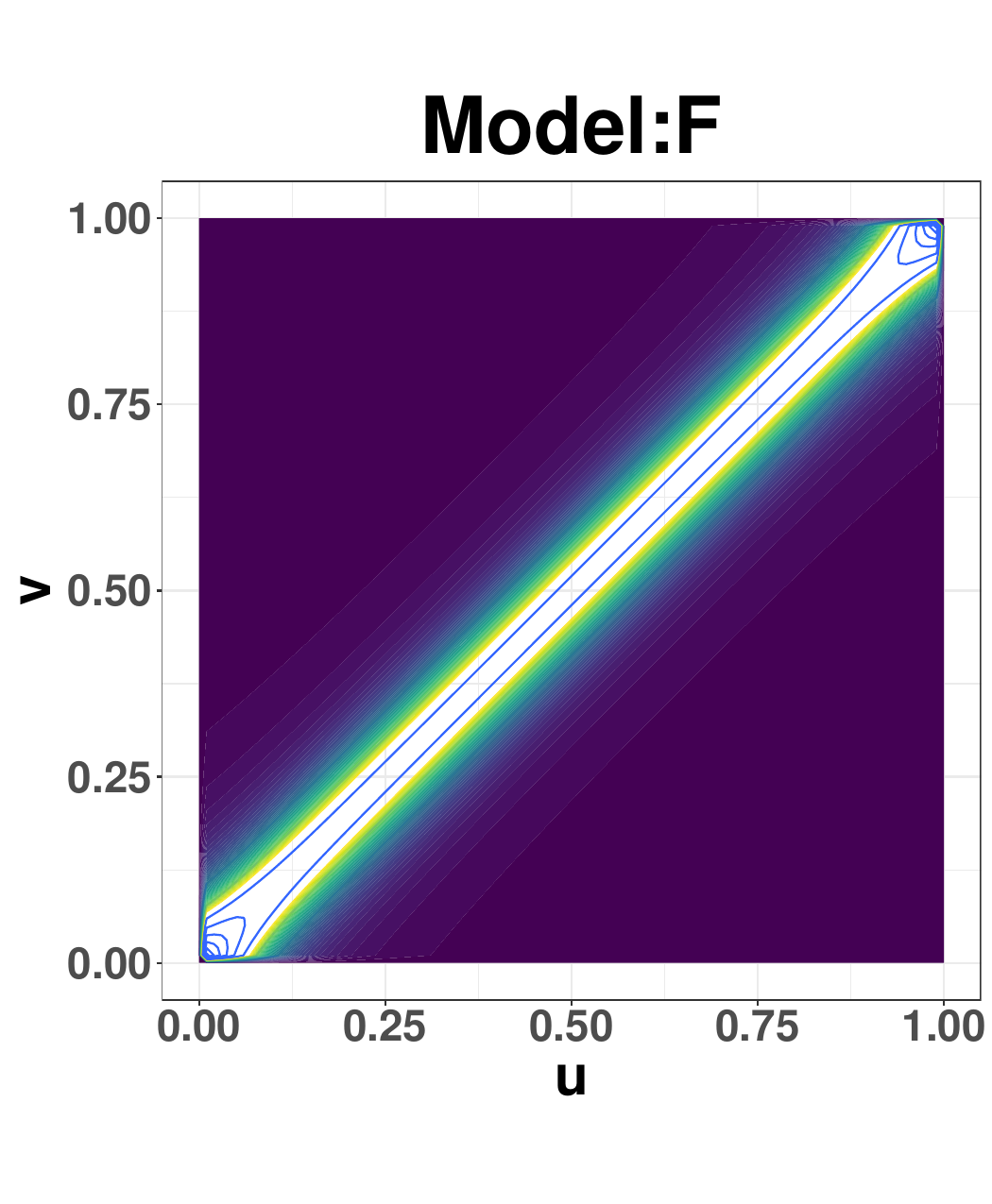}
\includegraphics[scale=0.15]{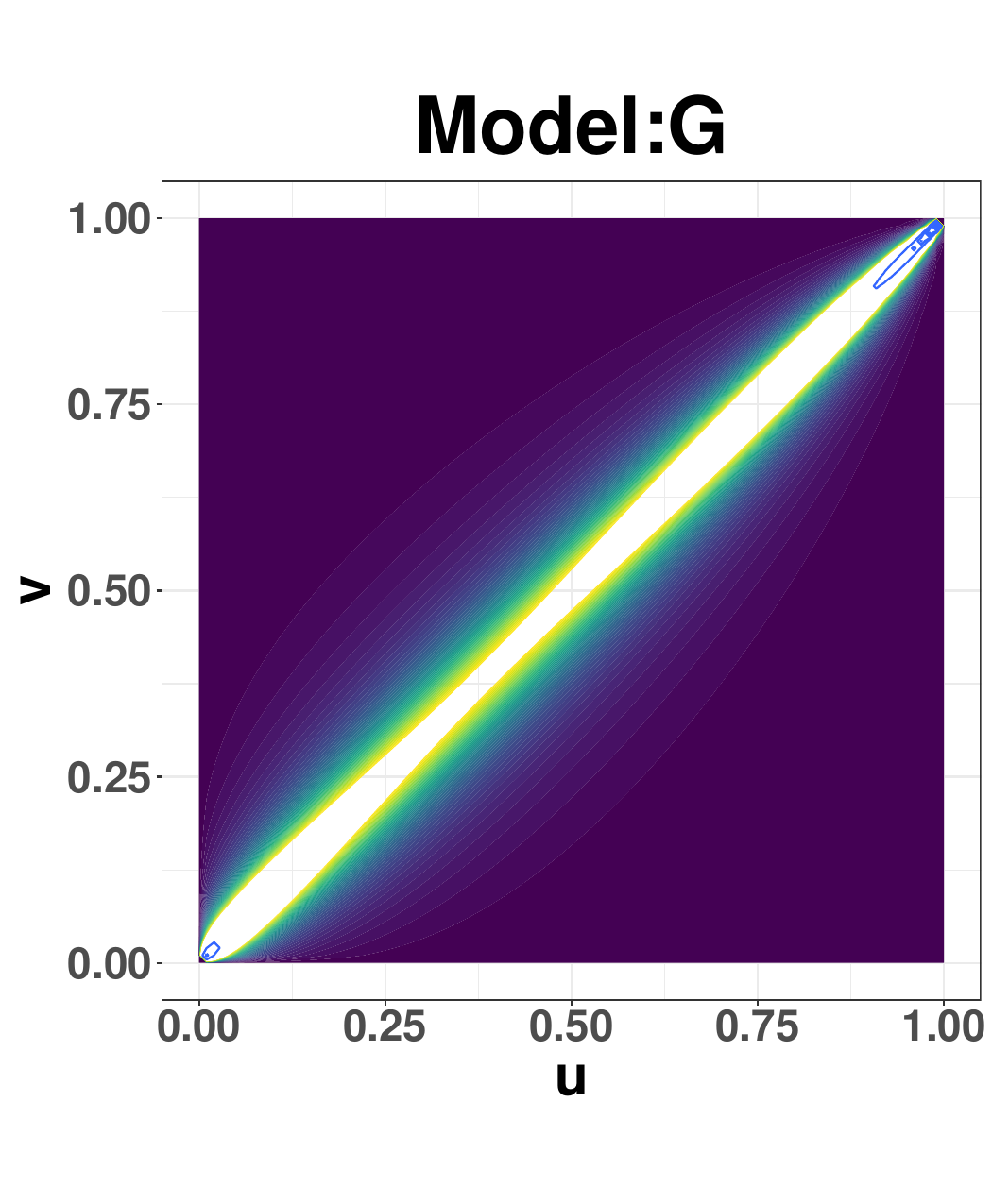}
\includegraphics[scale=0.15]{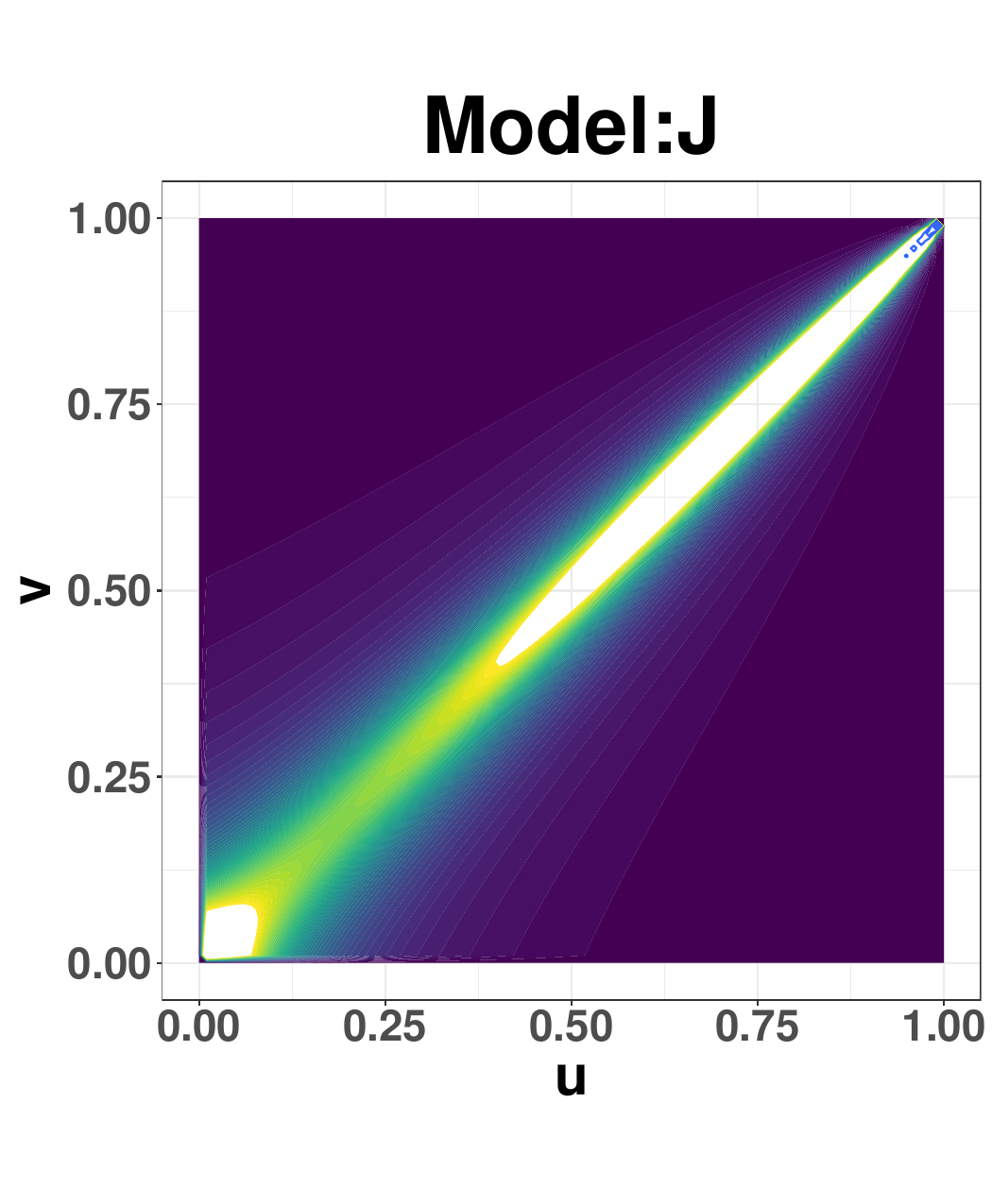}
}
\centerline{
\includegraphics[scale=0.15]{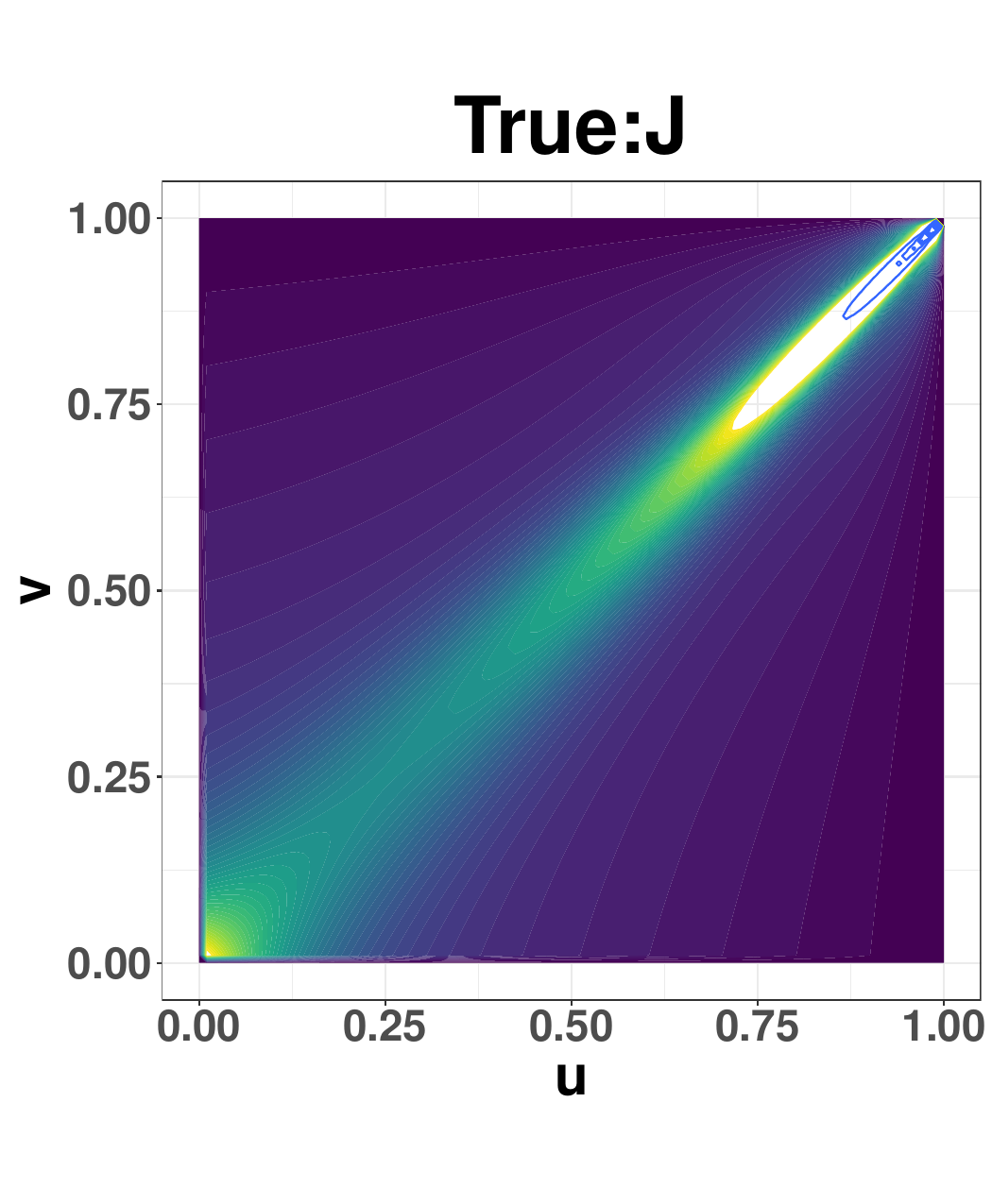}
\includegraphics[scale=0.15]{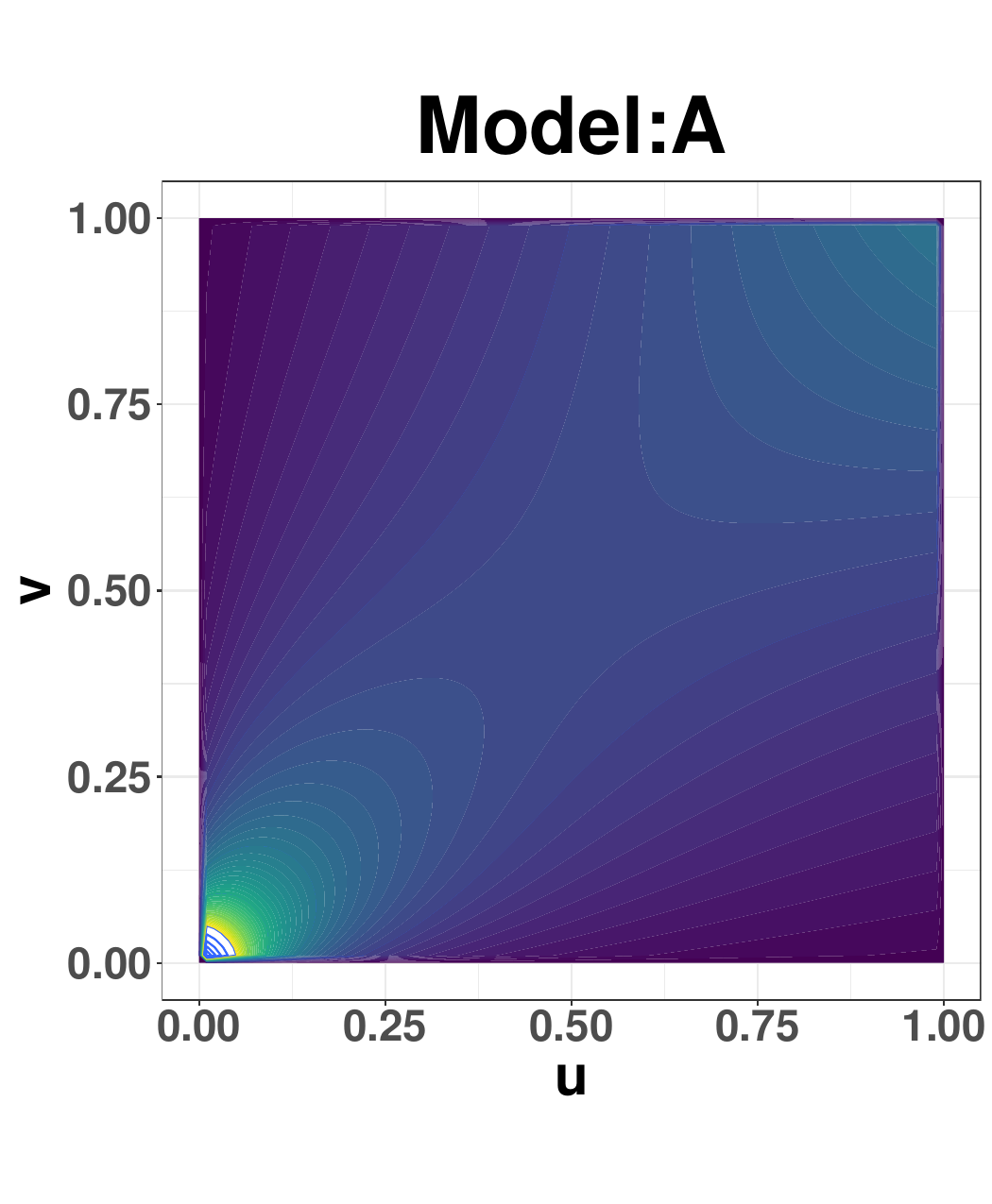}
\includegraphics[scale=0.15]{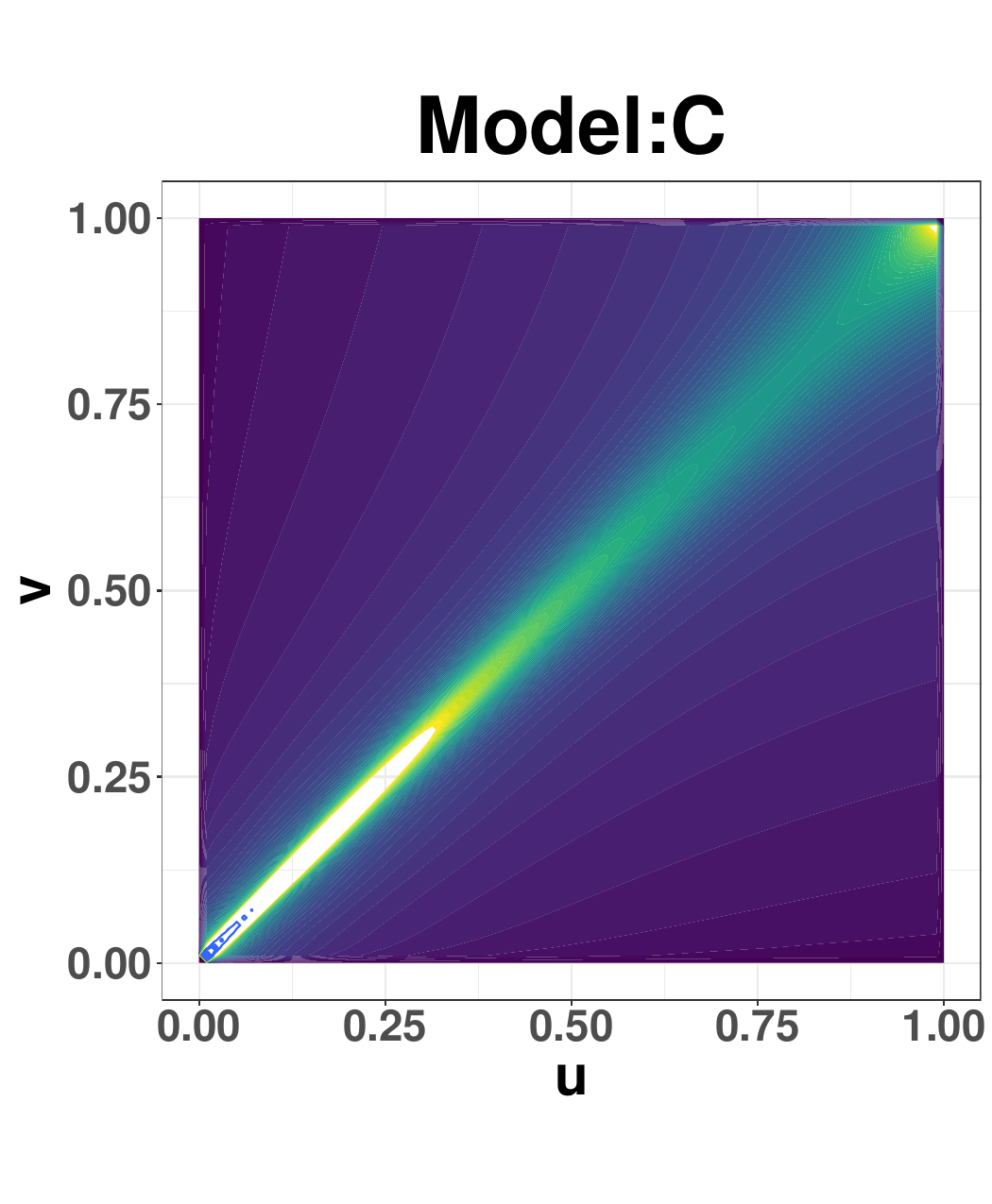}
\includegraphics[scale=0.15]{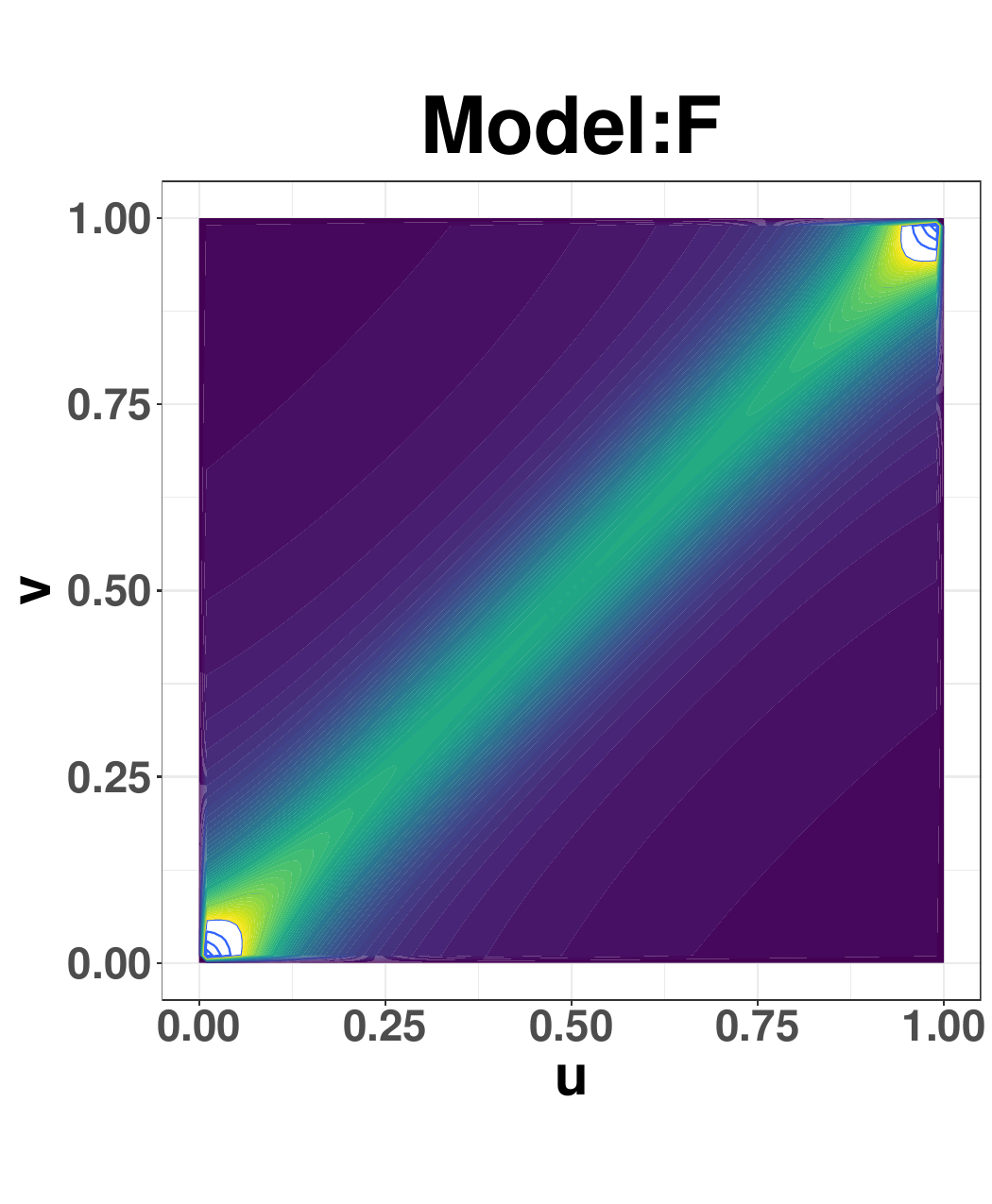}
\includegraphics[scale=0.15]{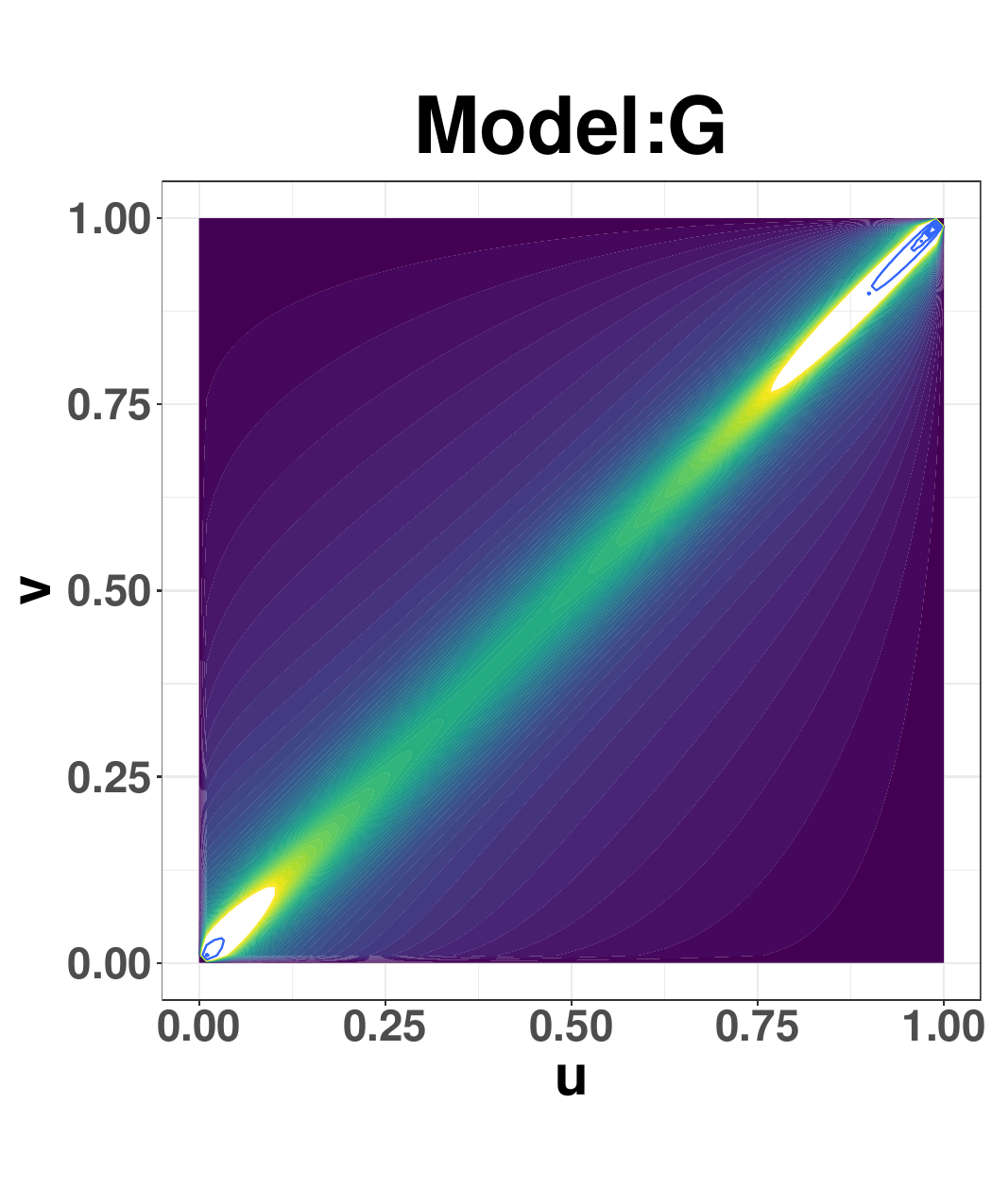}
\includegraphics[scale=0.15]{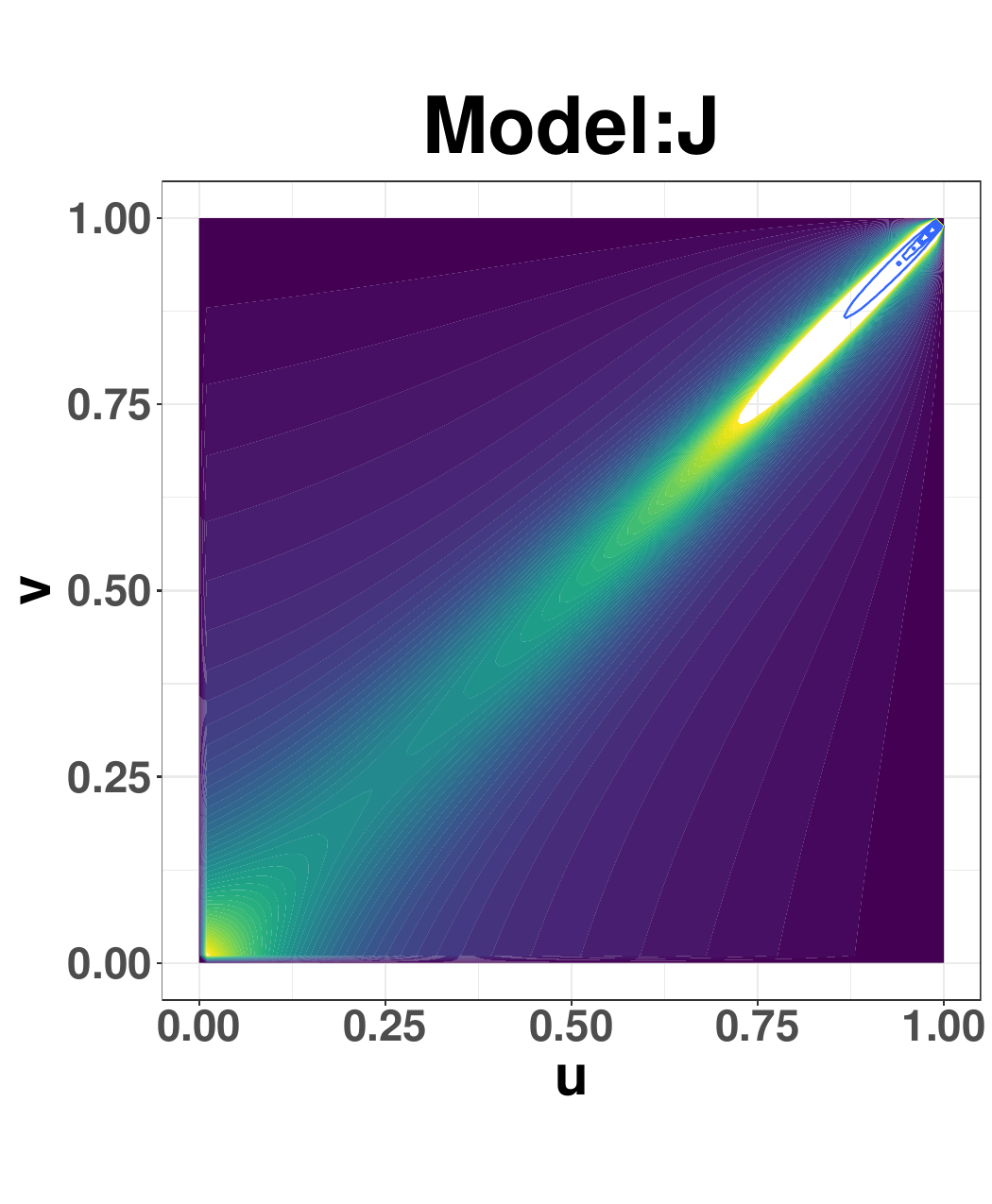}
}
\caption{Simulation study 1: Bivariate data from mixtures of Archimedean copulas with $n=500$. Density estimations. Across columns: True density (1st), AMH fitting (2nd), Clayton fitting (3rd), Frank fitting (4th), Gumbel fitting (5th), and Joe fitting (6th). Data generated model across rows: AMH (1st), Calyton (2nd), Frank (3rd), Gumbel (4th) and Joe (5th).}
\label{fig:simpred}
\end{figure}

In Figure \ref{fig:simpred} we present contour plots of the bivariate densities as heat maps for the true model (first column) and the 25 fits, columns two to six. The first row contains data generated from the AMH copula. The contours from the true model do not show much difference in color due to the weak dependence. The fitted models AMH and Clayton show similar contour patterns compared to the true one, whereas the fit based on the Joe kernel looks very dissimilar. Considering data from the mixture of Clayton copulas (second row), contours show a lower left strong positive dependence, combined with circular contours associated to the negative dependence obtained by the Clayton component with negative parameter. None of the models, apart from the Clayton one, does a good job in capturing the true dependence. 

When the data come from a mixture of Frank copulas (third row in Figure \ref{fig:simpred}), contours of the true density show a cross pattern in the corners. The fitting obtained with the Frank kernel is the only one that replicates the shapes of the contours. In the fourth row, we have the contours from a mixture of Gumbel copulas; apart from the Gumbel itself, the Clayton and Joe models seem to do a reasonable job. Finally, when data are coming from a mixture of Joe copulas, the heat map shows a strong right-upper dependence. Only models based on Joe and Gumbel seem able to capture it.  

\begin{figure}
\centerline{
\includegraphics[scale=0.2]{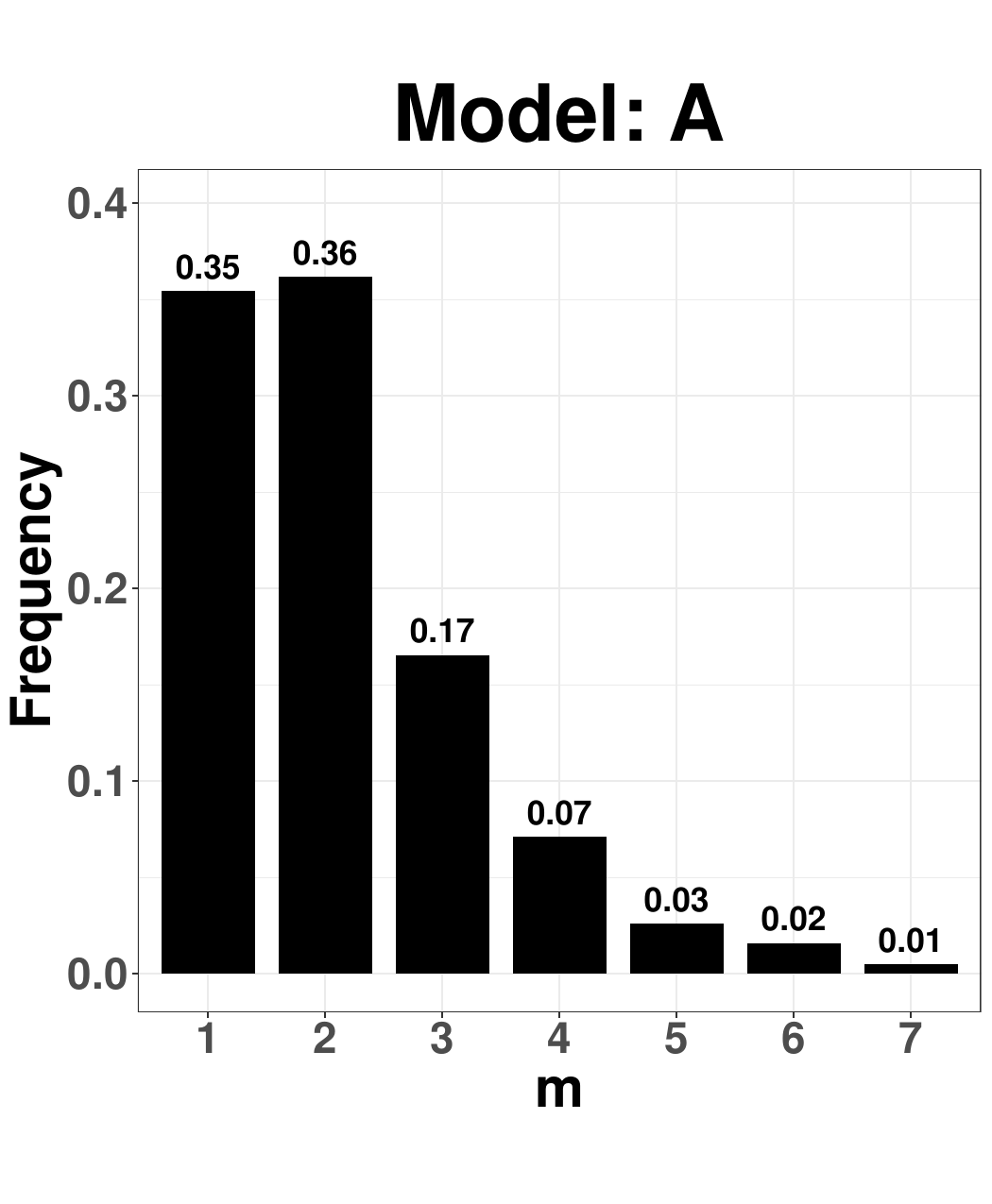}
\includegraphics[scale=0.2]{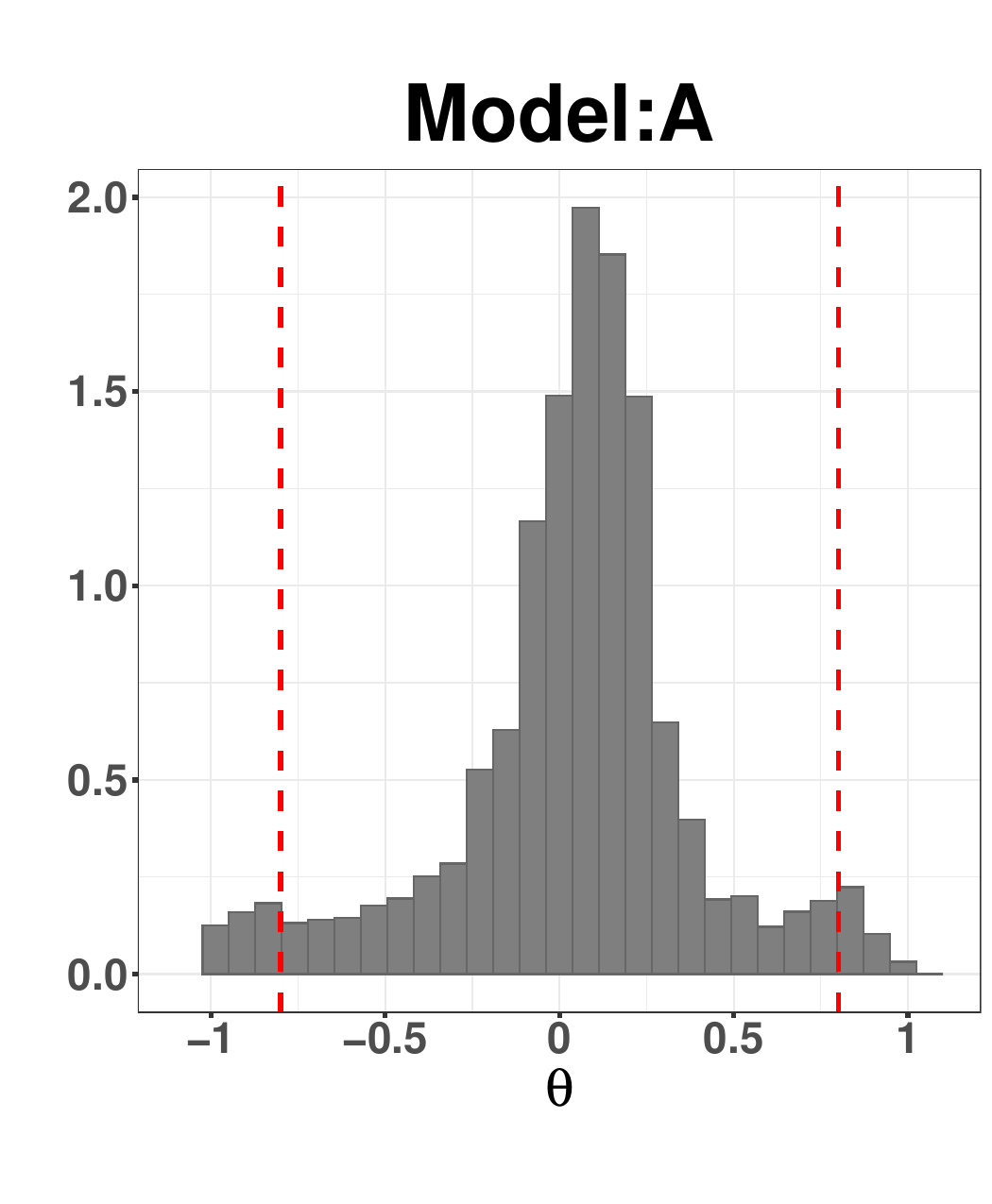}
}
\centerline{
\includegraphics[scale=0.2]{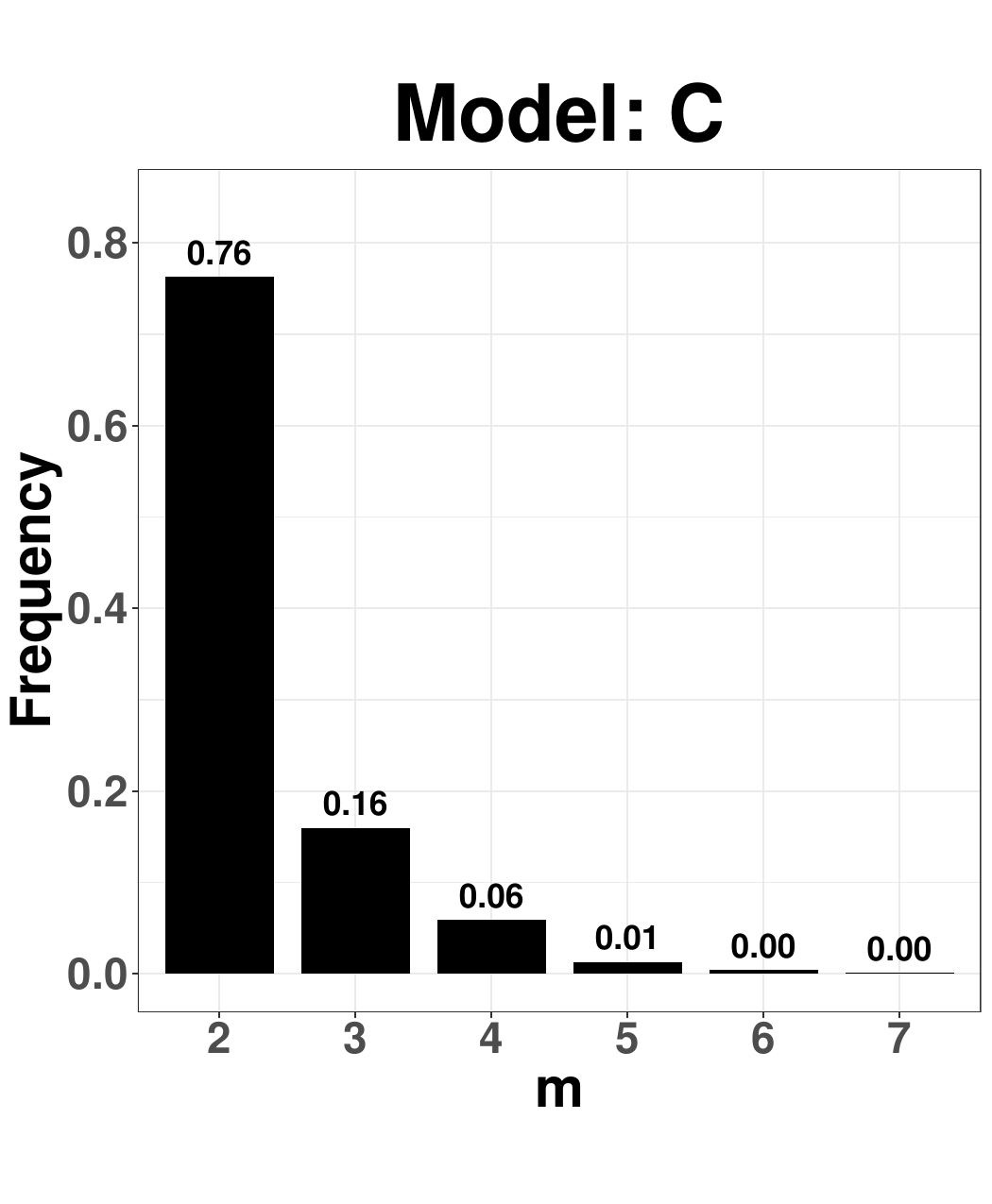}
\includegraphics[scale=0.2]{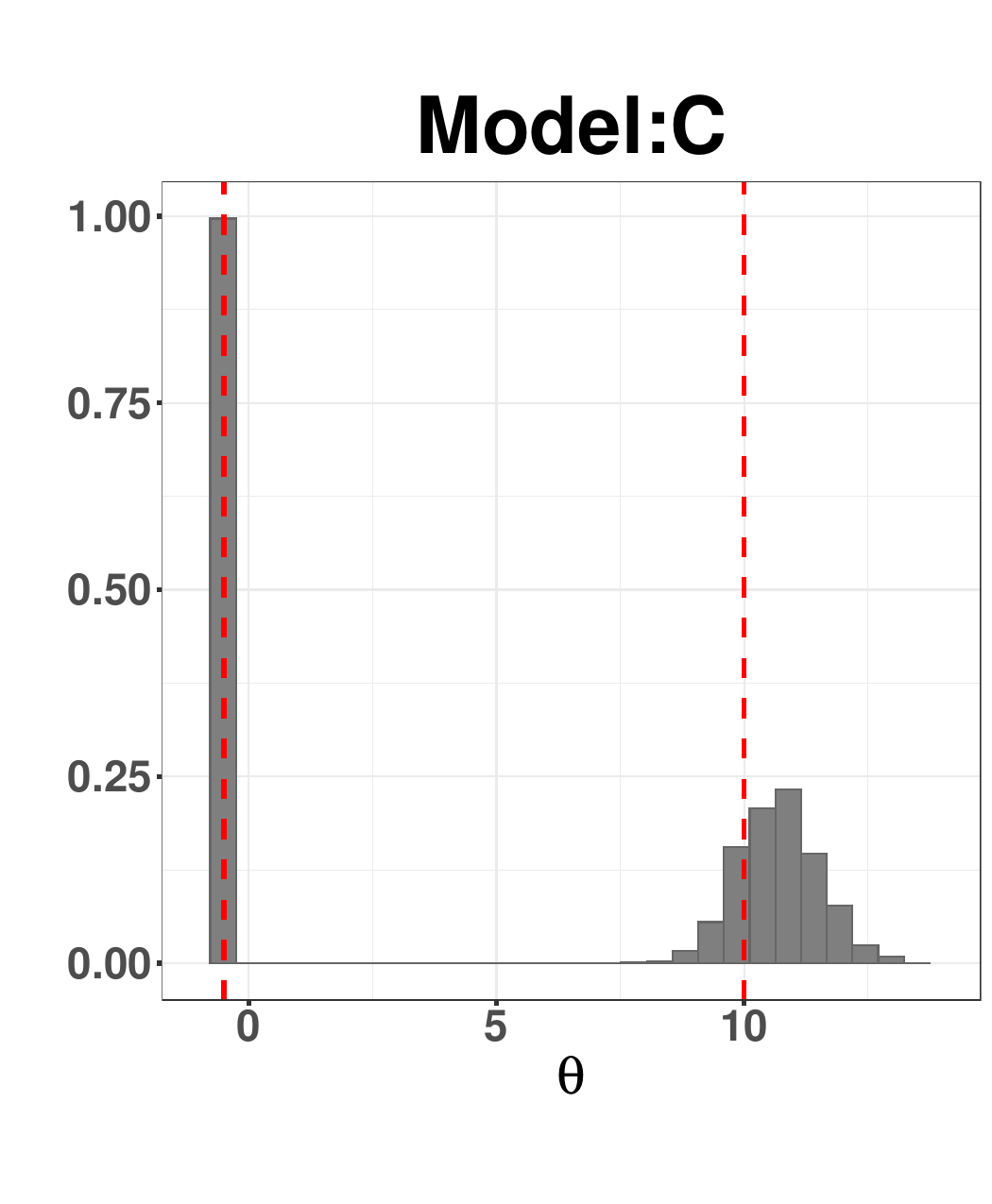}
}
\centerline{
\includegraphics[scale=0.2]{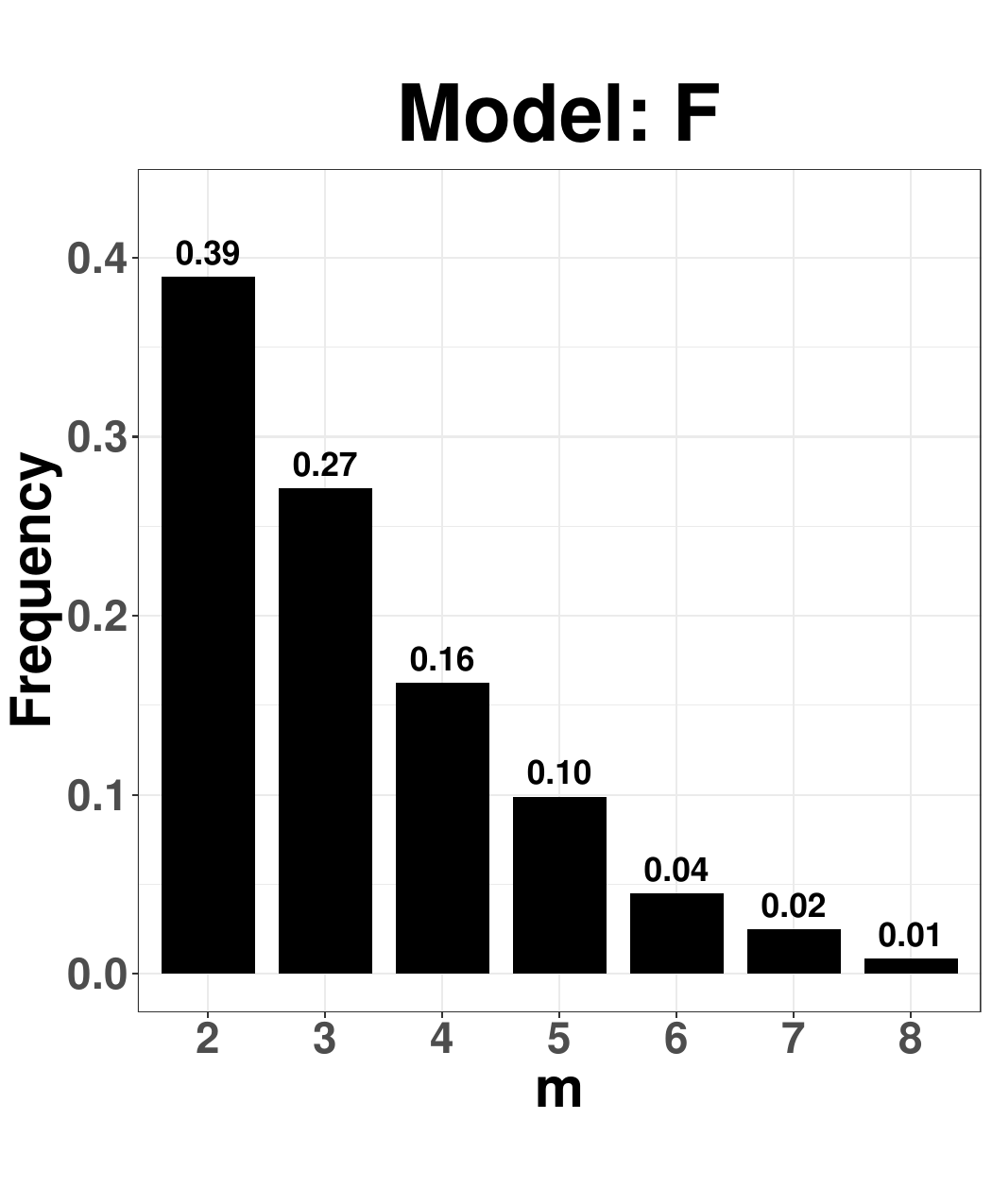}
\includegraphics[scale=0.2]{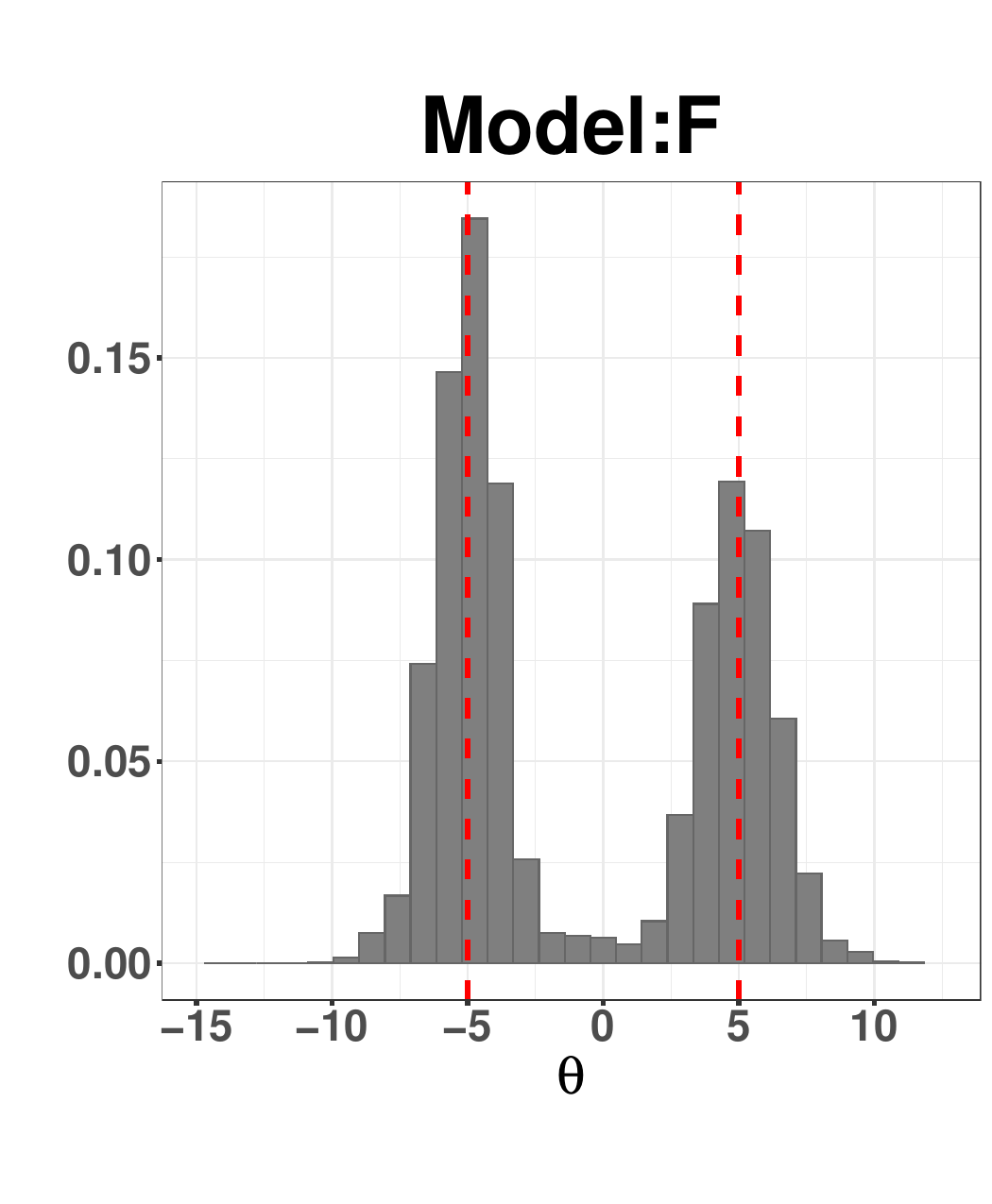}
}
\centerline{
\includegraphics[scale=0.2]{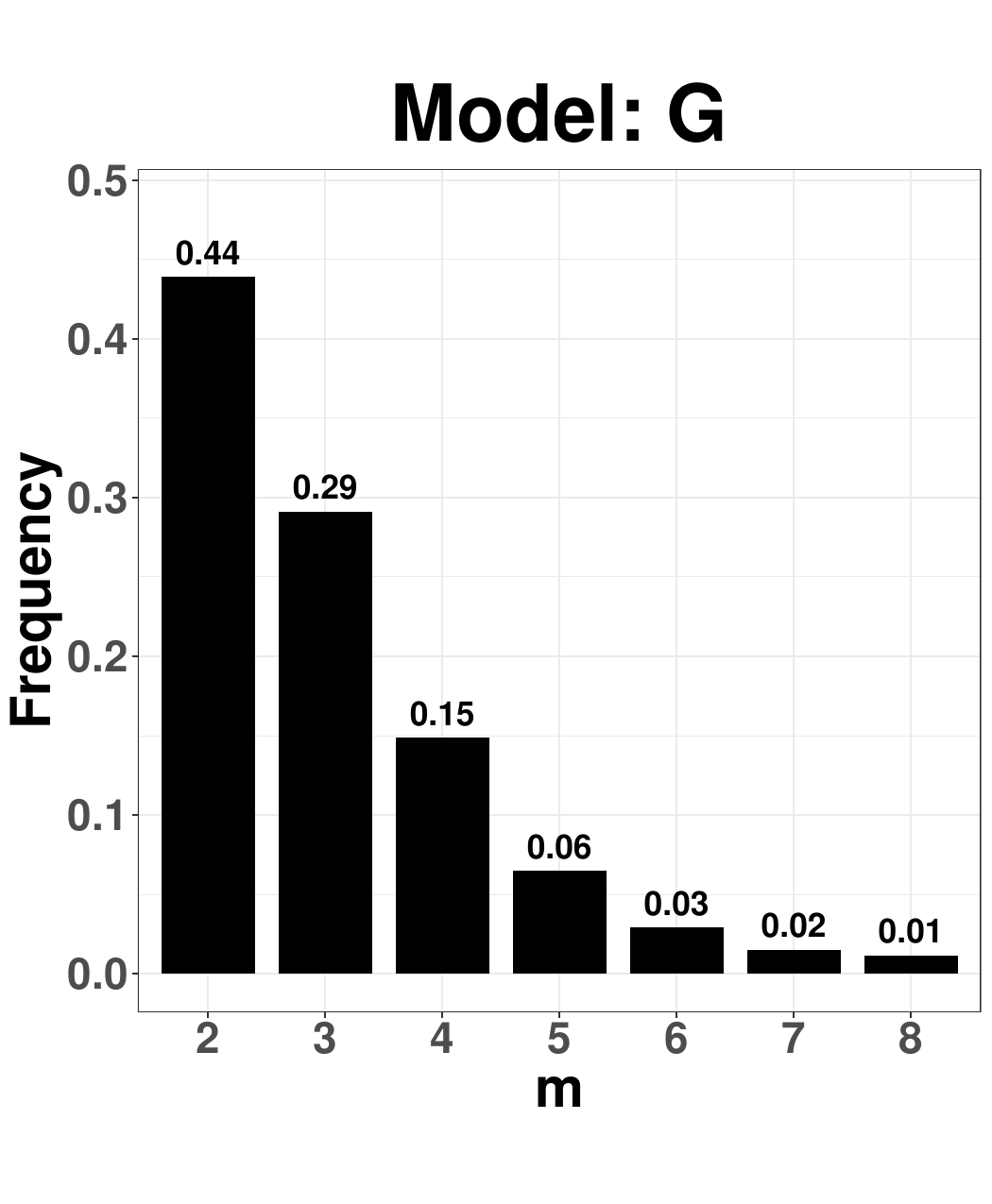}
\includegraphics[scale=0.2]{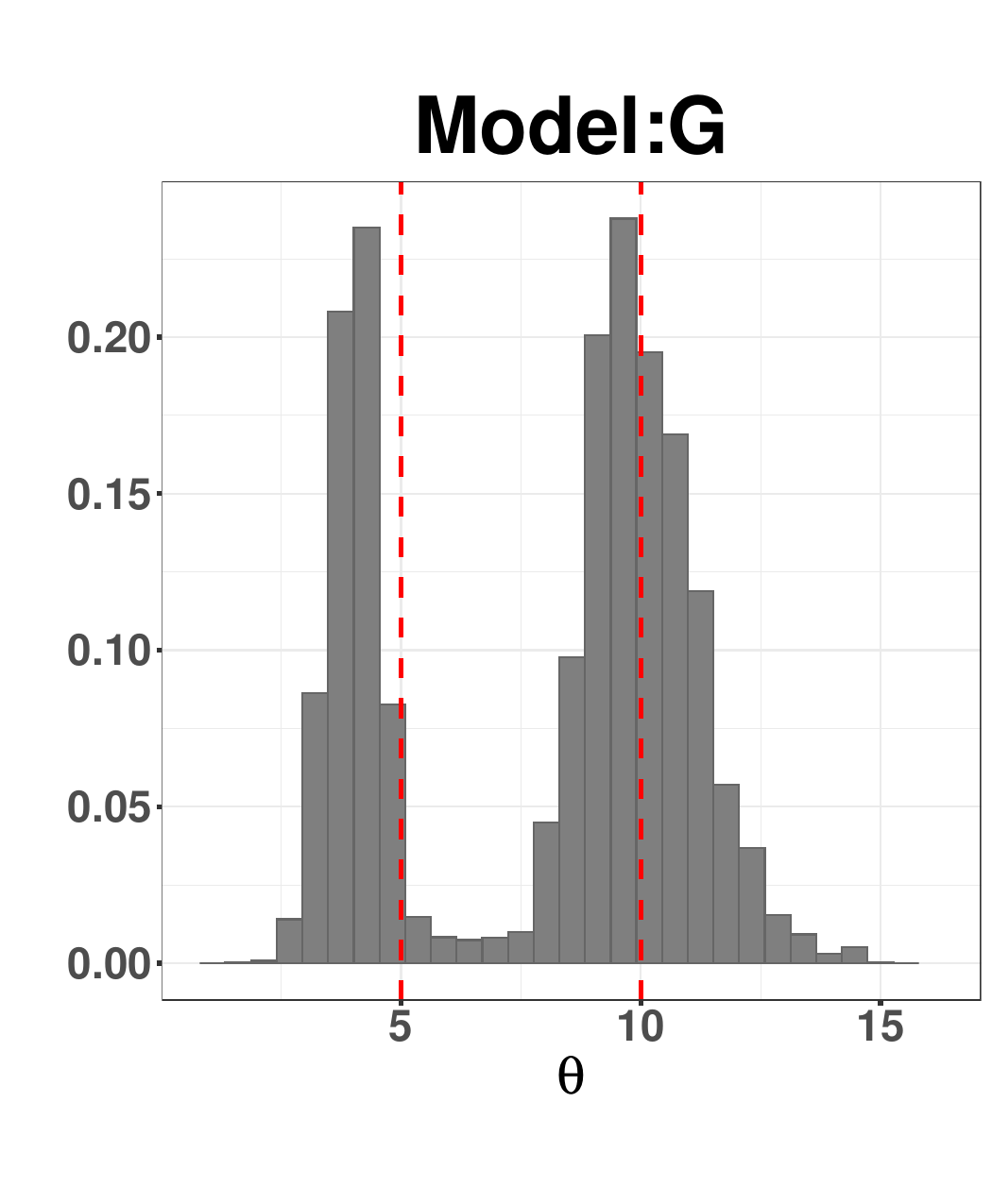}
}
\centerline{
\includegraphics[scale=0.2]{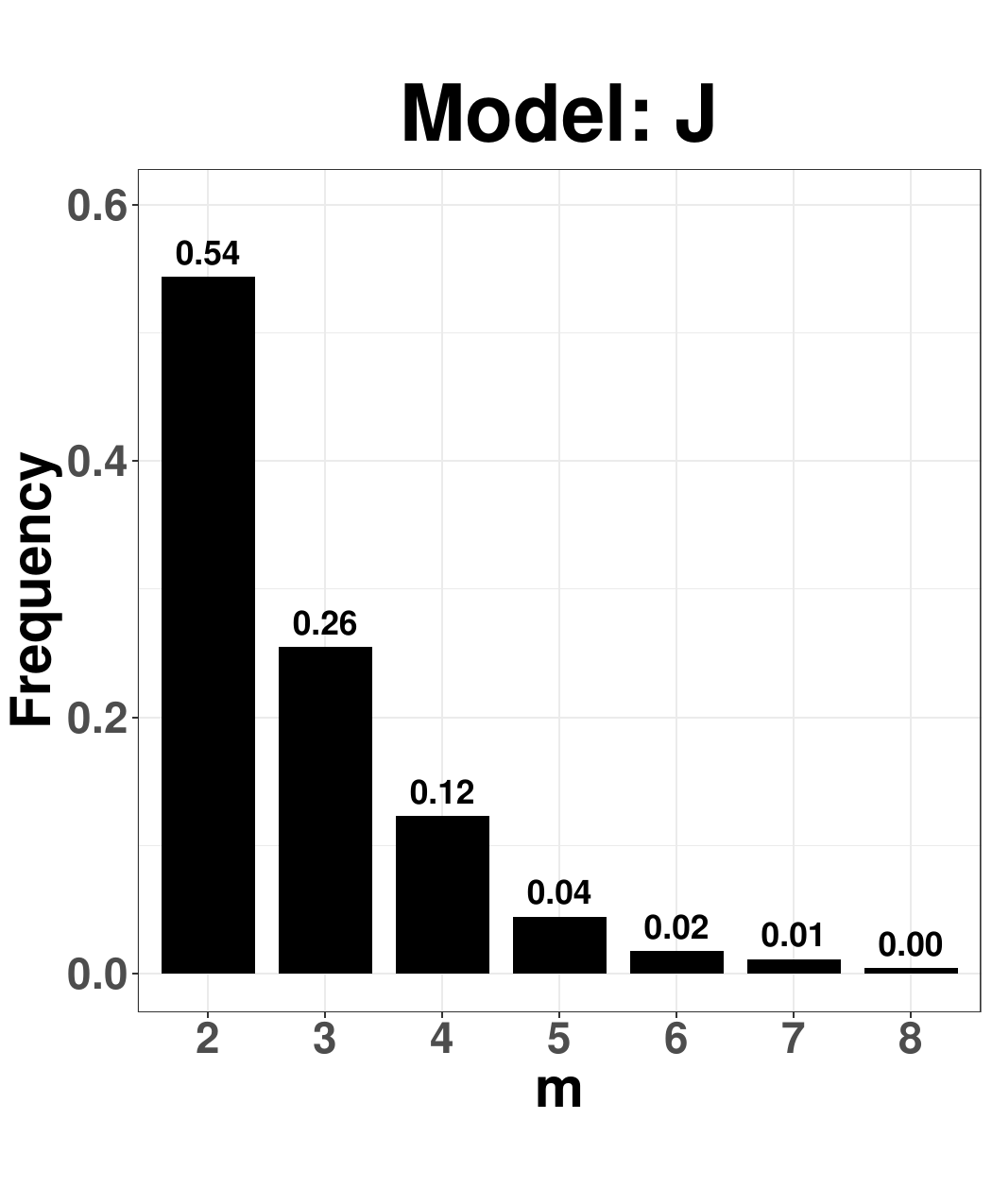}
\includegraphics[scale=0.2]{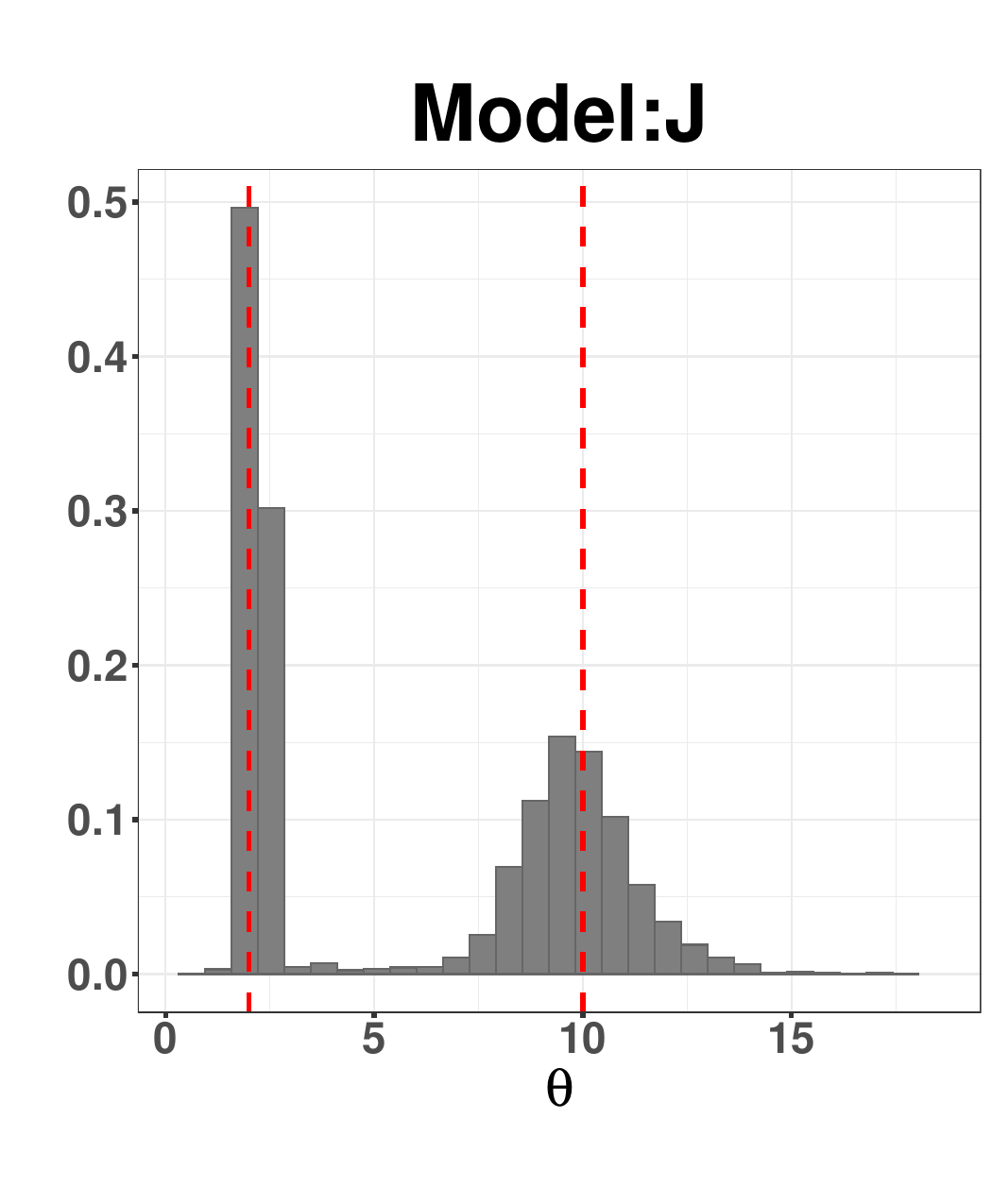}
}
\caption{Simulation study 1: Bivariate data from a mixture of Archimedean copulas with $n=500$. Posterior distributions when using the same kernel used to sample the data: the number of mixture components (1st column) and model parameters (2nd column). Models across rows: AMH (1st), Clayton (2nd), Frank (3rd), Gumbel (4th), and Joe (5th). Vertical dotted lines correspond to the true values.}
\label{fig:postnt}
\end{figure}

One feature of the Poisson-Dirichlet process mixture models is that we can assess the number of mixture components required to fit the data. The posterior distribution for the number of components with $n=500$ when using the same kernel as that used to sample the data is presented in the first column of Figure \ref{fig:postnt}. Furthermore, in the second column of the same figure, we present the histogram for all parameters of the mixture components $\theta_i$ combined. 

For all cases, the posterior distribution of the number of components is unimodal with a mode at $m=2$ components, which confirms that the mixing distribution given by the Poisson-Dirichlet process is able to recover the truth. Equally comforting, the posterior distribution of the parameters is also bimodal around the true values for all but one case, the AMH being the exception. This posterior distribution shows a high mode around zero with two tiny modes around the true parameters $(-0.8,0.8)$. Since the dependence induced by the AMH copula is very weak, the model is not able to detect the true values of the parameters. However, the density estimation is accurate. 

For the sample size of $n=500$, we also report the true, empirical, and estimated Kendall tau values (posterior mean) in Table \ref{tab:simktau}. Remarkably, most copulas do a good job of estimating the true association parameter. Exceptions are AMH and Gumbel when the true model is Clayton. To assess convergence of our estimates to the true values for larger sample sizes, 95\% credible intervals (CI), for the Kendall tau in the Gumbel case, are $(0.69,0.92)$ for $n=200$ and $(0.81,0.88)$ for $n=500$. Clearly, the intervals shrink to the true value of $0.85$. 

As a last inferential procedure, we select the best clustering configuration using the R package \texttt{salso}. In Table \ref{tab:postmcmc1} we report the post-MCMC summaries of the copula parameters, $\theta_j^*$, together with the posterior mean weight assigned to each component for $j=1,\ldots,\tilde{m}$. We only report the fit obtained when the copula family is the same as the one used to sample the data.

For the AMH case, we select one component whose copula parameter takes values in the 95\% CI $(-0.24,0.33)$. The model does not capture the true values of the parameters or the number of mixture components, most likely because of the weak dependence in the AMH copula. For the Clayton case, our model selects two components with 95\% CI's for copula parameters $(-0.504,-0.496)$ and $(9.896,12.329)$ which clearly contain the true values. 

In the Frank case, our clustering selection procedure chooses two mixture components;  and the estimated copula parameters for these two components are around the true values. For the Gumbel case, we also select two mixture components. The estimated copula parameters have 95\% CIs of $(8.64,10.09)$ and $(2.48,3.37)$, where the latter does not cover the true value of 5. Finally, for the Joe case, we select two mixture components. The estimated 95\% CIs for these two components contain the true value of 2 but not the true value of 10. 

As a final analysis, we compare our model with the BSA model of \cite{hoyos&nieto:20}, which relies on a generator defined in terms of a quadratic spline. The LPML statistics obtained with this competing model are reported in the last column of Tables \ref{tab:lpml200} and \ref{tab:lpml500}. Clearly, the proposed mixture model is superior for most of the kernels used. 

\subsection{Simulation study 2}

To better place our mixture model in context, we generate data from a non-Archimedean copula. In particular, we consider the bivariate Gaussian copula with parameter $\rho=0.7$ and a two-component mixture of Gaussian copulas with equal weights and parameters $\rho_1=-0.7$ and $\rho_2=0.7$. Recall that in a Gaussian copula, the relationship between Kendall $\tau$ association parameter and Pearson's correlation is $\tau=\frac{2}{\pi}\arcsin(\rho)$ \citep{fang2002meta}. Therefore, in the first case we have $\tau\approx0.49$ and in the mixture case we have an overall $\tau=0$. 

For both cases, we took a sample of size $n=500$. We fitted our Bayesian nonparametric mixture model using each of the five members of the Archimedean family as mixture densities. The prior specifications of our model are the same as in Simulation Study 1. Additionally, MCMC was run for 15,000 iterations with a burn-in of 5,000 and a thinning of 5. LPML statistic was computed to assess the fit of the model, together with point and 95\% CI estimates for the overall Kendall tau. These values are reported in Table \ref{tab:ss2}. 

For the single Gaussian copula (left half of Table \ref{tab:ss2}), according to the LPML, the best proposed model is the Gumbel followed by the Frank. These two models produce a Kendall tau point estimate that is very close to the true value $\tau=0.49$. However, all models, except the AMH produce a credible interval that contains the true value of $\tau$. For comparison, we also fitted a Gaussian copula benchmark model with a uniform prior for $\rho$. The fit statistics are reported in the last row of Table \ref{tab:ss2}. As expected, this model obtains the best fit, but the LPML statistic of the Gumbel mixture is not far off. 

For the mixture of Gaussians copula (right half of Table \ref{tab:ss2}), the best fit is obtained by the Clayton, followed by the Frank. Kendall tau point estimates for the five kernels are all close to the true value of $\tau=0$. Four of the five kernels produce credible intervals that contain the true value of $\tau$, the exception being the Gumbel, which allows only positive dependence. We also fitted a single Gaussian copula and although the Kendall tau point and interval estimates are good, the LPML is really bad, as expected. 

\begin{figure}
\centerline{
  \includegraphics[scale=0.3]{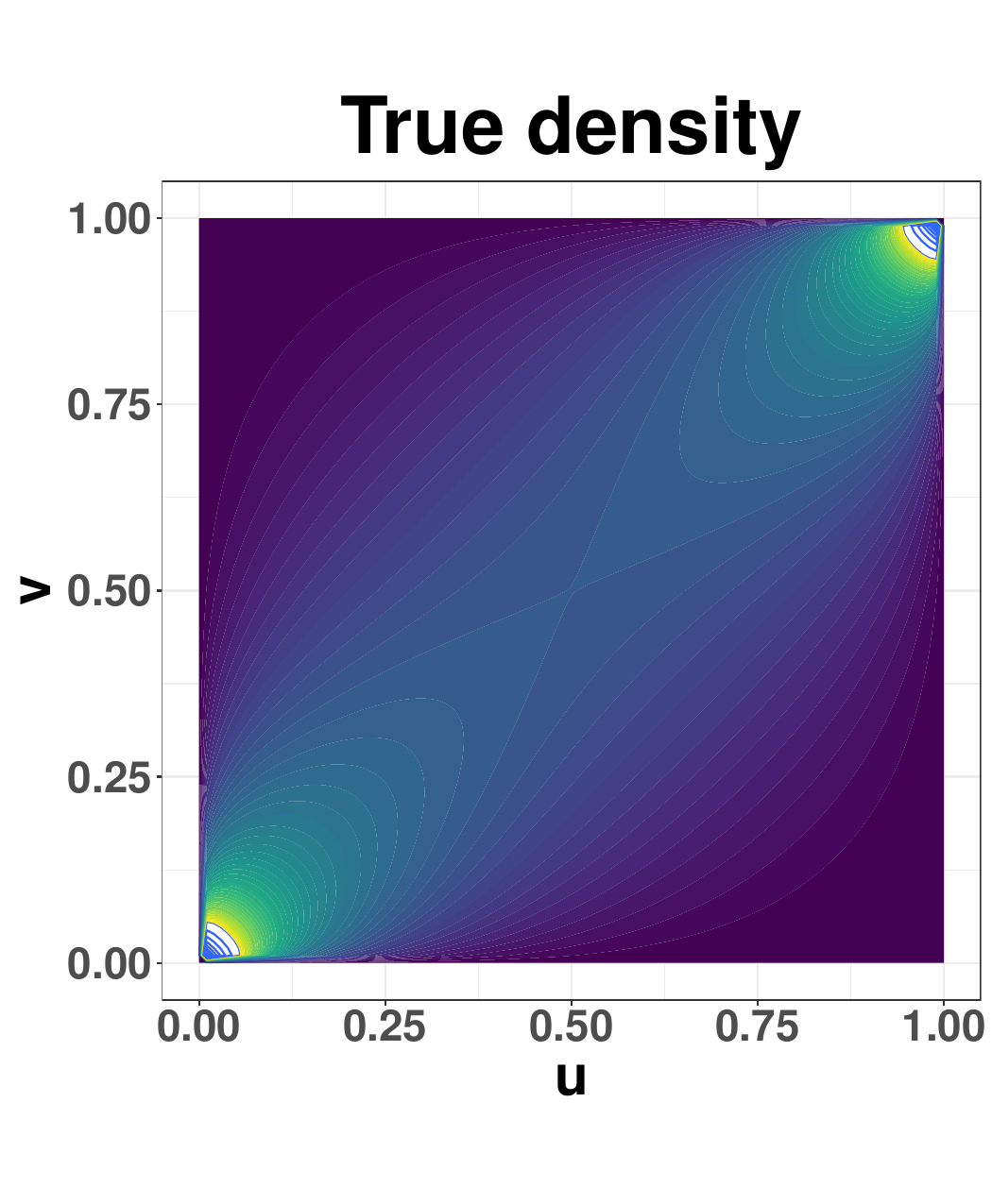}
  \includegraphics[scale=0.3]{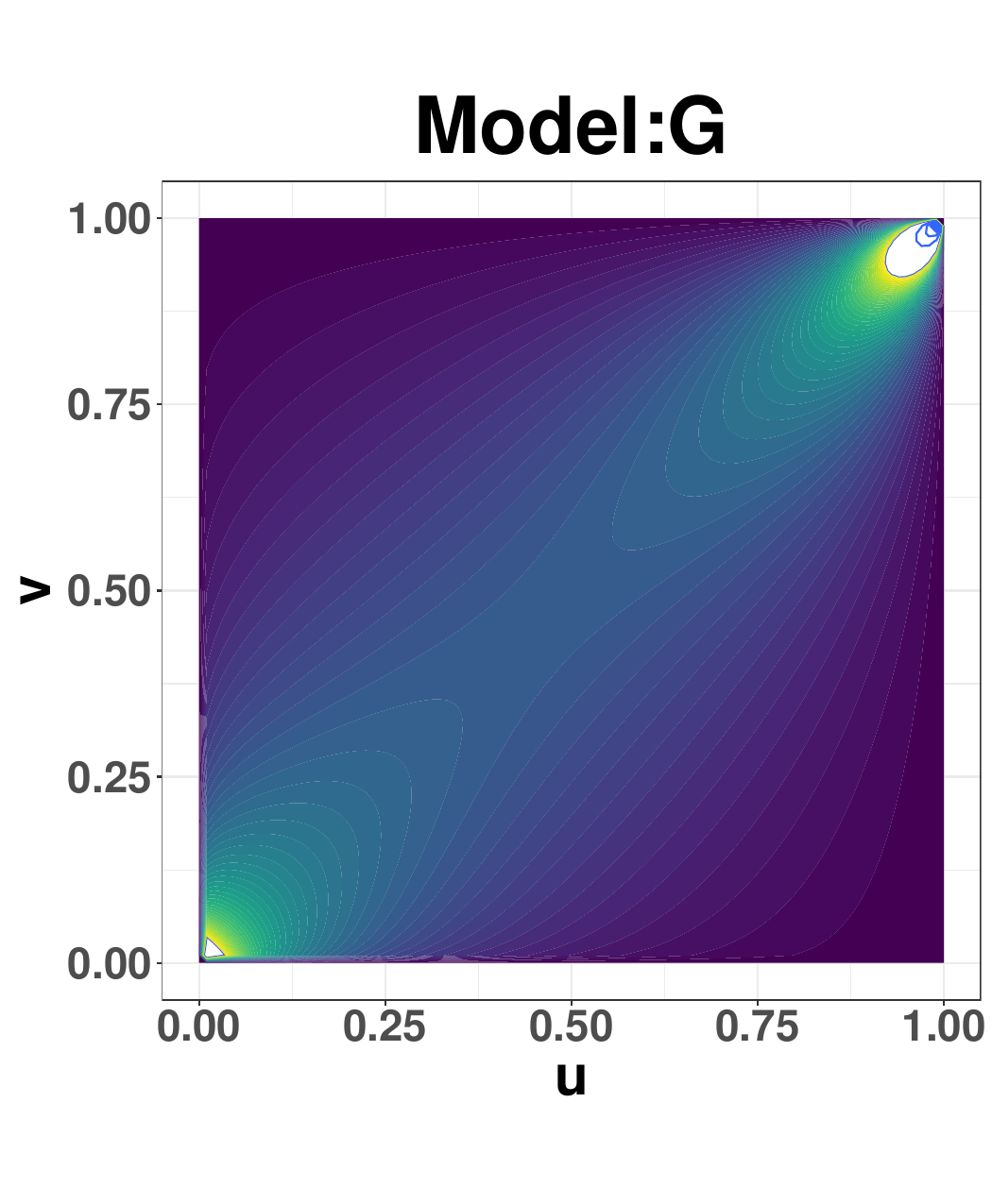}
  \includegraphics[scale=0.3]{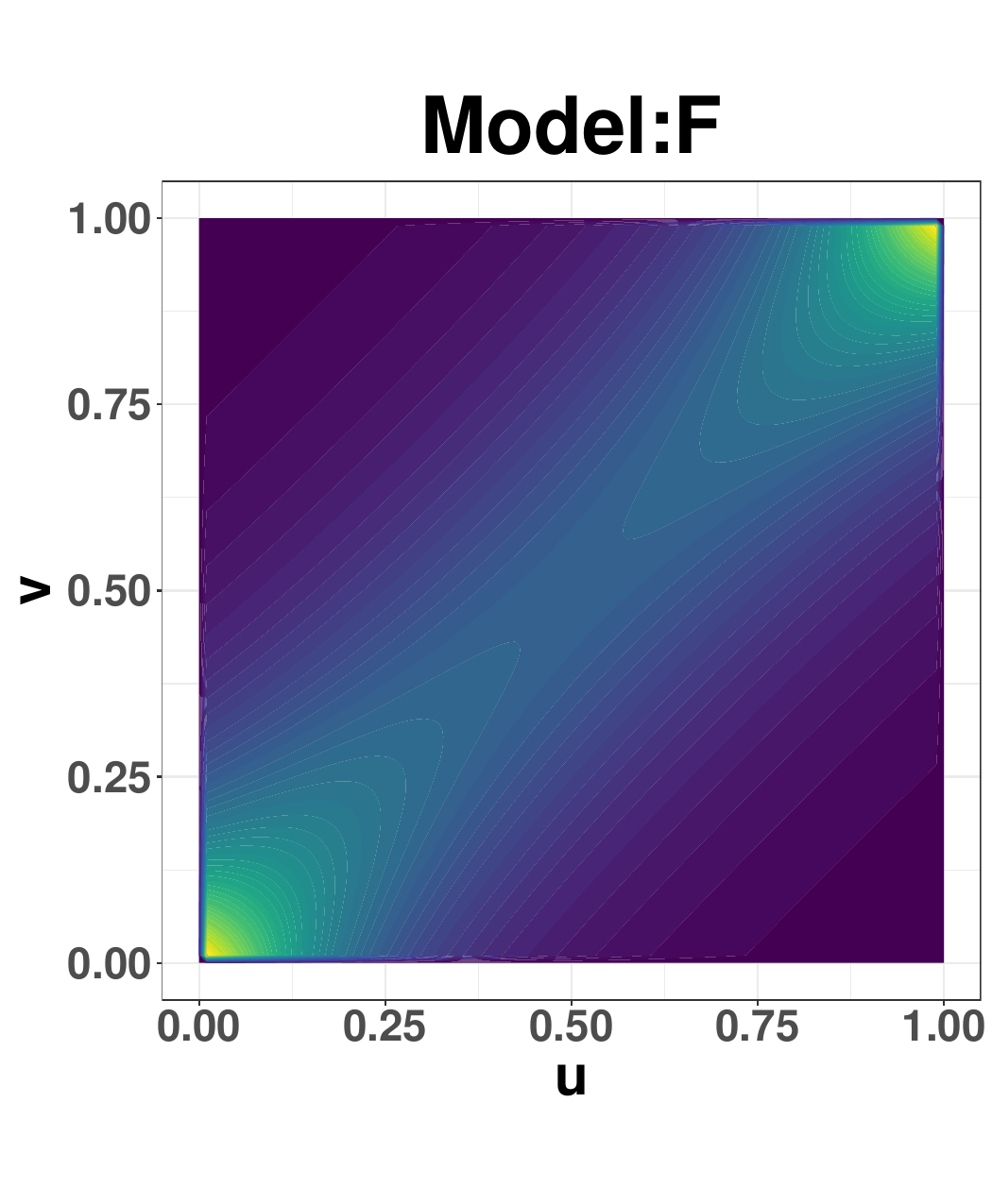}
}
\centerline{
  \includegraphics[scale=0.3]{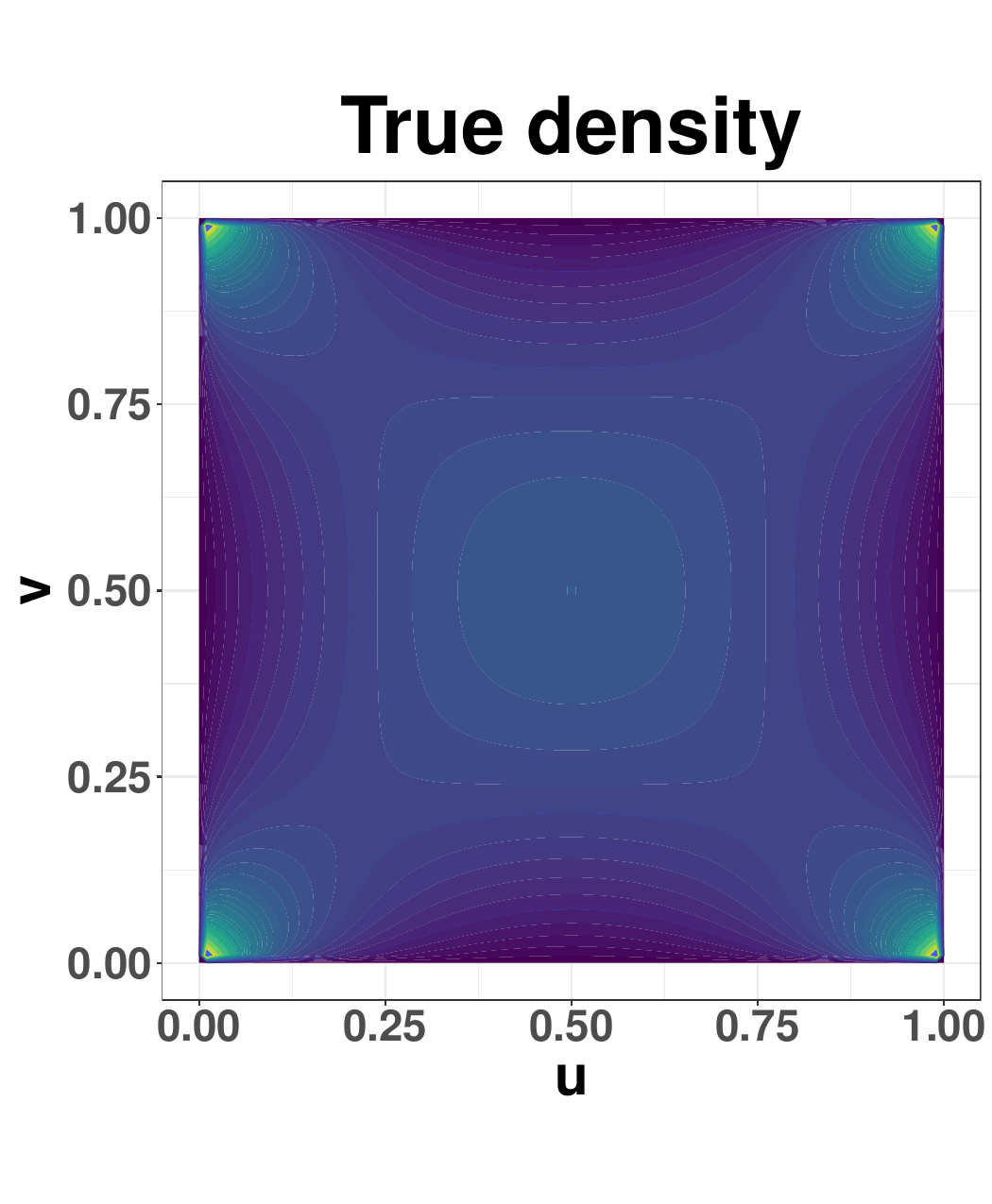}
  \includegraphics[scale=0.3]{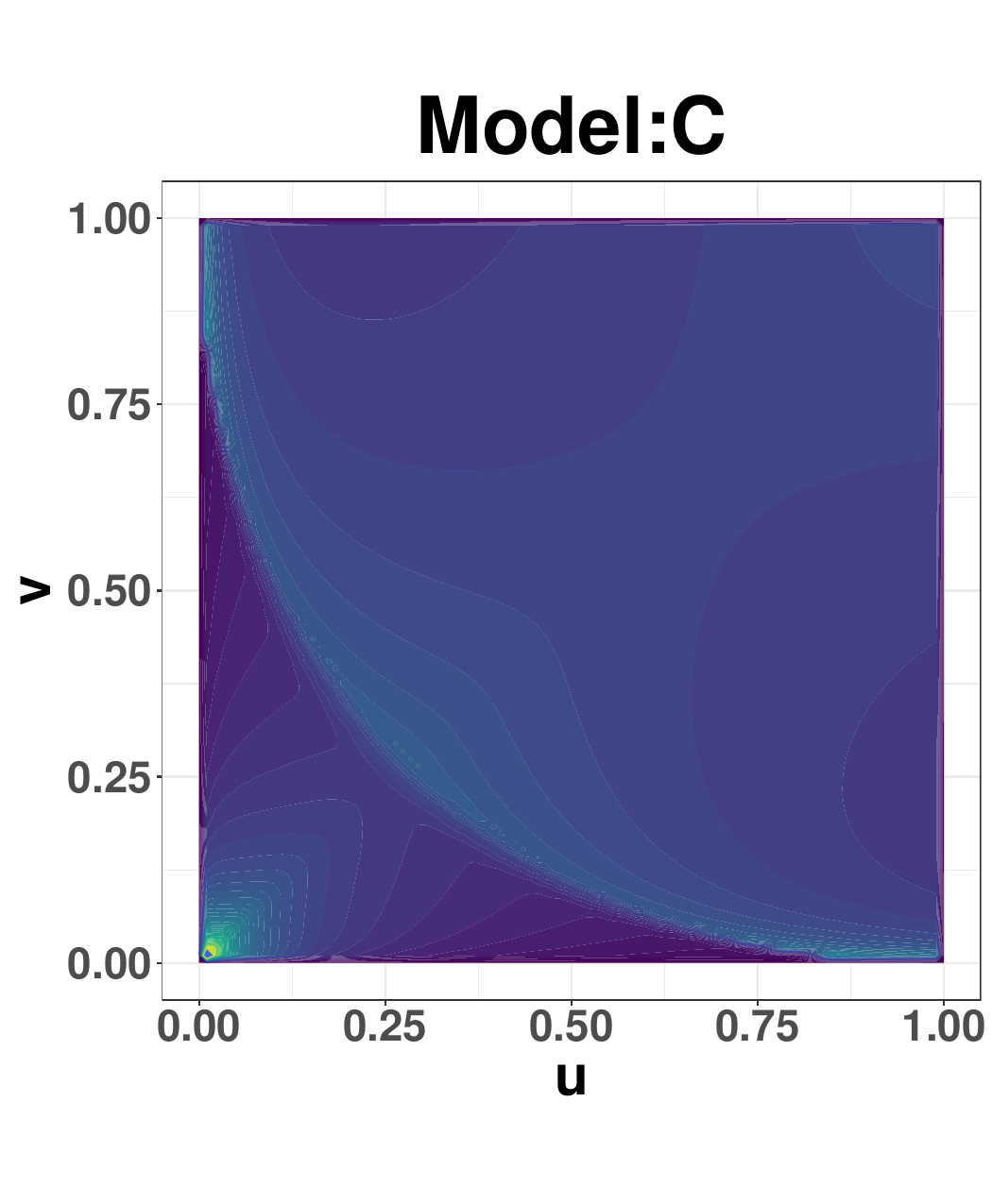}
  \includegraphics[scale=0.3]{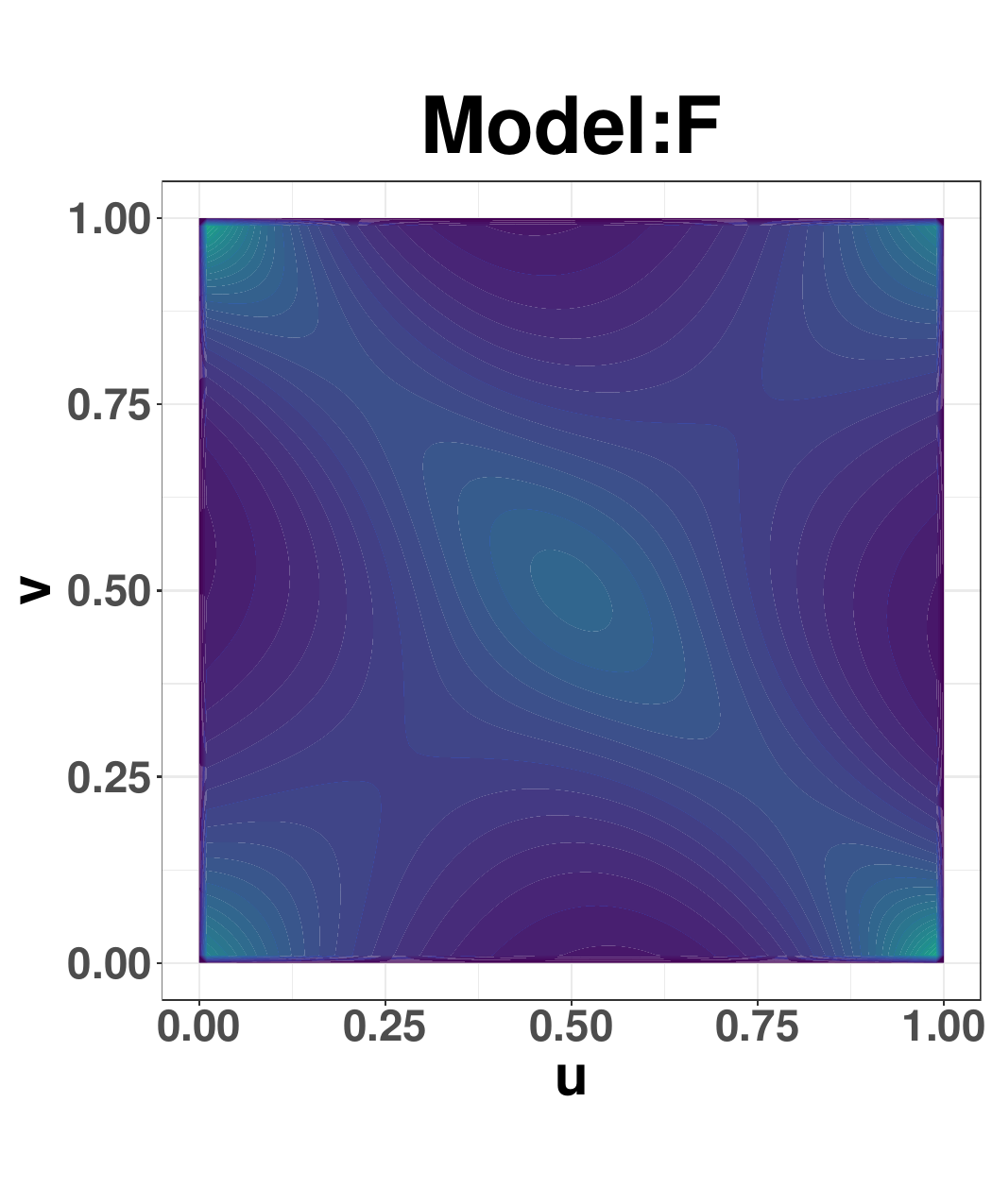}
}
\caption{Simulation study 2: First row: Single Gaussian copula true density (1st panel), mixture of Gumbels density estimate (2nd panel), and mixture of Franks density estimate (3rd panel). Second row: Mixture of Gaussians copula trues density (1st panel), mixture of Claytons density estimate (2nd panel), and mixture of Franks density estimate (3rd panel).}
\label{fig:postsim2}
\end{figure}

We finally compare the model fit by producing copula density estimates with the best two fitting models and compare them to the true density. These are depicted in Figure \ref{fig:postsim2}. For the single Gaussian case (top row), the Gumbel and Frank estimates are quite similar to the true density; perhaps the Gumbel estimates better capture the contours around the main diagonal. For the mixture of Gaussian case (bottom row), although we are able to produce good estimates for the overall dependence, the contours do not seem to be similar to the true ones. 

\subsection{Simulation study 3}

This section contains a multivariate simulation study. Specifically, we consider a vector of dimension four, that is, $\bU=(U_1,U_2,U_3,U_4)$ coming from a mixture of three Clayton copulas $f(\bu)=\sum_{j=1}^3\pi_j f_C(\bu\mid\vartheta_j)$, with $\bpi=(0.2,0.3,0.5)$. We consider three simulation settings for the copula parameters. Specifically, in setting 1: $\bvartheta=(1,5,15)$; in setting 2: $\bvartheta=(2,5,10)$; and in setting 3: $\bvartheta=(2,7,15)$. We sampled $n=1000$ data points from each of these three settings. 

We fitted our Bayesian nonparametric mixture model with the Clayton kernel. The hyperparameters are: $c_\theta=4,\ d_\theta=1$, $p_a=1$, $q_a=20$, $c_b=1$ and $d_b=20$. The MCMC was run for 15,000 iterations with a burn-in of 5,000 and a thinning of 5. 

\begin{figure}
\centerline{
\includegraphics[scale=0.3]{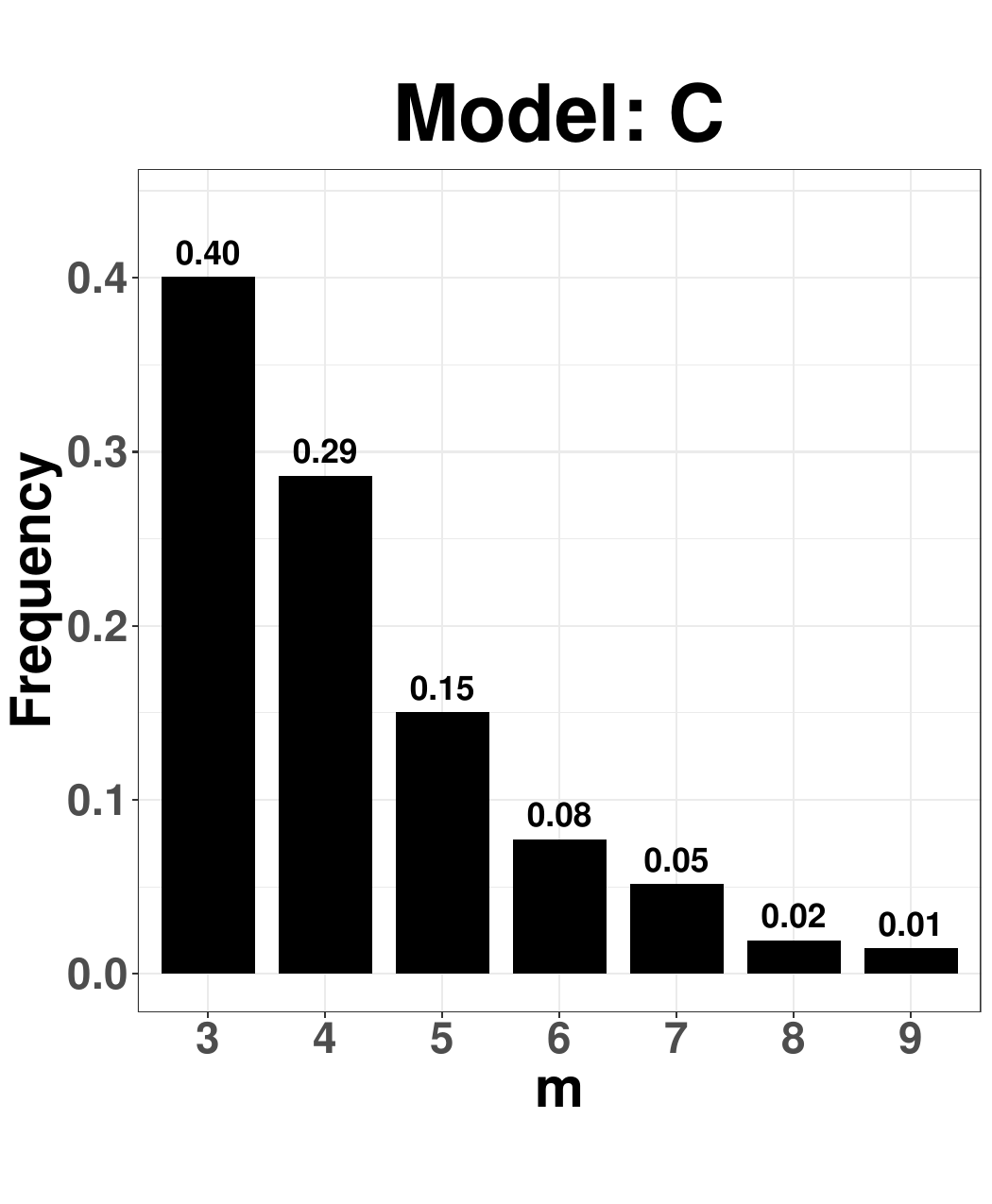}
\includegraphics[scale=0.3]{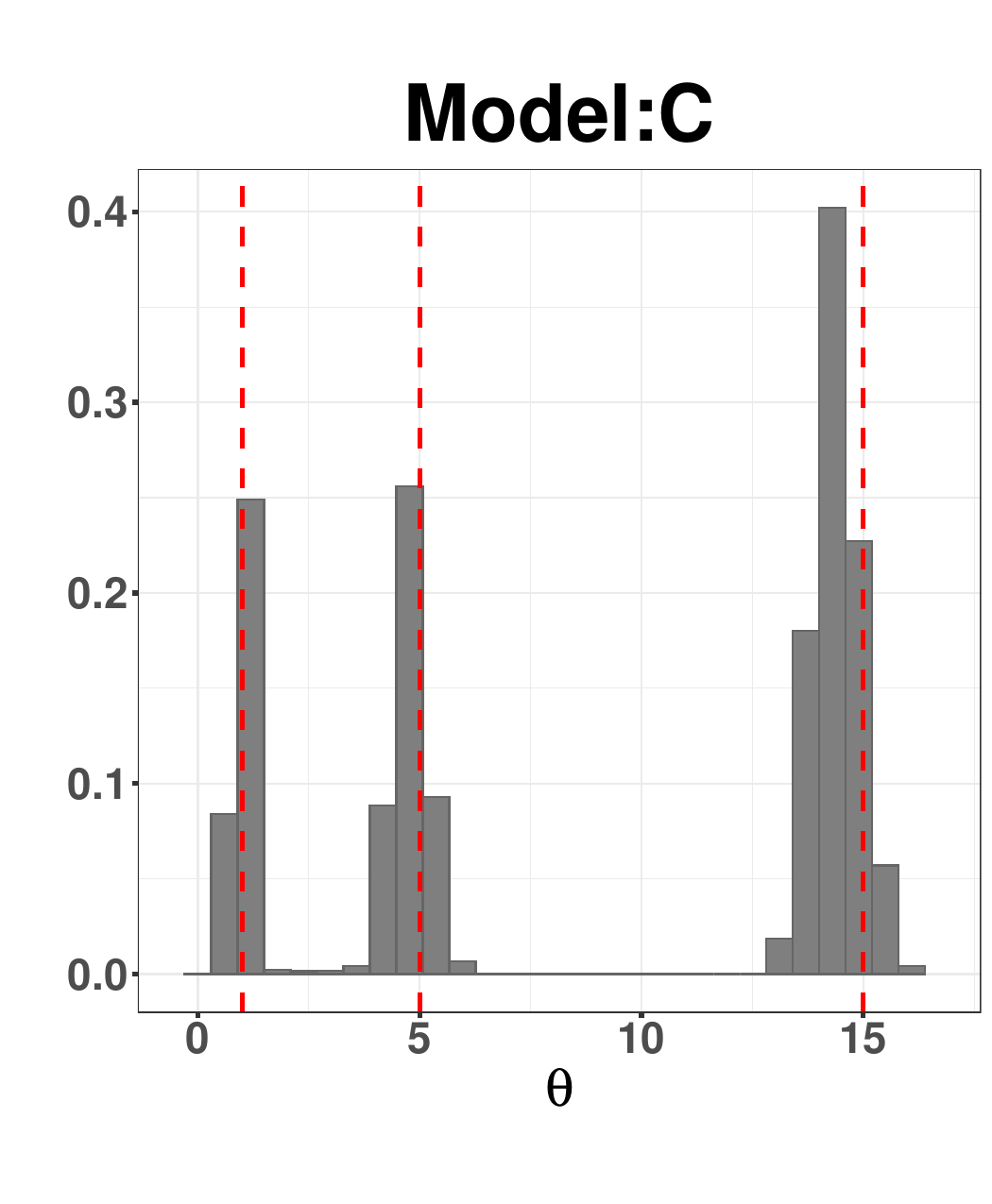}
}
\centerline{
\includegraphics[scale=0.3]{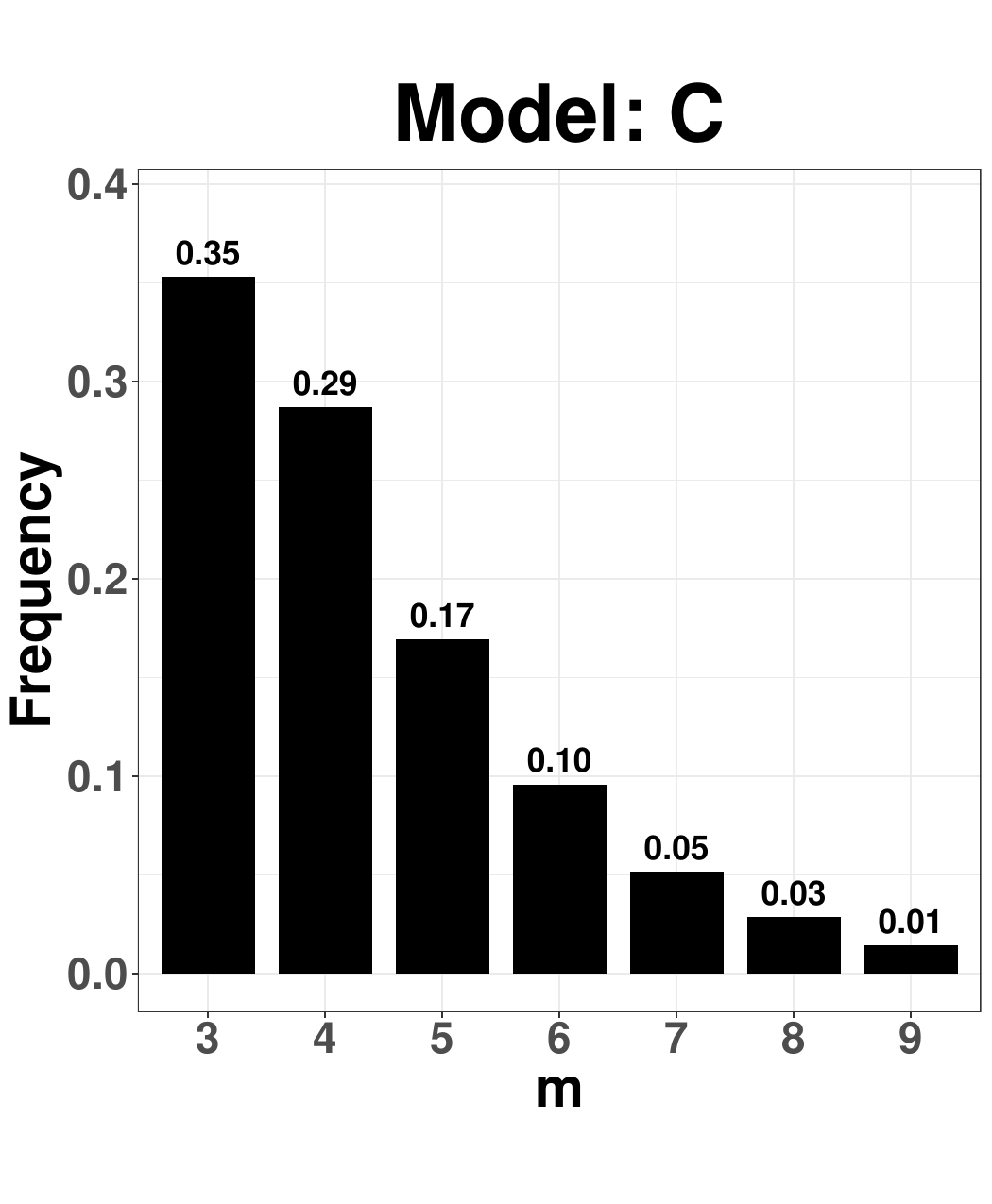}
\includegraphics[scale=0.3]{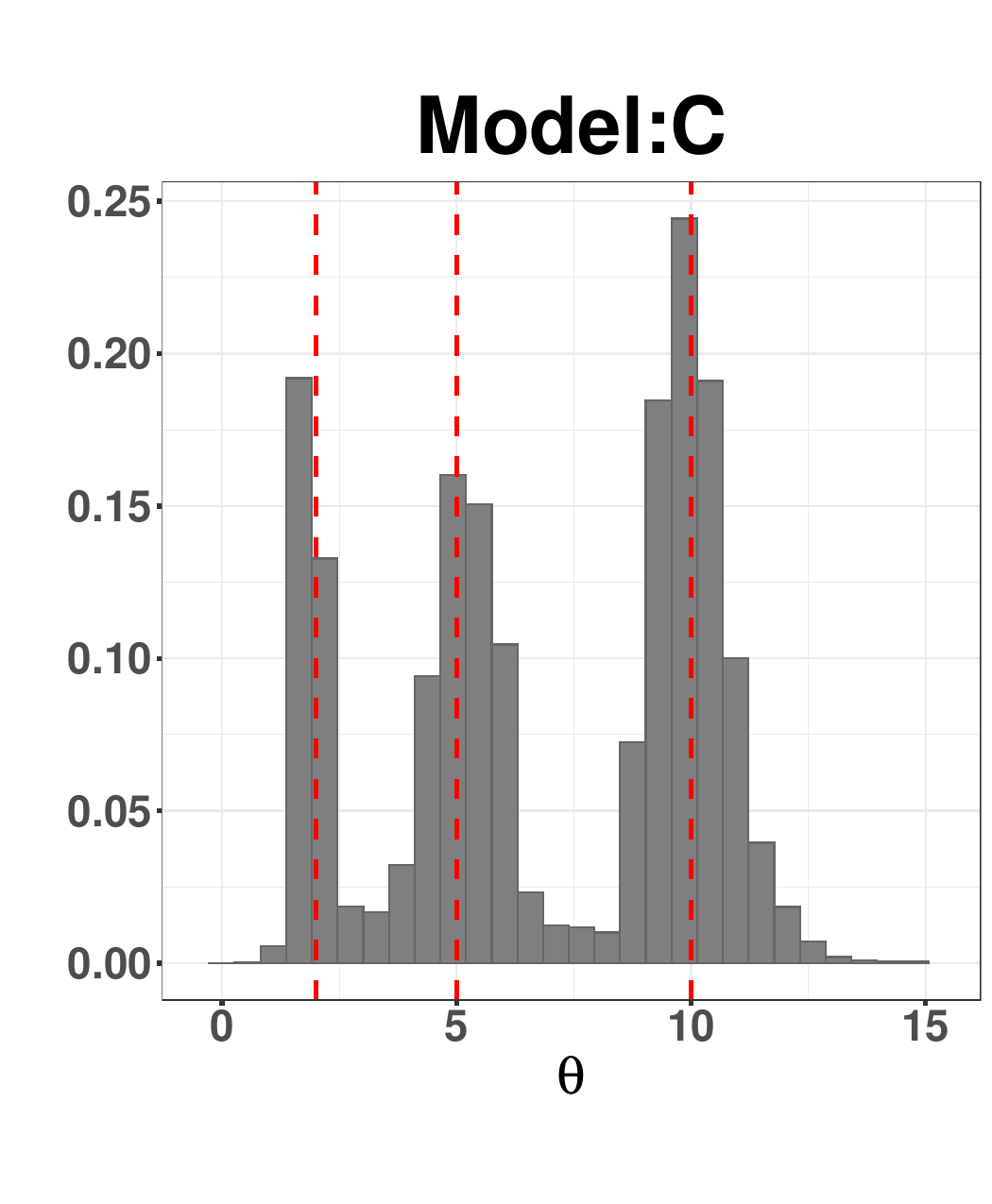}
}
\centerline{
\includegraphics[scale=0.3]{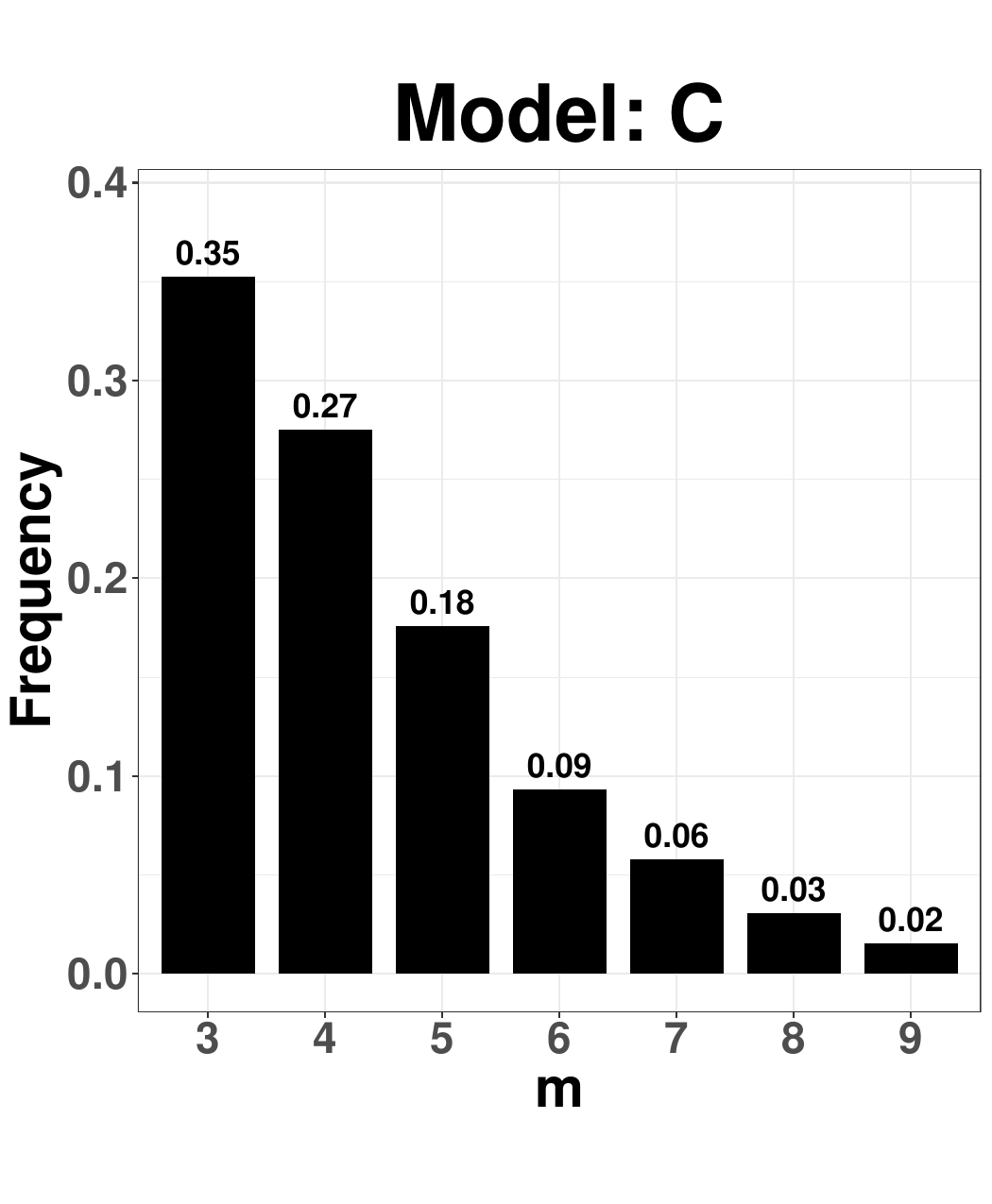}
\includegraphics[scale=0.3]{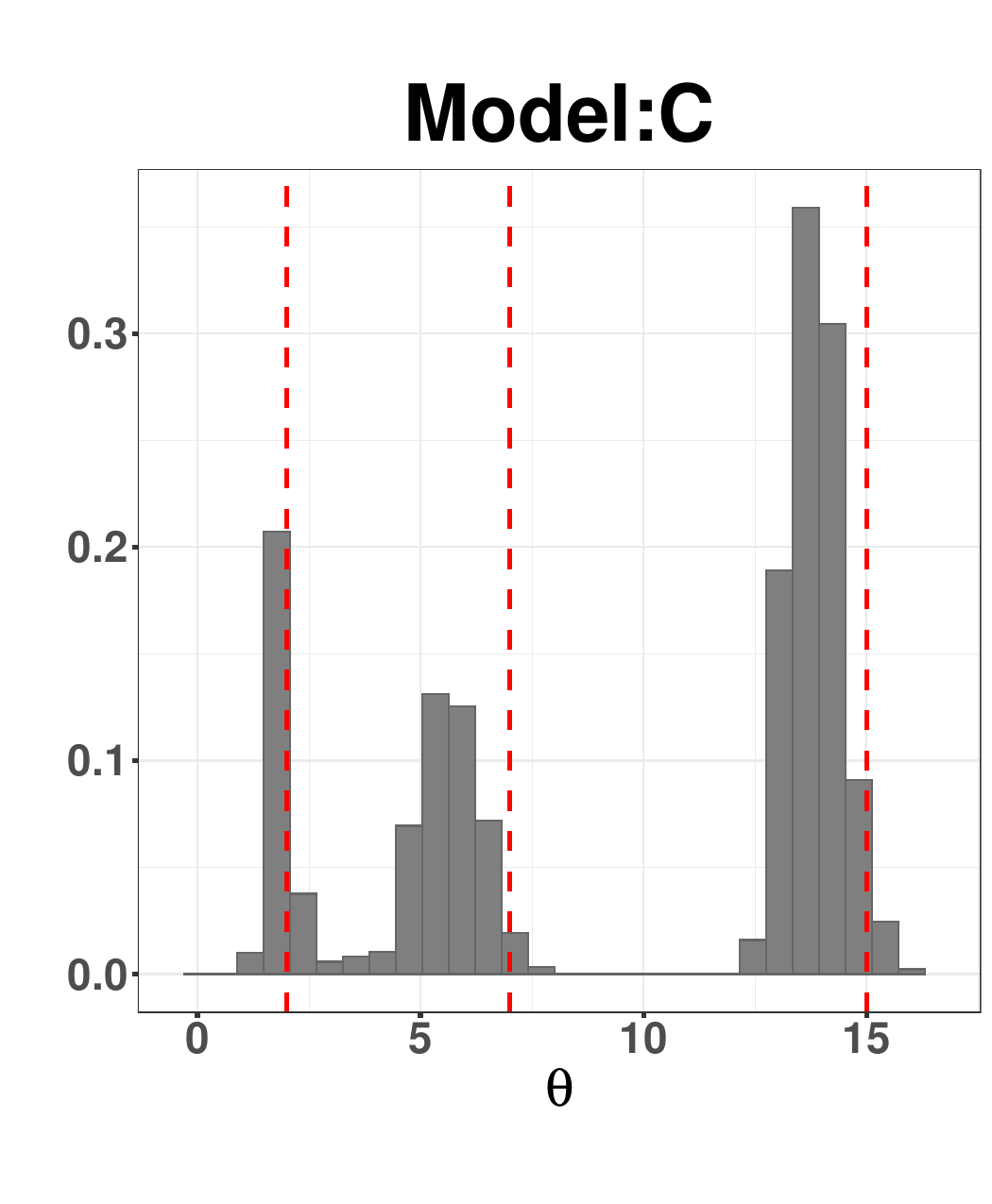}
}
\caption{Simulation study 3: Four-dimensions data from a mixture of 3 Clayton copulas with weights $\bpi=(0.2,0.3,0.5)$ and different copula parameters: $\bvartheta=(1,5,15)$ (top), $\bvartheta=(2,5,10)$ (middle) and $\bvartheta=(2,7,15)$ (bottom). Posterior distribution of the number of components (left) and model parameters (right). Vertical dotted lines correspond to the true values.}
\label{fig:clayton_hdim}
\end{figure}

The posterior distributions for the number of components and for the copula parameters are included in Figure \ref{fig:clayton_hdim}. The number of components has a mode at 3 for the three settings, and the posterior distribution for the copula parameters is three-modal in the three settings, with modes around the true values, shown as vertical dotted lines. 

The true and estimated Kendall tau values are reported in Table \ref{tab:Kendall_tau}. The point estimates (posterior mean) are very close to the true values and 95\% CI all contain the true association parameters. 

In order to report a single clustering, we use the R package \texttt{salso} with binder loss. The results are shown in Table \ref{tab:postmcmc3}. For all settings, the search procedure selects five groups. In setting 1, the first three groups account for 92\% of the total weight. We note a shrinkage towards the center values. The first group has a weight of 0.58, similar to the theoretical 0.5 with 95\% CI $(14.4,15.7)$, the second group has a weight of 0.18 with 95\% CI $(3.9,4.6)$, and the third has a weight of 0.17 with interval 95\% CI $(0.63,0.93)$. Apart from the first group, the other two parameter estimates are slightly lower than the true values. 

For setting 2, the first group is clearly associated with the third true mixture component with a weight of $0.54$ and a parameter 95\% CI of $(11.1,12.1)$ slightly larger than $10$, the second group with a weight of $0.14$ and a parameter CI of $(1.2,1.6)$, and the third group perfectly contains the parameter of the second true mixture component. Finally, for setting 3, the first group has the highest weight of 0.65, associated with the third true mixture component, and the last four groups are associated with the first two mixture components. 

In summary, a multivariate setting with several mixture components is quite challenging, and larger sample sizes are required to appropriately detect the number of components and dependence parameters. 

\subsection{Occupation data analysis}

\cite{candanedo&feldheim:16} presented a data set aimed at determining occupancy in a room. Original data contains, among other variables, information about carbon dioxide (CO2) and humidity ratio (HR), the latter defined as the ratio between temperature and relative humidity, and measurements were made every minute. 

\begin{figure}
\centering
\includegraphics[scale=0.7]{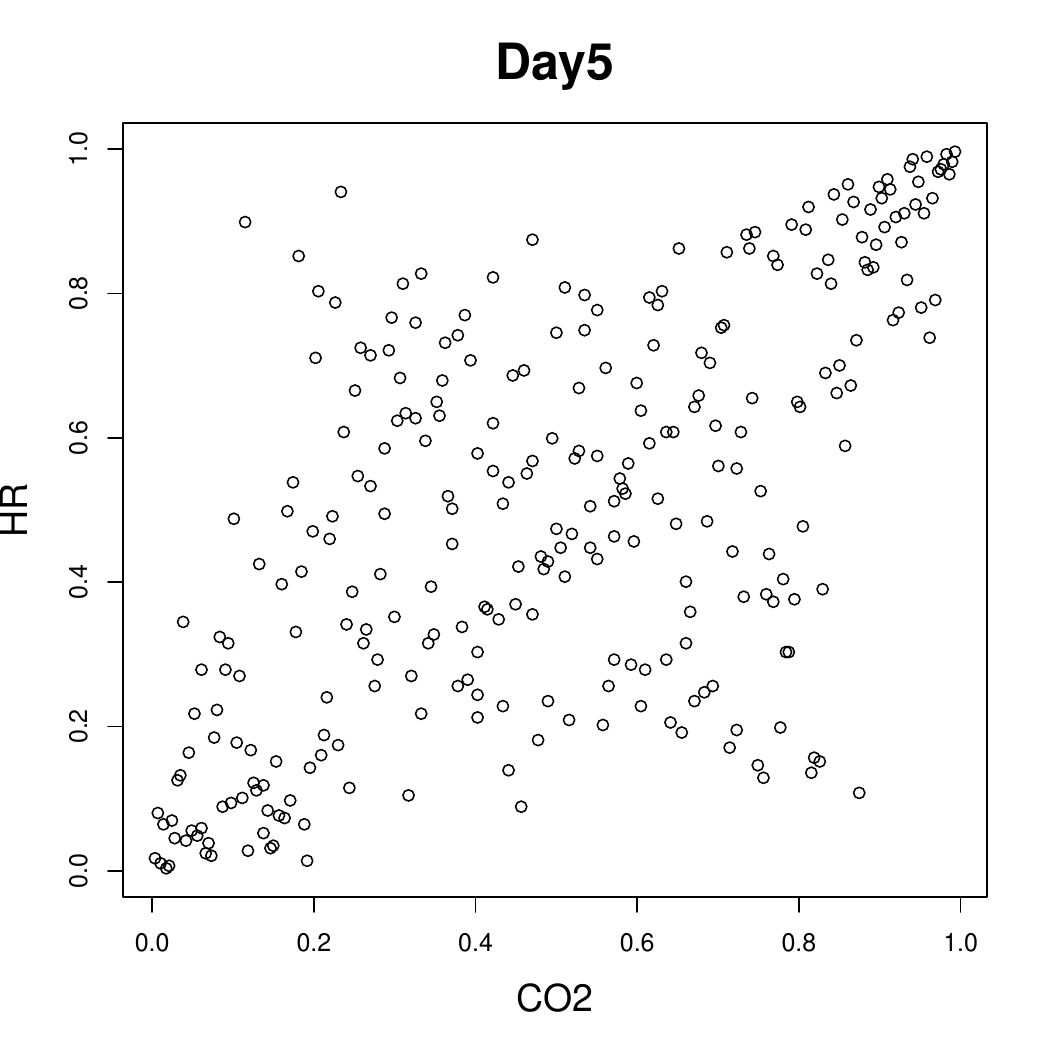}
\caption{Occupancy data. Scatterplot of carbon dioxide (CO2) versus humidity ratio (HR).}
\label{fig:day5}
\end{figure}

As suggested by \cite{candanedo&al:17}, the data are preprocessed by making averages of five minutes and taking the first differences. To study the dependence in these two variables, we further apply the modified rank transformation (inverse empirical cdf) to produce data in the interval $[0,1]$. Figure \ref{fig:day5} shows a dispersion diagram for 5 February 2015. Data points form a star with possible positive and negative dependencies. 

We fitted our Bayesian nonparametric mixture model to these data using the five common Archimedean generators. Prior specifications were defined as in the simulation studies, and MCMC had 15,000 iterations, 5,000 as burn-in, and a thinning of 5. 

\begin{figure}
\centerline{
  \includegraphics[scale=0.3]{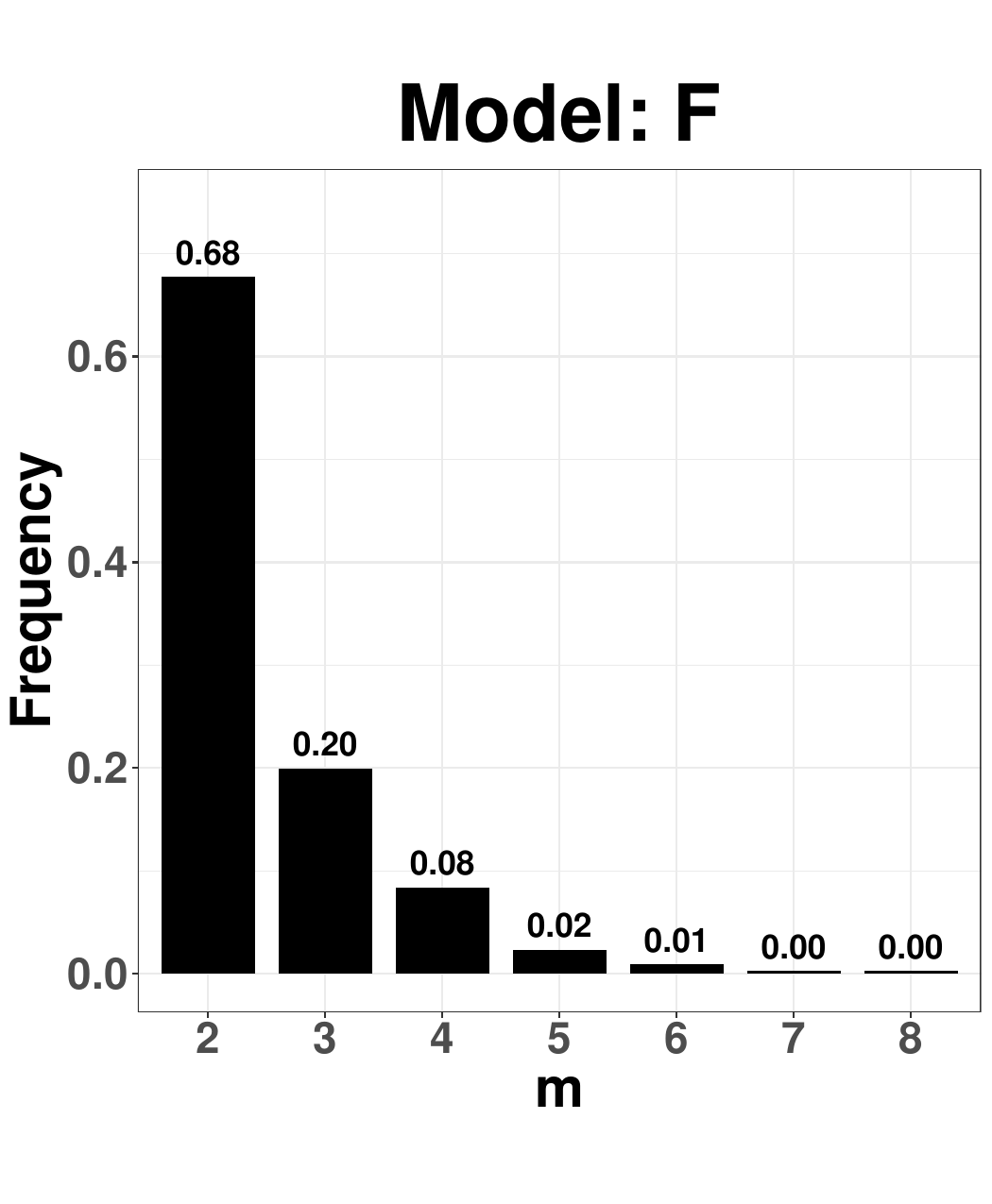}
  \includegraphics[scale=0.3]{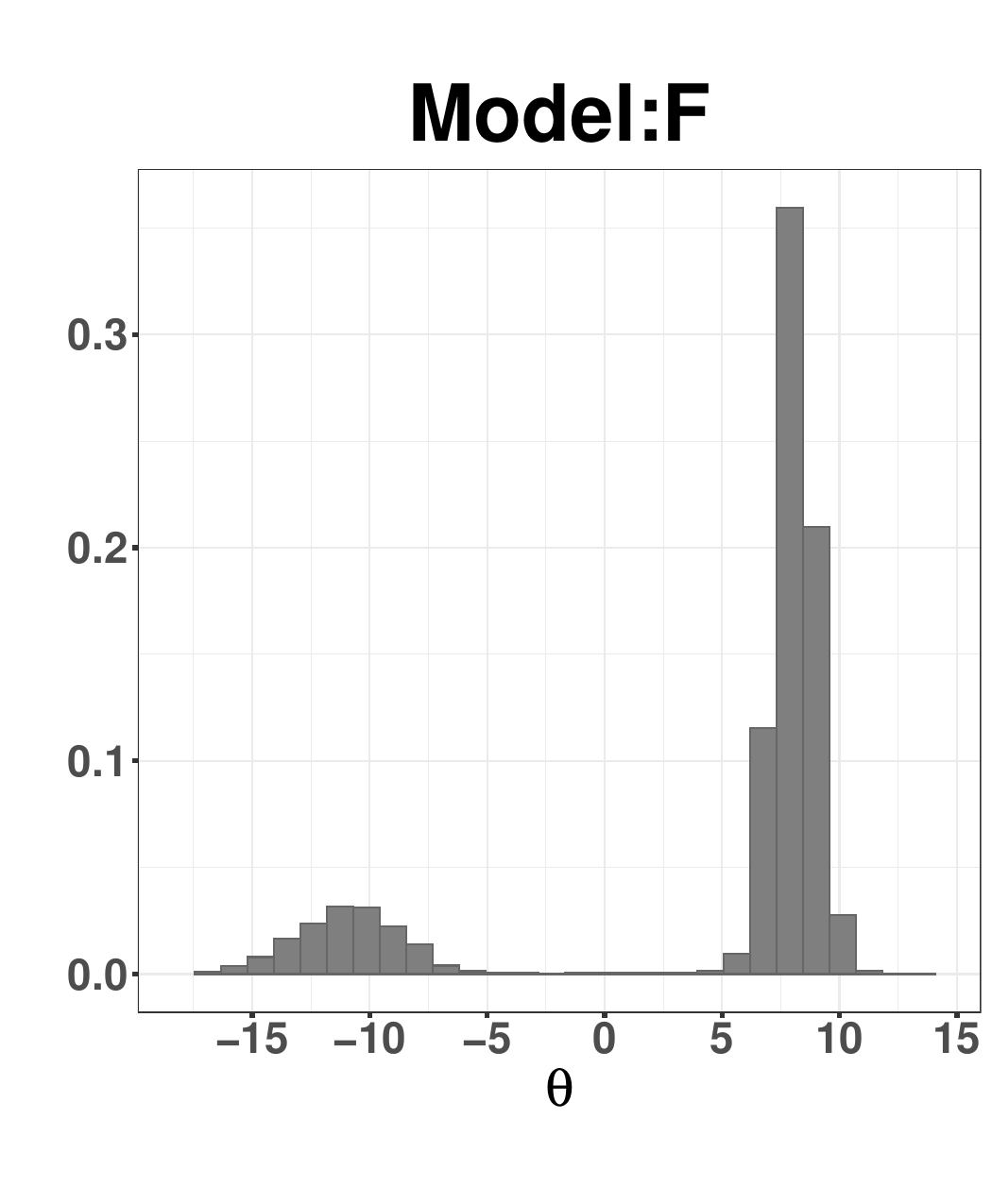}
  \includegraphics[scale=0.3]{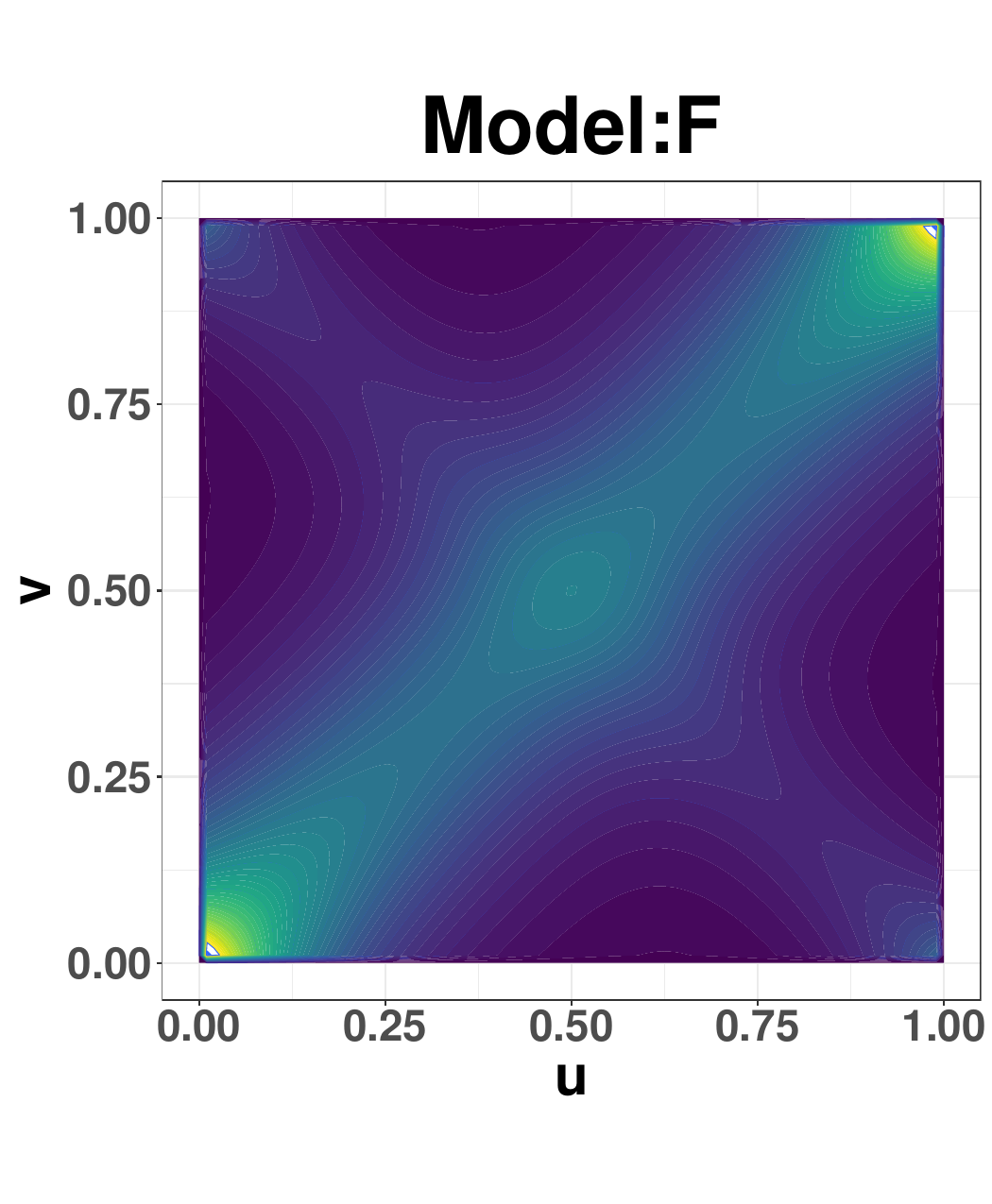}
}
\centerline{
  \includegraphics[scale=0.3]{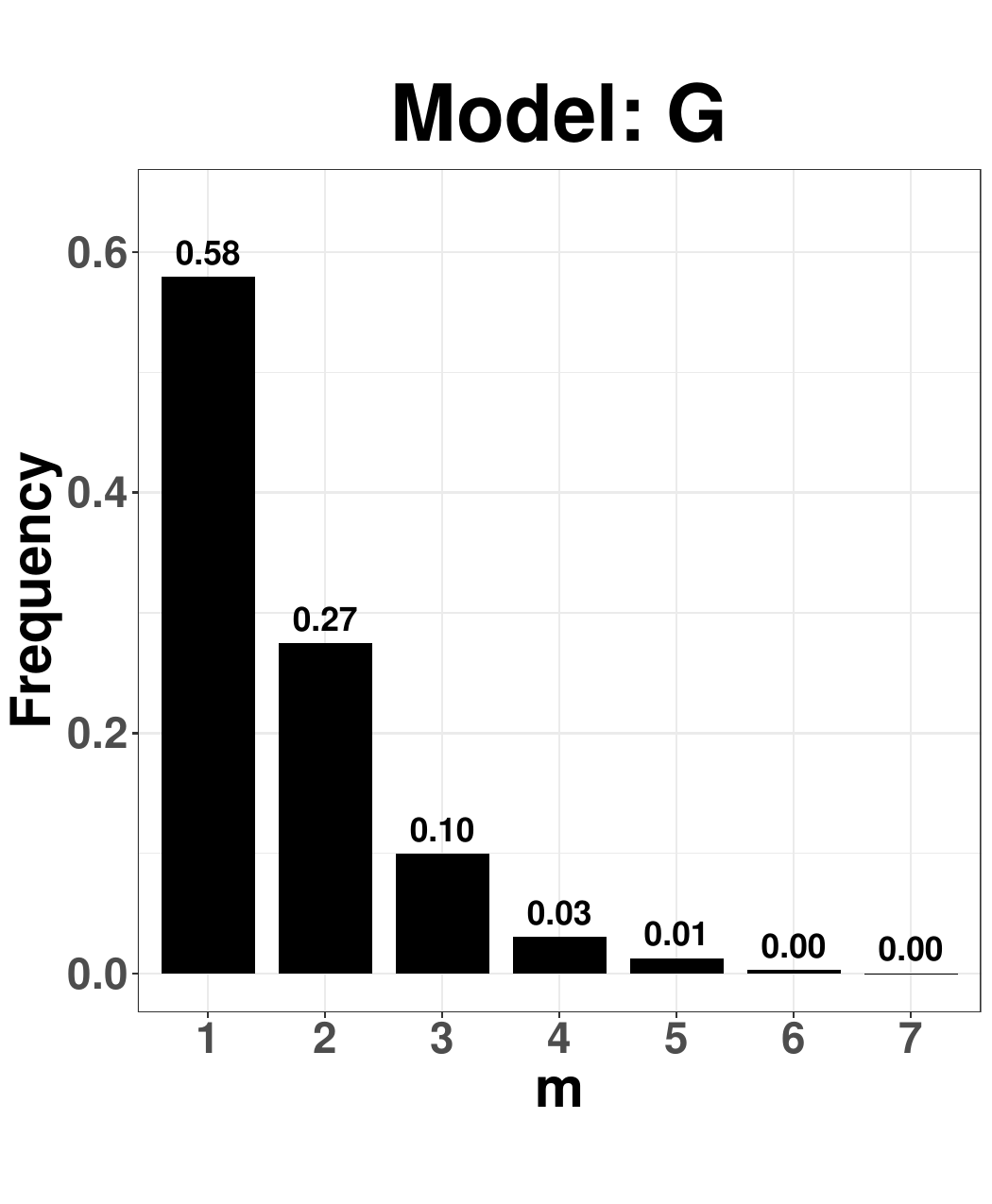}
  \includegraphics[scale=0.3]{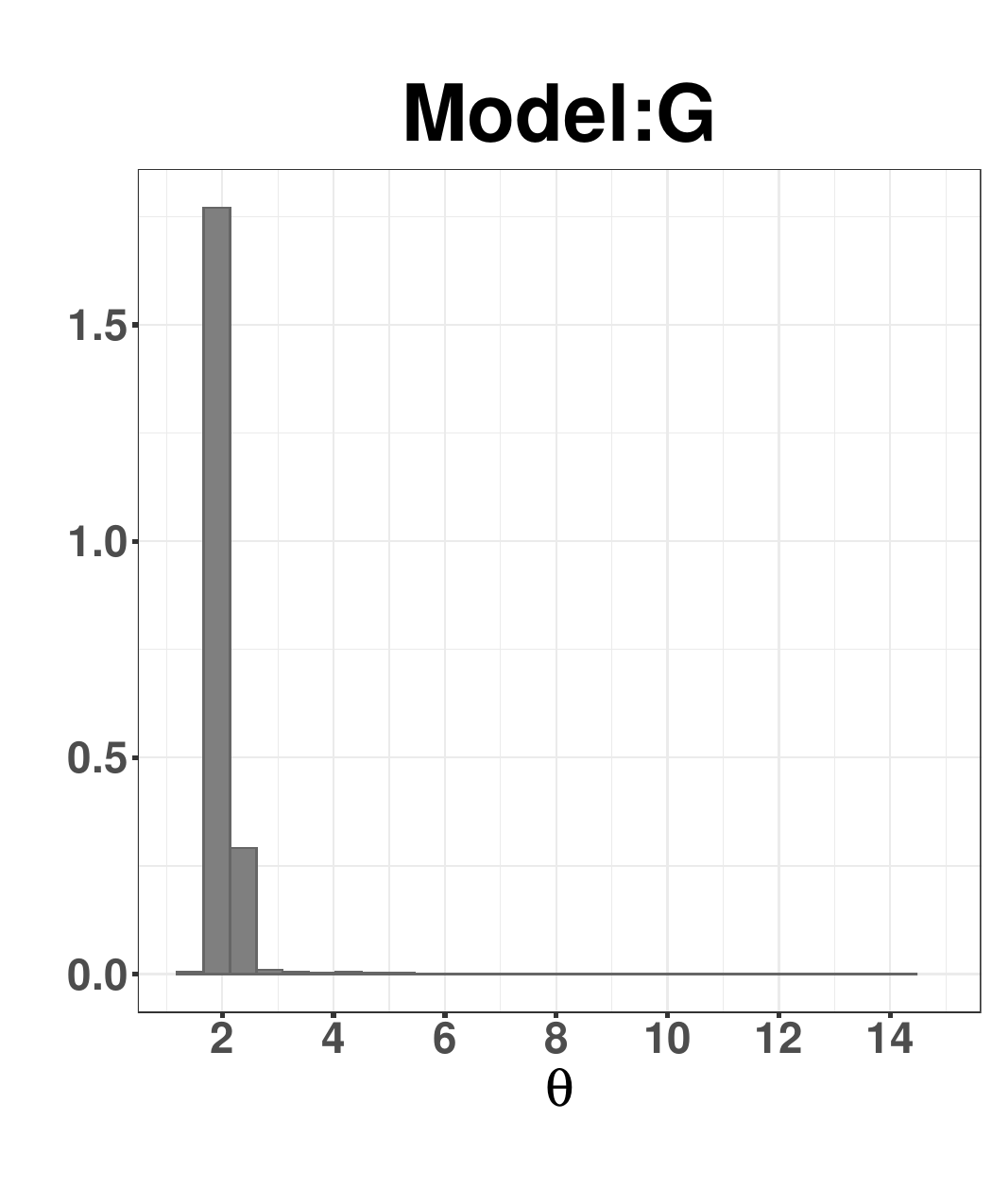}
  \includegraphics[scale=0.3]{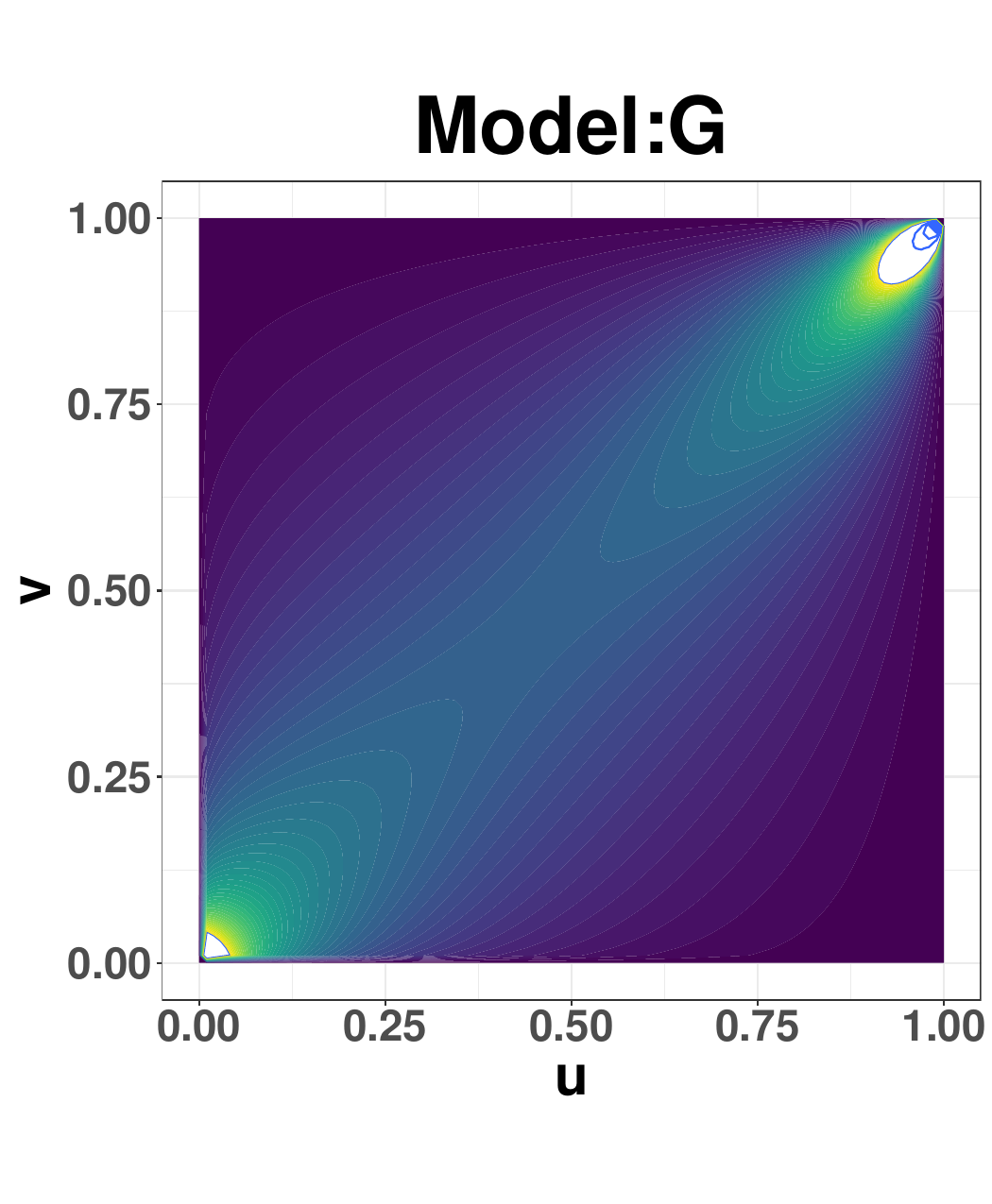}
}
\caption{Occupancy data. Posterior fittings for top two best models. Frank (top row) and Gumbel (bottom row). Across columns: Number of components (1st), $\theta_i$'s (2nd), and bivariate density (3rd).}
\label{fig:day5fit}
\end{figure}

To assess the fit of the model, we computed the LPML statistic and obtained the following values: 73 for the AMH, 84 for the Clayton, 103 for the Frank, 105 for the Gumbel and 91 for Joe. Clearly, the two best models are the Frank and the Gumbel. In Figure \ref{fig:day5fit} we report posterior inferences for these two models. The number of components obtained with the Frank model has a mode at 2 and the histogram of the posterior values of the parameters $\theta_i$ is bimodal with a heavier mode in a positive value around $8$ and a lighter mode in a negative value around $-11$. The density estimate shows the cross shape of the original data. On the other hand, with the Gumbel model, the number of components has a mode at 1 with the copula parameters $\theta_i$ concentrated around 2. The density estimate shows the positive dependence with wide contours in the center that resemble the negative dependence.  

The empirical Kendall tau for the data is $0.462$ and the corresponding estimates with the Frank mixture model is $0.4$ with a 95\% CI of $(-0.74,0.66)$; and for the Gumbel model we get $0.5$ with a 95\% CI of $(0.45,0.57)$. Both interval estimates contain the empirical value; however, there is a large difference in the width of the intervals. As mentioned in the previous paragraph, the posterior values of $\theta_i$ in the Frank case show a bimodal behavior with modes in positive and negative values, which combined induce high uncertainty in the overall dependence. 

Using the search procedure to estimate the number of components, in Table \ref{tab:postoccup} we report two groups for the Frank copula with estimated copula parameters at $8.6$ with a weight of $86\%$ for the first component and at $-13.9$ with a weight of $14\%$ for the second component. For the Gumbel copula, a single cluster is determined with a parameter value estimated between $1.86$ and $2.24$ with probability $95\%$. An advantage of the Gumbel model is that it has upper right-tail dependence and the data seem to support this. 

Again, we also compare with the competing model BSA. The LPML statistic obtained is 78.40, which is better than the fit with the AMH kernel, but worse than the fit obtained with the other four kernels. 

\subsection{Red wine data analysis}

We now consider the red wine data of \cite{cortez&al:09} which consists of several physicochemical tests of the red variants of $1,599$ Portuguese wine of the \emph{Vinho Verde} region. We concentrate on three variables: fixed acidity ($X_1)$, citric acid ($X_2$) and density ($X_3$). The idea is to characterize the dependence among these three variables. 

\begin{figure}
\centerline{
  \includegraphics[scale=0.5]{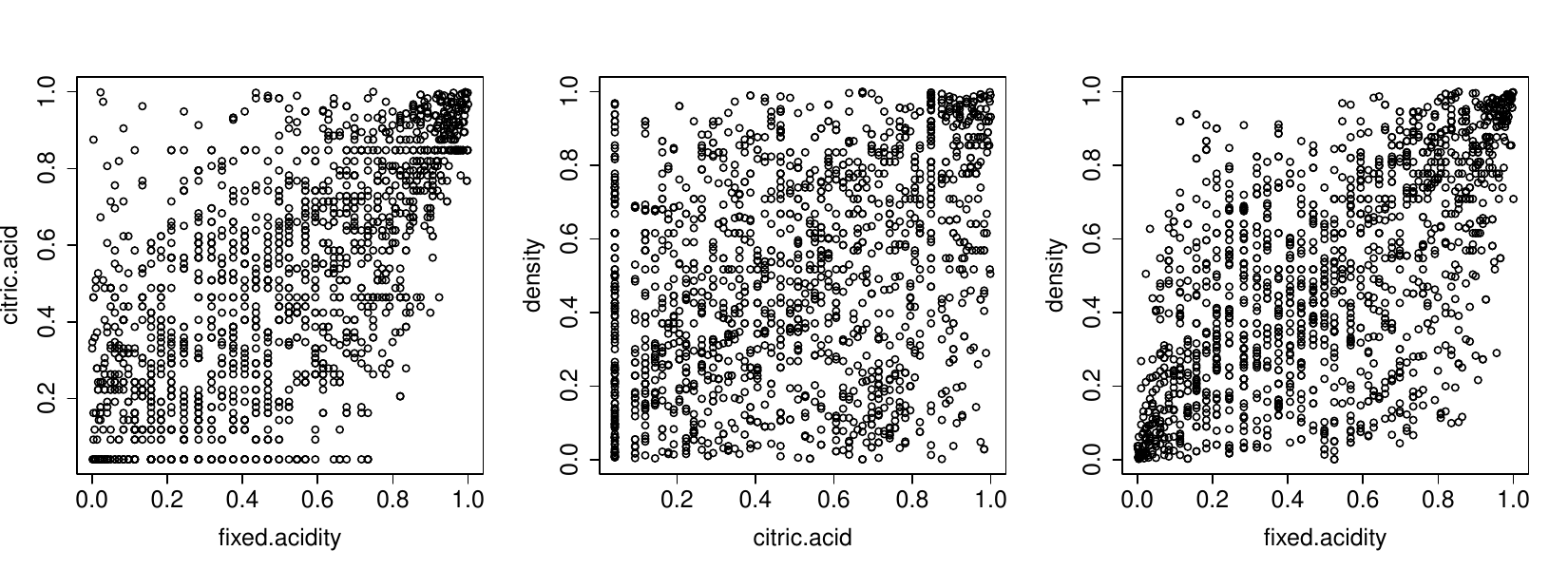}
}
\caption{Red wine data: fixed acidity, citric acid, and density. Pairwise scatterplots of the three variables.}
\label{fig:dispersion3}
\end{figure}

As in the previous analysis, we applied the modified rank transformation to produce data in the interval $[0,1]$. A simple exploratory graphical analysis (see Figure \ref{fig:dispersion3}) shows that these three variables exhibit a positive dependence, with pairwise empirical Kendall tau values $\tau_{1,2}=0.484$, $\tau_{1,3}=0.457$, and $\tau_{2,3}=0.245$, which do not differ much from each other, suggesting that a mixture of Archimedean copulas is appropriate for these data. 

We fitted our Bayesian nonparametric mixture model with different copulas. The hyperparameters are the same as in the multivariate simulation study. The MCMC was run for 15,000 iterations with a burn-in of 5,000 and a thinning of 5. The LPML measures are: 279 for AMH, 604 for the Clayton, 685 for the Frank, 711 for the Gumbel, and 413 for the Joe. 

\begin{figure}
\centerline{
  \includegraphics[scale=0.3]{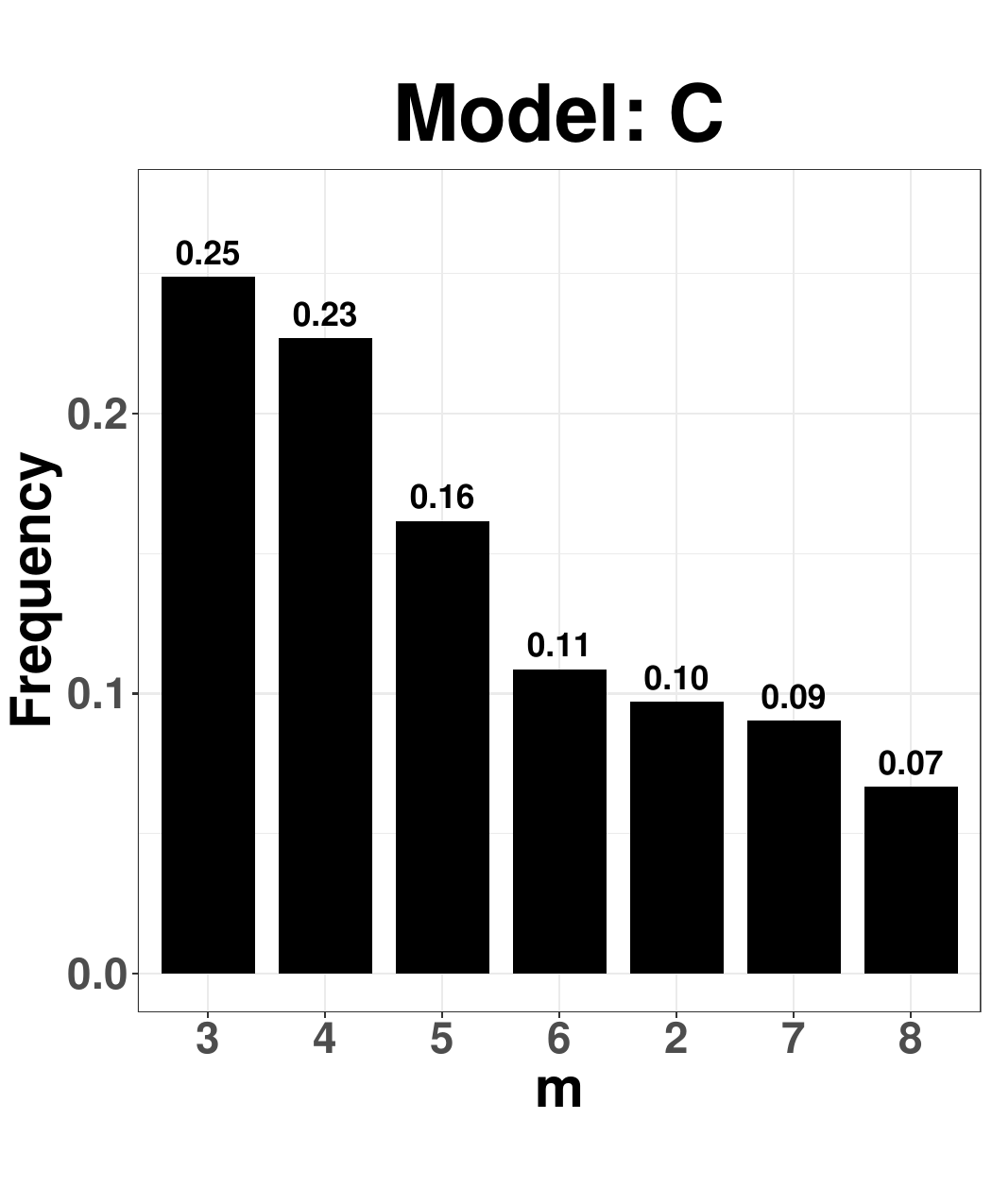}
  \includegraphics[scale=0.3]{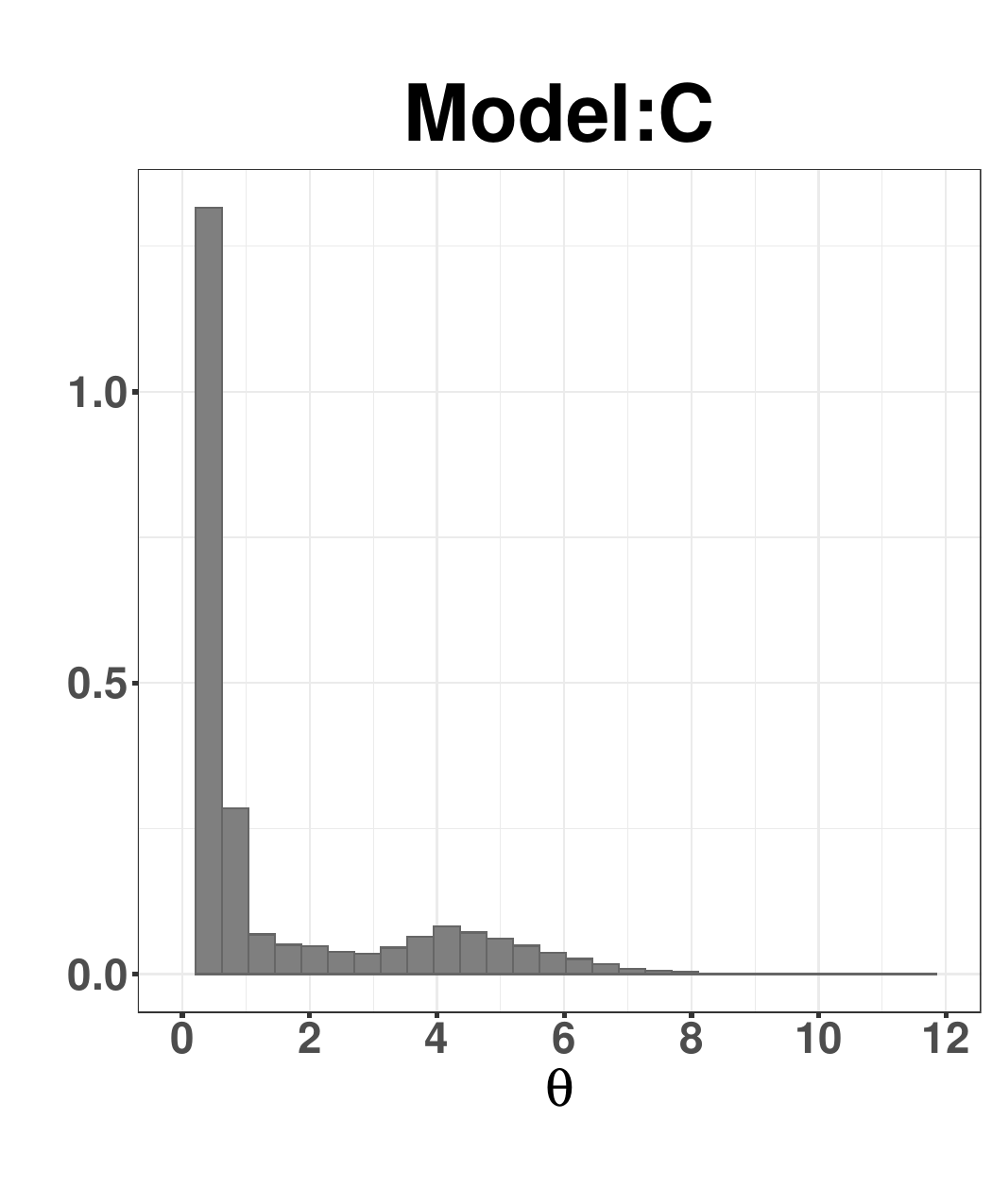}
  \includegraphics[scale=0.3]{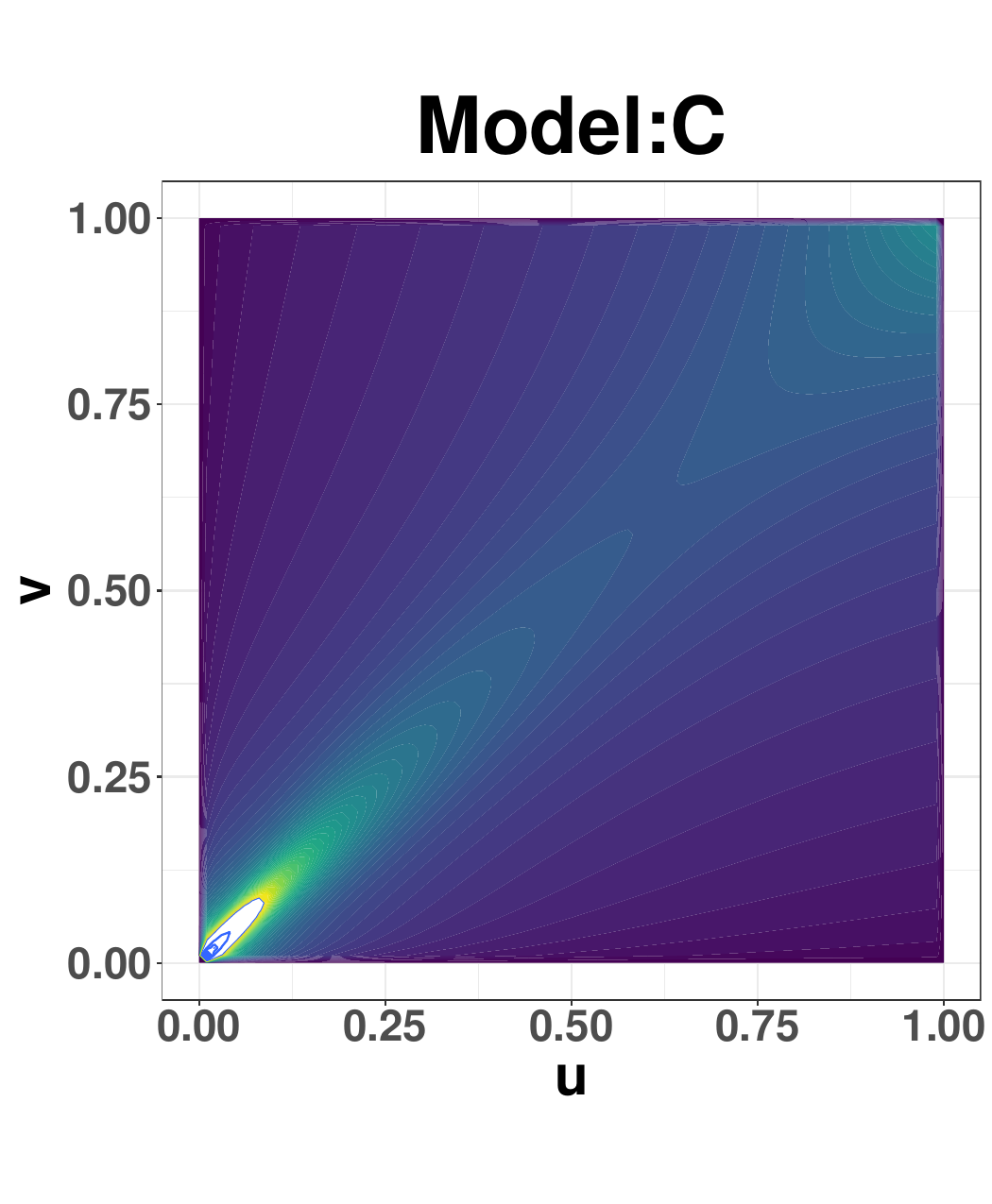}
}
\centerline{
  \includegraphics[scale=0.3]{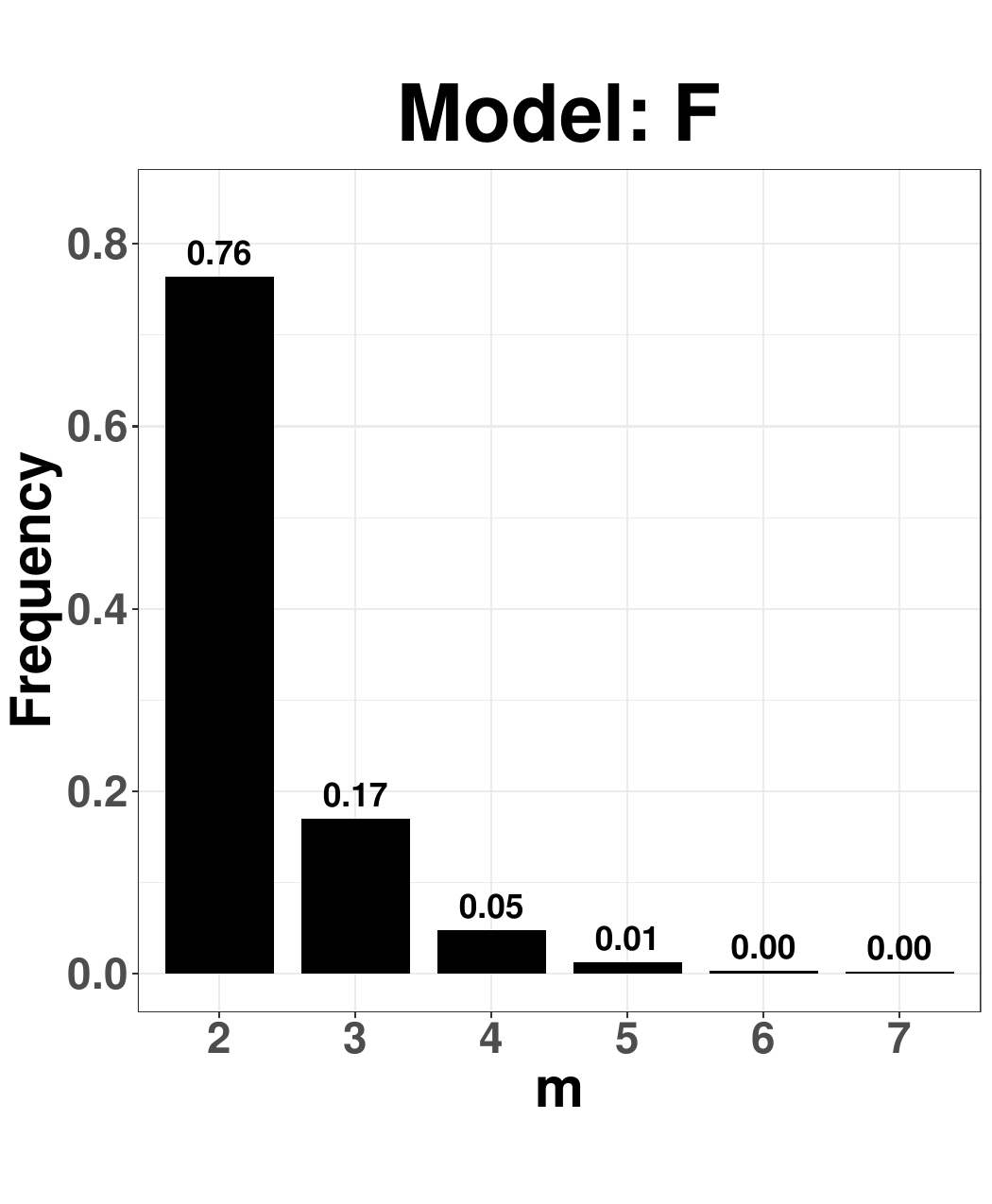}
  \includegraphics[scale=0.3]{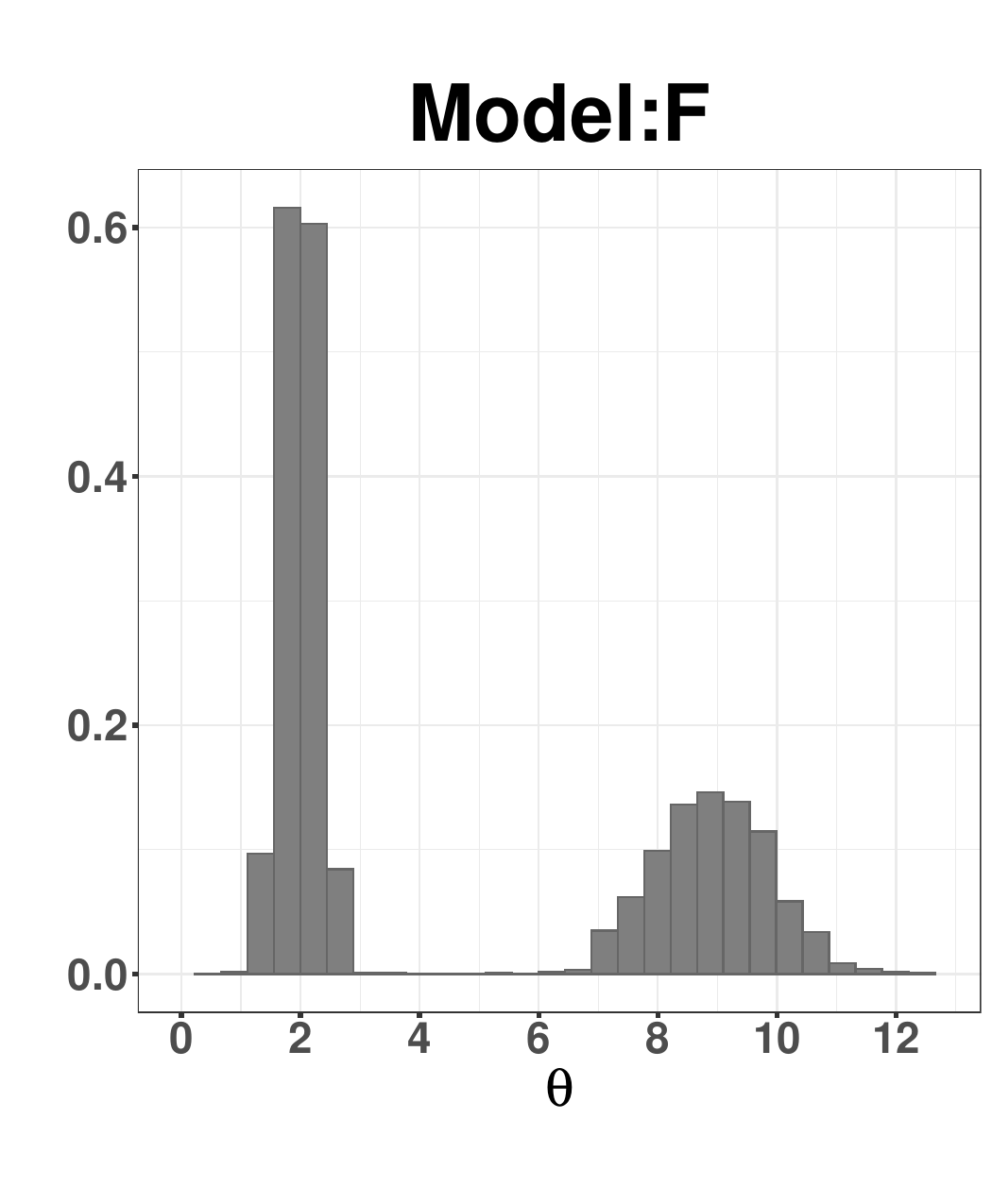}
  \includegraphics[scale=0.3]{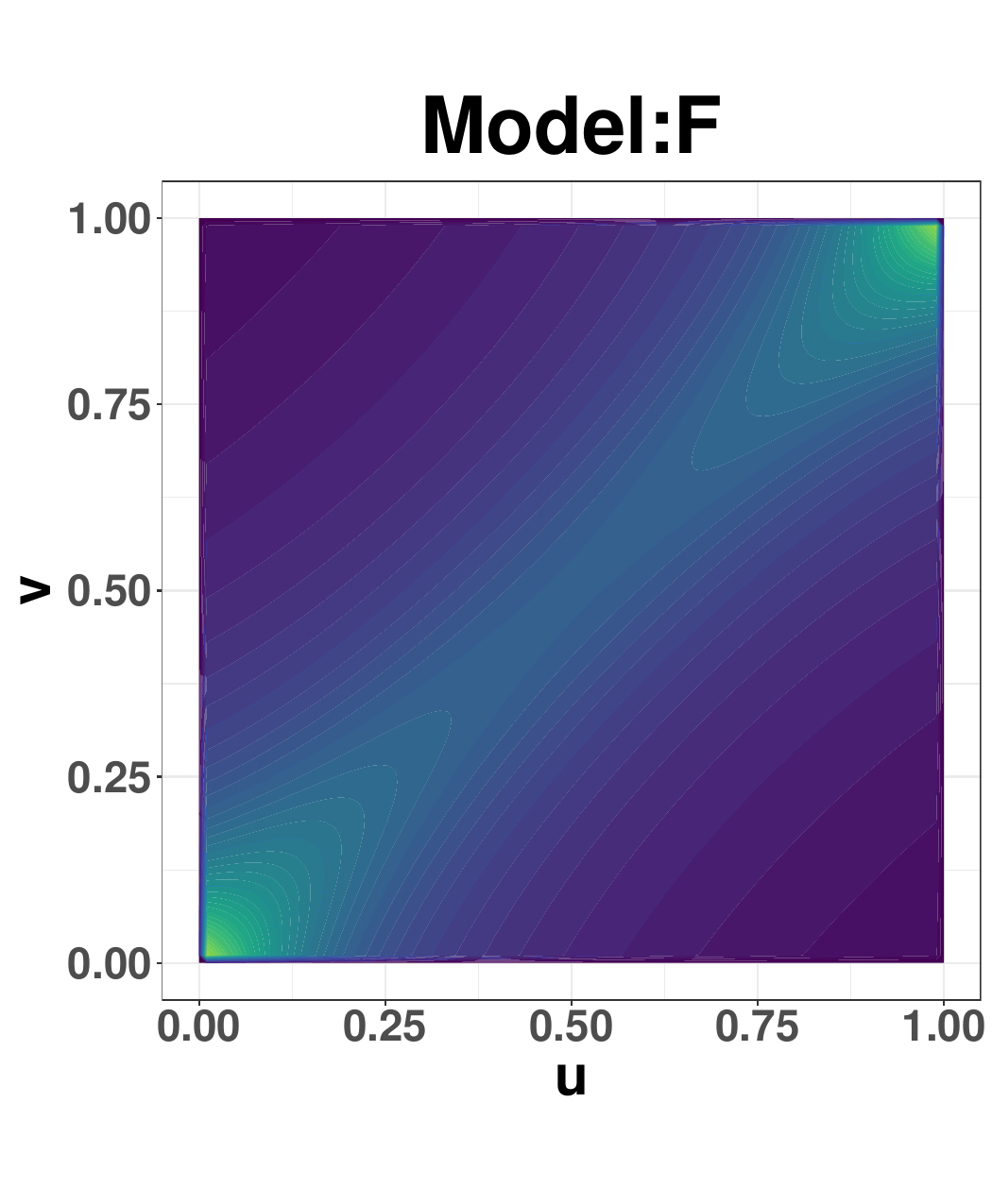}
}
\centerline{
  \includegraphics[scale=0.3]{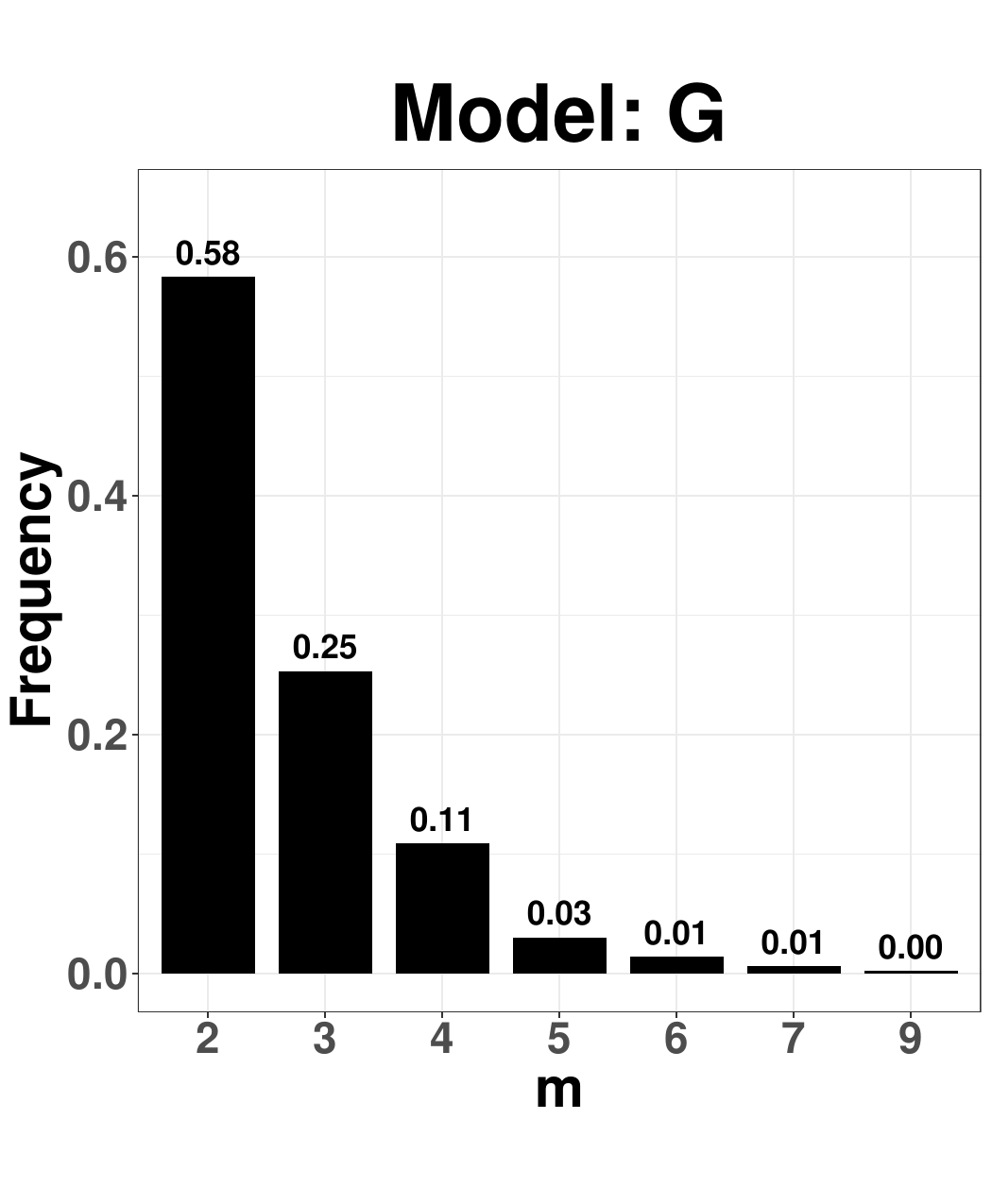}
  \includegraphics[scale=0.3]{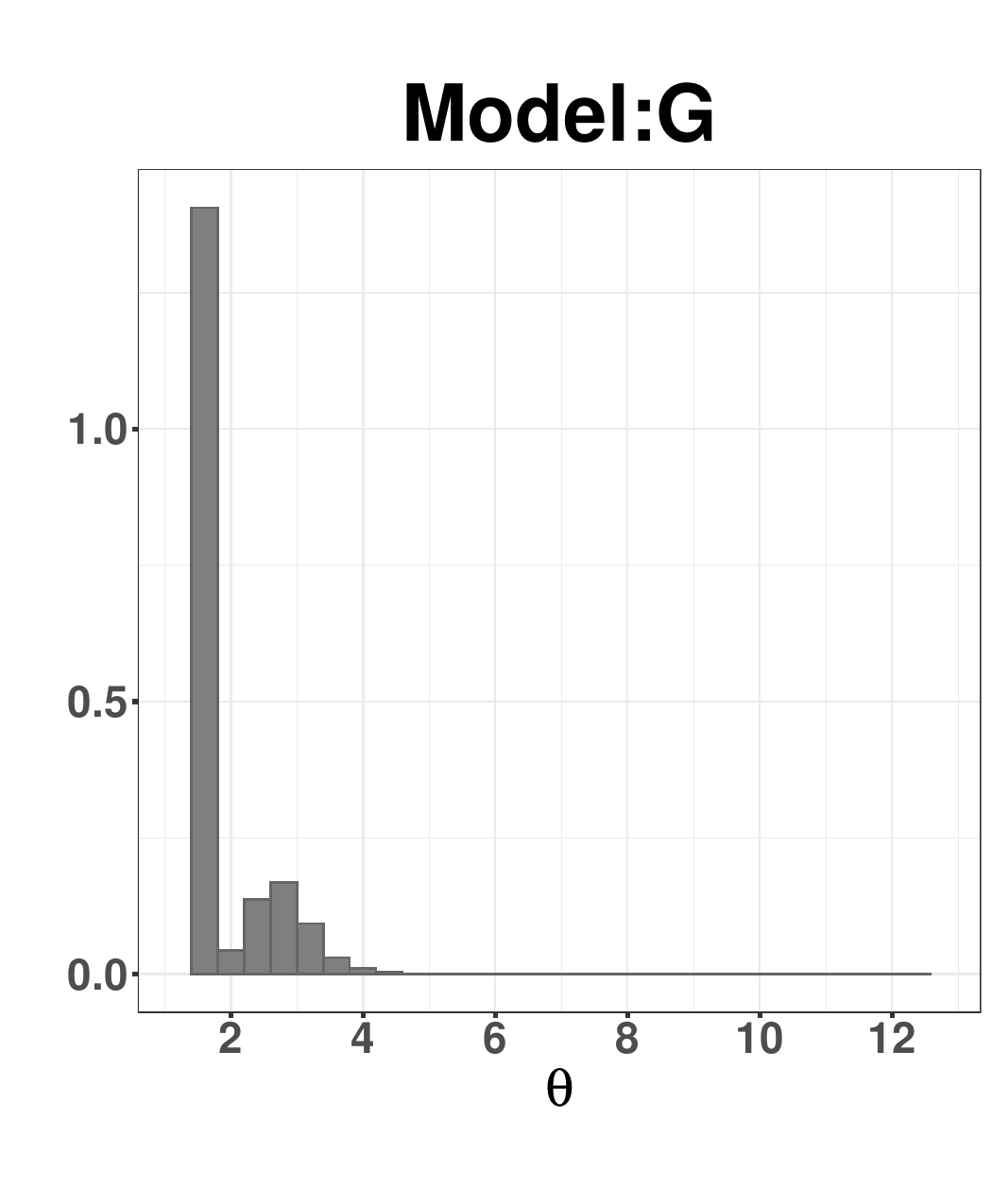}
  \includegraphics[scale=0.3]{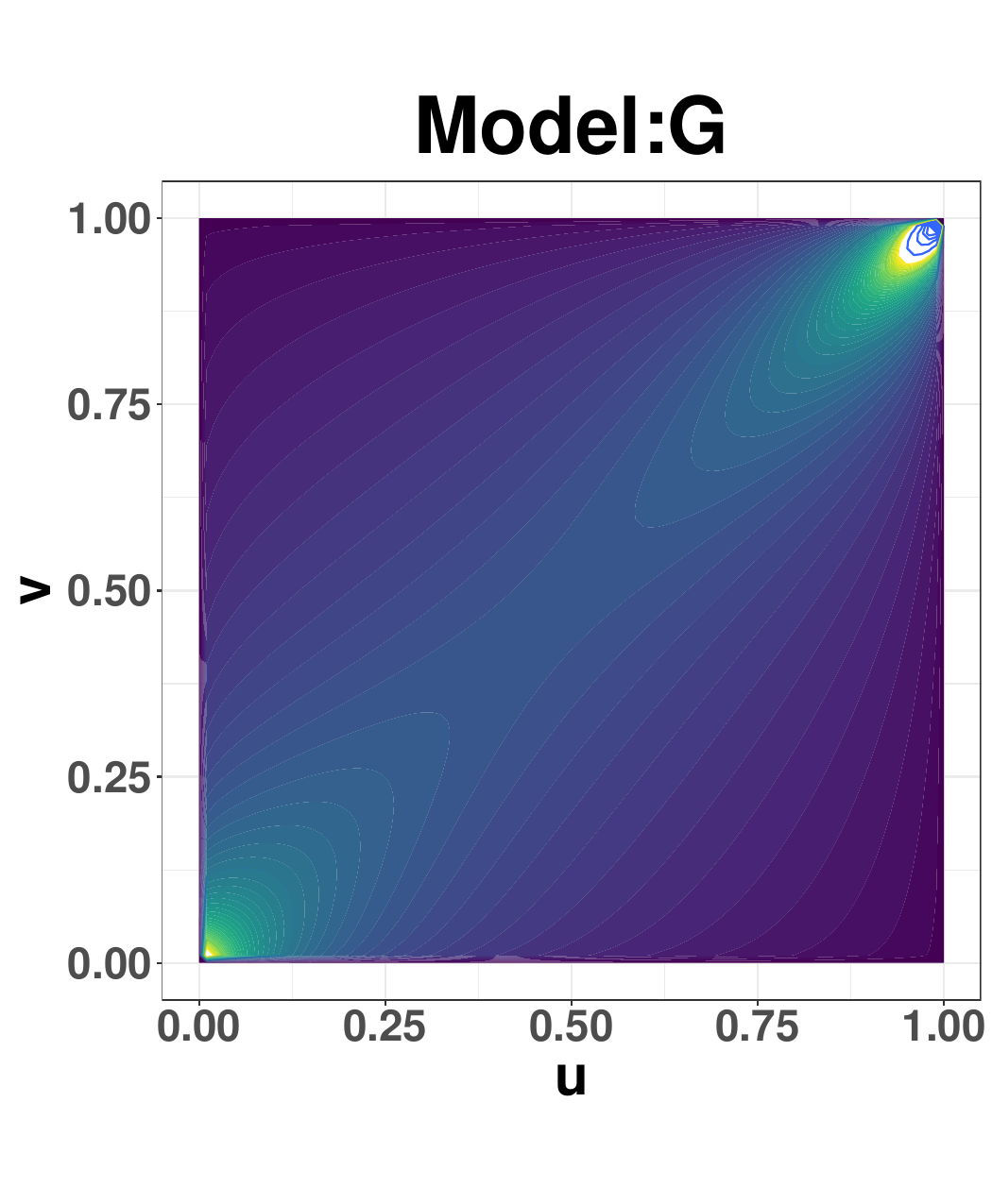}
}
\caption{Red wine data. Posterior distributions for the number of components, copula parameters, and bivariate density estimates. Estimated pairwise densities are the same for any pair of variables.}
\label{fig:postreal3}
\end{figure}

The posterior distributions for the number of components $m$, model parameters $\theta$ and bivariate density heat maps, for the three best fitting models, are shown in Figure \ref{fig:postreal3}. Note that dependence induced in an Archimedean copula is symmetric for any pair of variables; therefore, only one bivariate density estimate (heatmap) is reported. 

The posterior mode for the number of components is 3 for the Clayton, 2 for the Frank, and 2 for the Gumbel. The model parameters are estimated close to zero and slightly above 4 for the Clayton, around 2 and 9 for the Frank, and close to zero and slightly above 2 for the Gumbel. Bivariate densities are very similar with the three models, but with a strong lower left-tail dependence characteristic of the Clayton and an upper right-tail dependence in the Gumbel. Looking at the dispersion diagrams of the data, the upper right-tail dependence of the Gumbel seems to be more appropriate and is also supported by the LPML values. 

The posterior estimation of Kendall tau for the best-fitting model, Gumbel, is $0.361$  with a 95\% CI of $(0.24,0.70)$. Clearly, the CI contains the three empirical Kendall tau values that correspond to pairwise association parameters among the three characteristics of the wine. Implementing our clustering selection procedure, we obtain two groups with copula parameters estimated at: $1.52$ with a 95\% CI $(1.48,1.56)$ and weight $0.98$; and at $6.08$ with 95\% CI $(4.94,7.41)$ and weight $0.02$.

\section{Concluding remarks}
\label{sec:conclusion}

We propose Bayesian inference for a nonparametric mixture model of Archimedean copulas. Our model depends on the multivariate Archimedean copula densities, which require as many derivatives as the dimension of the data. Depending on the specific Archimedean family, some of the required derivatives are more difficult to compute than others. For instance, derivatives for the Clayton, Gumbel and Frank families are comparatively straightforward to obtain using the R codes that are available at \textit{copula} R package by \cite{hofert&al:23} . 

The runtime to fit our model significantly depends on the number of clusters chosen in each iteration. This in turn depends on the size and the specific data. For 2 to 4 clusters and with approximately 200 observations, in a 2-dimensional space, and over 15,000 iterations, the computation completes in 50 minutes. The computer specifications are: Intel i9 processor at 2.3 GHz with 8 cores and 32 GB of RAM. 

Our proposed model exhibited good performance in capturing dependence when the data were generated from an Archimedean or a mixture of Archimedean families; however, it fell short when the true copulas were outside the Archimedean family, like a mixture of Gaussian copulas. 

Due to the construction of the Archimedean copulas, pairwise Kendall tau coefficients for the elements in a vector are the same. To allow for different Kendall tau coefficients in different pairs of variables, an extension such as hierarchical Archimedean copulas could be used (\citet{li&al:21, hofert&pham:13}). We leave this for future work.

\section*{Acknowledgements}
We thank the editor, associate editor, and referees for their insightful comments that helped us improve the article. This work was supported by \textit{Asociaci\'on Mexicana de Cultura, A.C.} while the second author was visiting the Department of Statistical Sciences at the University of Toronto. The third author was supported by NSERC of Canada RGPIN-2024-04506 Discovery Grant.

\section*{Ethics declarations}
Conflict of interest: The authors declare no conflict of interest.

\section*{Supplementary material}
Code and data to replicate the numerical experiments presented here can be found at: https://github.com/RuyiPan/BNP-CopulaMixtures/tree/main

\bibliographystyle{natbib}

\newpage

\begin{table}
\centering
\begin{tabular}{cccc} \hline \hline \\[-3mm]
Copula & $\phi_\theta(t)$ & $\Theta$ & $\tau_\theta$ \\[1mm]  \hline \\[-3mm]
AMH & $\log\left\{\frac{1-\theta(1-t)}{t}\right\}$ & $[-1,1]$ & $1-2\{\theta+(1-\theta^2)\log(1-\theta)\}/(3\theta^2)$ \\[2mm]
CLA & $t^{-\theta}-1$ & $[-1,\infty)$ & $\theta/(\theta+2)$ \\[2mm]
FRA & $-\log\left\{\frac{\exp(-\theta t)-1}{\exp(-\theta)-1}\right\}$ & $\RB$ & $1-4\{1-D_1(\theta)\}/\theta$ \\[2mm]
GUM & $(-\log t)^{\theta}$ & $[1,\infty)$ & $1-1/\theta$ \\[2mm]
JOE & $-\log\left\{1-(1-t)^\theta\right\}$ & $[0.238734,\infty)$ & $1-4\sum_{k=1}^\infty 1/\{k(\theta k+2)\{\theta(k-1)+2\}\}$ \\[2mm]
\hline \hline
\end{tabular}
\caption{Most common Archimedean families and properties. Copula generator $\phi_{\theta}(t)$, parameter space $\Theta$ and Kendall tau $\tau_\theta$. The Debye function of order one is given by $D_1(\theta)=(1/\theta)\int_0^\theta t/(e^t-1)\d t$.}
\label{tab:afam}
\end{table}

\begin{table}
\centering
\begin{tabular}{ccc} \hline \hline \\[-3mm]
Copula & $\phi_\theta^{(1)}(t)$ & $\phi_\theta^{(2)}(t)$ \\[1mm]  \hline \\[-3mm]
AMH & $\frac{\theta-1}{t\{1-\theta(1-t)\}}$ & $\frac{(1-\theta)(1-\theta+2\theta t)}{\left\{t(1-\theta(1-t))\right\}^2}$ \\[2mm]
CLA & $-\theta t^{-(\theta+1)}$ & $\theta(\theta+1) t^{-(\theta+2)}$ \\[2mm]
FRA & $\frac{\theta\exp(-\theta t)}{\exp(-\theta t) - 1}$ & $\frac{\theta^2\exp(-\theta t)}{\{\exp(-\theta t) -1\}^2}$ \\[2mm]
GUM & $-\frac{\theta}{t}(-\log t)^{\theta-1}$ & $\frac{\theta}{t^2} (-\log t)^{\theta -1}+\frac{\theta(\theta-1)}{t^2}(-\log t)^{\theta -2}$ \\[2mm]
JOE & $\frac{-\theta(1-t)^{\theta-1}}{1-(1-t)^\theta}$ & $\frac{\theta(\theta-1)(1-t)^{\theta-2}+\theta(1-t)^{2\theta-2}}{\{1-(1-t)^\theta\}^2}$ \\[2mm]
\hline \hline
\end{tabular}
\caption{First and second derivatives of the generator for the five most common Archimedean families.}
\label{tab:derivatives}
\end{table}

\begin{table}
\centering
\begin{tabular}{ccccccc} \hline\hline \\[-3mm]
Data / Model & AMH & CLA & FRA & GUM & JOE & BSA \\[1mm] \hline \\[-3mm]
AMH & 0.34 & -0.08 & 0.02 & -1.07 & \textbf{2.48} & -4.32 \\[1mm] 
CLA & 14.75 & \textbf{147.42} & 74.91 & 78.37 & 53.23 & 7.4\\[1mm] 
FRA & 0.13 & 7.11 & \textbf{10.00} & 1.62 & 3.09 & -3.01 \\[1mm] 
GUM & 77.85 & 199.23 & 238.08 & \textbf{276.27} & 270.27 & 110.14\\[1mm] 
JOE & 37.76 & 75.85 & 98.33 & 118.10 & \textbf{120.91} & 55.33\\[1mm] 
\hline\hline
\end{tabular}
\caption{Simulation study 1: Bivariate data from a mixture of Archimedean copulas. LPML statistics when taking a sample of size $n=200$ and fitting the five models. Bayesian semiparametric Archimedean copula competing model in the last column.} 
\label{tab:lpml200}
\end{table}

\begin{table}
\centering
\begin{tabular}{ccccccc} \hline\hline \\[-3mm]
Data / Model & AMH & CLA & FRA & GUM & JOE & BSA \\[1mm] \hline \\[-3mm]
AMH &-0.52 & 0.51  &-1.00  &-3.81  & \textbf{2.36} & -4.01 \\[1mm] 
CLA & 29.25 & \textbf{352.86} & 203.95 & 164.52 & 141.49 & 1.83 \\[1mm] 
FRA & 2.04 & 2.69 & \textbf{22.63} & -2.17 & 22.67 & 6.59 \\[1mm] 
GUM & 226.44 & 586.62 & 647.14 & \textbf{715.18} & 671.68 & 172.86 \\[1mm] 
JOE & 95.57 & 209.54 & 230.50 & 290.03 & \textbf{296.43} & 97.60 \\[1mm] 
\hline\hline
\end{tabular}
\caption{Simulation study 1: Bivariate data from a mixture of Archimedean copulas. LPML statistics when taking a sample of size $n=500$ and fitting the five models. Bayesian semiparametric Archimedean copula competing model in the last column.} 
\label{tab:lpml500}
\end{table}

\begin{table}
\centering
\begin{tabular}{cccrrrrr} \hline\hline \\[-3mm]
Data / Model & True & Emp & AMH & CLA & FRA & GUM & JOE \\ 
\hline \\[-3mm]
AMH & 0.04 & 0.01 & 0.02  & 0.04 & 0.02 & 0.05 & -0.04 \\[1mm]
CLA & 0.25 & 0.14 & 0.12 & 0.22 & 0.22 & 0.33 & 0.19\\[1mm]
FRA & 0.00 & -0.07 & -0.05 & -0.01 & -0.07 & 0.05 & -0.08\\[1mm]
GUM &0.85 & 0.84 & 0.33 & 0.78 & 0.83 & 0.85 & 0.81\\[1mm]
JOE &  0.59 & 0.54 & 0.30 & 0.48 & 0.58 & 0.61 & 0.59\\[1mm]
\hline\hline
\end{tabular}
\caption{Simulation study 1: Bivariate data from a mixture of Archimedean copulas. True, empirical and estimated Kendall's tau values with $n=500$.} 
\label{tab:simktau}
\end{table}

\begin{table}
\centering
\begin{tabular}{rrr} \hline\hline \\[-3mm]
& $\theta_1$ & $\theta_2$ \\[1mm] \hline \\[-3mm]
 \multicolumn{3}{l}{\textbf{Model: AMH (-0.8, 0.8)}}\\   
 mean & 0.066\\
  $q_{2.5}$ & -0.237 \\ 
  $q_{97.5}$ & 0.328\\
  weight & 1  \\[1mm] \hline \\[-3mm]
 \multicolumn{3}{l}{\textbf{Model: CLA (-0.5, 10)}}\\ 
 mean & 11.108 & -0.502\\
  $q_{2.5}$ & 9.896 & -0.504\\
  $q_{97.5}$ & 12.329 & -0.496\\
  weight & 0.514 & 0.486\\[1mm] \hline \\[-3mm]
   \multicolumn{3}{l}{\textbf{Model: FRA (-5, 5)}}\\  
 mean &-5.996 & 5.956\\
  $q_{2.5}$ & -6.940 & 4.927\\
  $q_{97.5}$ &-5.089 & 7.017\\ 
  weight &0.568 & 0.432 \\[1mm] \hline \\[-3mm]
  \multicolumn{3}{l}{\textbf{Model: GUM (5, 10)}}\\ 
 mean &9.344 & 2.897\\
  $q_{2.5}$ & 8.641 & 2.482\\
  $q_{97.5}$ & 10.094 & 3.371\\
  weight & 0.822 & 0.178\\[1mm] \hline \\[-3mm]
  \multicolumn{3}{l}{\textbf{Model: JOE (2, 10)}}\\ 
 mean & 2.151 & 12.825\\
  $q_{2.5}$ & 1.907 & 11.383\\
  $q_{97.5}$ &2.425 & 14.239\\
  weight &0.542 & 0.458\\[1mm] 
  \hline \hline
\end{tabular}
\caption{Simulation study 1: Bivariate data from a mixture of Archimedean copulas ($n=500$). Post MCMC summaries given chosen cluster configuration.} 
\label{tab:postmcmc1}
\end{table}

\begin{table}
\centering
\begin{tabular}{c|ccc|ccc} \hline\hline \\[-3mm]
Model & $\widehat\tau$ & 95\% CI & LPML & $\widehat\tau$ & 95\% CI & LPML \\ \hline \\[-3mm]
AMH & 0.314 & (0.294, 0.330) & 113.12 & 0.019 & (-0.180, 0.329) & 8.16 \\[1mm]
CLA & 0.380 & (0.253, 0.723) & 112.96 &  0.011 & (-0.368, 0.494) & 39.25 \\[1mm]
FRA & 0.492 & (0.449 , 0.530) & 152.71 & -0.029 & (-0.661, 0.524) & 38.50 \\[1mm]
GUM & 0.474 & (0.436, 0.511) & 157.64 & 0.075 & (0.018, 0.633) & 4.17 \\[1mm]
JOE & 0.407 & (0.295, 0.705) & 130.83 & -0.036 & (-0.489, 0.640) & 21.27 \\[1mm]
\hline \\[-3mm]
GAUSS & 0.492 & (0.458, 0.522) & 162.93 & -0.030 & (-0.085, 0.035) & -1.35 \\[1mm]
\hline
\end{tabular}
\caption{Simulation study 2: Bivariate data from a single Gaussian (left) and a mixture of Gaussians (right). Posterior estimates for $\tau$ and LPML fit measures.} 
\label{tab:ss2}
\end{table}

\begin{table}
\centering
\begin{tabular}{cccc} \hline\hline
Setting & $\tau$ & $\widehat\tau$ & 95\% CI \\ \hline
1 & 0.722 & 0.719 & (0.300, 0.884)\\
2 & 0.731 & 0.731 & (0.451, 0.851) \\
3 & 0.775 & 0.780 & (0.456, 0.882) \\
\hline\hline
\end{tabular}
\caption{Simulation study 3: Multivariate data from a mixture of Archimedean copulas with 3 settings. Reported are true and estimated Kendall tau.} 
\label{tab:Kendall_tau}
\end{table}

\begin{table}
\centering
\begin{tabular}{rrrrrr} \hline\hline \\[-3mm]
 & $\theta_1$ & $\theta_2$ & $\theta_3$ & $\theta_4$ & $\theta_5$ \\[1mm] \hline \\[-3mm]
\multicolumn{6}{l}{\textbf{Setting 1: $\bpi=(0.2,0.3,0.5), \btheta=(1,5,15)$}}\\
 mean & 15.049 & 4.270 & 0.773 & 7.157 & 3.515\\
  $q_{2.5}$ & 14.437 & 3.915 & 0.625 & 5.971 & 2.490\\
  $q_{97.5}$ & 15.664 & 4.602 & 0.927 & 8.416 & 4.532\\
  weight & 0.576 & 0.179 & 0.167 & 0.046 & 0.032\\[1mm] \hline \\[-3mm]
\multicolumn{6}{l}{\textbf{Setting 2: $\bpi=(0.2,0.3,0.5), \btheta=(2,5,10)$}}\\
 mean & 11.628 & 1.367 & 4.865 & 6.125 & 3.331\\
  $q_{2.5}$ & 11.140 & 1.161 & 4.381 & 5.471 & 2.938\\
  $q_{97.5}$ & 12.118 & 1.580 & 5.381 & 6.856 & 3.761\\
  weight & 0.544 & 0.141 & 0.112 & 0.105 & 0.098\\[1mm] \hline \\[-3mm]
\multicolumn{6}{l}{\textbf{Setting 3: $\bpi=(0.2,0.3,0.5), \btheta=(2,7,15)$}}\\
 mean & 14.585 & 1.525 & 4.522 & 7.242 & 5.755\\
  $q_{2.5}$ & 13.981 & 1.301 & 4.093 & 6.196 & 4.669\\
  $q_{97.5}$ & 15.199 & 1.736 & 4.945 & 8.299 & 6.819\\
  weight & 0.647 & 0.125 & 0.122 & 0.058 & 0.048\\[1mm] \hline \\[-3mm]
\end{tabular}
\caption{Simulation study 3: Multivariate data from a mixture of Archimedean copulas. Post MCMC summaries given chosen cluster configuration.} 
\label{tab:postmcmc3}
\end{table}

\begin{table}
\centering
\begin{tabular}{ccccr} \hline \hline \\[-3mm] 
 & $\theta_1$ & $\theta_2$ & $\theta_1$ \\[1mm] \hline \\[-3mm]
 Model & \multicolumn{2}{c}{FRA} & GUM \\
  mean & 8.60 & -13.92 & 2.05\\
  $q_{2.5}$ & 7.50 & -17.80 &  1.86\\
  $q_{97.5}$ &  9.78 & -10.38 & 2.24\\ 
  weight & 0.86 & 0.14 & 1 \\[1mm] \hline \hline
\end{tabular}
\caption{Occupancy data. Post MCMC summaries given chosen cluster configuration.} 
\label{tab:postoccup}
\end{table}

\newpage

\end{document}